\RequirePackage{rotating}
\documentclass[traditabstract, longauth]{aa} 

\usepackage{amssymb}
\usepackage{amsmath}
\usepackage{hyperref}
\usepackage{graphicx}
\usepackage{natbib}
\usepackage{booktabs}
\usepackage{caption}
\usepackage{placeins}
\usepackage[toc,page]{appendix} 
\usepackage{gensymb}

\usepackage{comment}
\usepackage{threeparttable}
\usepackage{multicol}
\usepackage{tabularx}
\usepackage{lscape}
\usepackage{rotating}

\usepackage{orcidlink}

\def\macs{MACS J0138$-$2155}
\def\hst{HST}
\def\jwst{JWST}

\def\glafic{\textsc{glafic}}
\def\GLEE{\textsc{GLEE}}
\def\lenstool{\textsc{Lenstool}}
\def\lenstoolone{\textsc{Lenstool I}}
\def\lenstooltwo{\textsc{Lenstool II}}
\def\zitrin{\textsc{Zitrin-analytic}}
\newcommand{\mrmartian}{\scalebox{0.8}{MrMARTIAN}}

\def\wslap{\textsc{WSLAP+}}

\usepackage{color}
\usepackage{xcolor}

\newcommand{\zd}{z_{\rm d}}
\newcommand{\zs}{z_{\rm s}}
\newcommand{\Hc}{H_0}
\newcommand{\Hcfc}{H_{0,{\rm fc}}}
\newcommand{\Om}{\Omega_{\rm m}}
\newcommand{\OL}{\Omega_{\rm \Lambda}}
\newcommand{\kmsMpc}{{\rm km}\,{\rm s}^{-1}\,{\rm Mpc}^{-1}}
\newcommand{\kms}{{\rm km}\,{\rm s}^{-1}}

\newcommand{\tbaEncore}{\Delta t_{\rm 1b,1a}}
\newcommand{\tcaEncore}{\Delta t_{\rm 1c,1a}}
\newcommand{\tdaEncore}{\Delta t_{\rm 1d,1a}}

\newcommand{\tbaRequiem}{\Delta t_{\rm 2b,2a}}
\newcommand{\tcaRequiem}{\Delta t_{\rm 2c,2a}}
\newcommand{\tdaRequiem}{\Delta t_{\rm 2d,2a}}

\newcommand{\modpar}{\boldsymbol{\eta}}
\newcommand{\chiimsq}{{\chi_{\rm im}^2}}
\newcommand{\chiimsqmin}{{\chi_{\rm im,min}^2}}
\newcommand{\chiimsqboost}{{\chi_{\rm im,boost}^2}}
\newcommand{\Ndof}{{N_{\rm dof}}}
\newcommand{\hoEncore}{66.9^{+11.2}_{-8.1}}
\newcommand{\tbaEncoreMeasured}{-39.8^{+3.9}_{-3.3}}
\newcommand{\magEncoreA}{-26.9^{+2.6}_{-2.5}}
\newcommand{\magEncoreB}{40.7^{+7.0}_{-6.9}}

\newcounter{figsave}

\begin{document}

\title{Cosmology with supernova Encore in the strong lensing cluster MACS J0138$-$2155}
\subtitle{Lens model comparison and $H_0$ measurement}

\titlerunning{Lens model comparison and $H_0$ from SN Encore}

\author{S.~H.~Suyu\inst{\ref{tum},\ref{mpa}}\thanks{suyu@mpa-garching.mpg.de}
        \orcidlink{0000-0001-5568-6052} \and A.~Acebron\inst{\ref{unican},\ref{inafmilano}}\orcidlink{0000-0003-3108-9039} 
        \and C.~Grillo\inst{\ref{unimi},\ref{inafmilano}}\orcidlink{0000-0002-5926-7143} \and P.~Bergamini\inst{\ref{unimi},\ref{inafbologna}}\orcidlink{0000-0003-1383-9414}
        \and
        G.~B.~Caminha\inst{\ref{tum},\ref{mpa}} \orcidlink{0000-0001-6052-3274}
        \and   
        S.~Cha\inst{\ref{yonsei}}\orcidlink{0000-0001-7148-6915} 
        \and
        J.~M.~Diego\inst{\ref{unican}}\orcidlink{0000-0001-9065-3926}
        \and
        S.~Ertl\inst{\ref{mpa},\ref{tum}}\orcidlink{0000-0002-5085-2143} 
        \and
        N.~Foo\inst{\ref{arizonastateuni}} 
        \and
        B.~L.~Frye\inst{\ref{arizona}}\orcidlink{0000-0003-1625-8009} 
        \and Y.~Fudamoto\inst{\ref{chiba},\ref{arizona}}\orcidlink{0000-0001-7440-8832} 
        \and G.~Granata\inst{\ref{unimi},\ref{uferrara},\ref{icg}}\orcidlink{0000-0002-9512-3788} 
        \and
        A.~Halkola\inst{\ref{tuusula}} 
        \and
        M.~J.~Jee\inst{\ref{yonsei},\ref{ucd}}\orcidlink{0000-0002-5751-3697}
        \and
        P.~S.~Kamieneski\inst{\ref{asu}} 
        \and
        {A.~M.~Koekemoer}\inst{\ref{stsci}}\orcidlink{0000-0002-6610-2048} 
        \and {A.~K.~Meena}\inst{\ref{BenGurion},\ref{IISc}}\orcidlink{0000-0002-7876-4321} 
        \and
        {A.~B.~Newman}\inst{\ref{carnegie}} 
        \and
        S.~Nishida\inst{\ref{chibaphys}} 
        \and
        M.~Oguri\inst{\ref{chiba},\ref{chibaphys}}\orcidlink{0000-0003-3484-399X} 
        \and P.~Rosati\inst{\ref{uferrara},\ref{inafbologna}}\orcidlink{0000-0002-6813-0632} 
        \and S.~Schuldt\inst{\ref{unimi},\ref{inafmilano}}\orcidlink{0000-0003-2497-6334} 
        \and
        {A.~Zitrin}\inst{\ref{BenGurion}}\orcidlink{0000-0002-0350-4488} 
        \and
        R.~Ca\~nameras\inst{\ref{aix}} 
        \and
        E.~E.~Hayes\inst{\ref{cambridge}}\orcidlink{0000-0003-3847-0780}
        \and
        C.~Larison\inst{\ref{stsci}}\orcidlink{0000-0003-2037-4619}
        \and        E.~Mamuzic\inst{\ref{mpa},\ref{tum}}
        \and
        M.~Millon\inst{\ref{eth}}\orcidlink{0000-0001-7051-497X}
        \and        
        J.~D.~R.~Pierel\inst{\ref{stsci},\ref{ef}}\orcidlink{0000-0002-2361-7201} 
        \and
        {L.~Tortorelli}\inst{\ref{LMU}}
        \and
        {H.~Wang}\inst{\ref{mpa},\ref{tum}}\orcidlink{0000-0002-1293-5503}
}

\institute{
    Technical University of Munich, TUM School of Natural Sciences, Physics Department,  James-Franck-Stra{\ss}e 1, 85748 Garching, Germany \label{tum}
    \and
        Max-Planck-Institut f{\"u}r Astrophysik, Karl-Schwarzschild Stra{\ss}e 1, 85748 Garching, Germany\\
    e-mail: \href{mailto:suyu@mpa-garching.mpg.de}{\tt suyu@mpa-garching.mpg.de} \label{mpa}
    \and 
        Instituto de F\'isica de Cantabria (CSIC-UC), Avda.~Los Castros s/n, 39005 Santander, Spain 
    \label{unican}
    \and
    INAF -- IASF Milano, via A.~Corti 12, I-20133 Milano, Italy
    \label{inafmilano}
    \and    Dipartimento di Fisica, Universit\`a  degli Studi di Milano, via Celoria 16, I-20133 Milano, Italy
    \label{unimi}
    \and
    INAF -- OAS, Osservatorio di Astrofisica e Scienza dello Spazio di Bologna, via Gobetti 93/3, I-40129 Bologna, Italy
    \label{inafbologna}
    \and
    Department of Astronomy, Yonsei University, 50 Yonsei-ro, Seoul 03722, Korea
    \label{yonsei}
    \and
    School of Earth and Space Exploration, Arizona State University, Tempe, AZ 85716, USA
    \label{arizonastateuni}
    \and
    Department of Astronomy/Steward Observatory, University of Arizona, 933 N. Cherry Avenue, Tucson, AZ 85721, USA
    \label{arizona}
    \and
    Center for Frontier Science, Chiba University, 1-33 Yayoi-cho, Inage-ku, Chiba 263-8522, Japan
    \label{chiba}
    \and
    Dipartimento di Fisica e Scienze della Terra, Università degli Studi di Ferrara, via Saragat 1, 44122 Ferrara, Italy\label{uferrara}
    \and
    Institute of Cosmology and Gravitation, University of Portsmouth, Burnaby Rd, Portsmouth PO1 3FX, UK \label{icg}
    \and
    Py\"orrekuja 5 A, 04300 Tuusula, Finland
    \label{tuusula}
    \and
    Department of Physics and Astronomy, University of California, Davis, One Shields Avenue, Davis, CA 95616, USA \label{ucd}
    \and
    School of Earth and Space Exploration, Arizona State University, PO Box 876004, Tempe, AZ 85287-6004, USA
    \label{asu}
    \and
    Space Telescope Science Institute, 3700 San Martin Drive, Baltimore, MD 21218, USA
    \label{stsci}
    \and
    Department of Physics, Ben-Gurion University of the Negev, P.O. Box 653, Be'er-Sheva 84105, Israel
    \label{BenGurion}
    \and
    Department of Physics, Indian Institute of Science, Bengaluru 560012, India
    \label{IISc}
    \and
    Observatories, Carnegie Science, Pasadena, CA 91101, USA
    \label{carnegie}
    \and
    Department of Physics, Graduate School of Science, Chiba University, 1-33 Yayoi-Cho, Inage-Ku, Chiba 263-8522, Japan 
    \label{chibaphys}
    \and
    Aix-Marseille Université, CNRS, CNES, LAM, Marseille, France
    \label{aix}
    \and
    Institute of Astronomy and Kavli Institute for Cosmology, University of Cambridge, Madingley Road, Cambridge CB3 0HA, UK
    \label{cambridge}
    \and
    Institute for Particle Physics and Astrophysics, ETH Zurich, Wolfgang-Pauli-Strasse 27, CH-8093 Zurich, Switzerland
    \label{eth}
    \and
    NASA Einstein Fellow \label{ef}
    \and
    University Observatory, Ludwig-Maximilians-Universit\"at M\"unchen, Scheinerstraße 1, 81679, M\"unchen, Germany
    \label{LMU}
}

\authorrunning{Suyu, Acebron, Grillo et al.}

\date{Received --; accepted --}

\abstract{Robust mass modeling of strong-lensing galaxy clusters is crucial for studying cosmology and galaxy evolution. We present and compare seven mass models of the galaxy cluster \macs\,, constructed using six independent modeling software programs, including parametric and free-form approaches. By conducting a blind analysis where all the mass-modeling teams constructed their models independently without exchanging results, we quantified uncertainties arising from modeling software and assumptions. \macs\ is unique as the only cluster found to strongly lens two supernovae (SNe), Requiem and Encore, from the same host galaxy at a redshift of $z=1.949$, providing an excellent probe of cosmology through time delays between their multiple images. Through the Hubble Space Telescope, James Webb Space Telescope, and Multi Unit Spectroscopic Explorer observations, we assembled high-quality data products, including  eight sets of ``gold" lensed-image systems consisting of 23 multiple images with secure spectroscopic redshifts. We further identified one ``silver" lensed-image system with a likely but nonsecure redshift measurement. By restricting ourselves to high-quality gold images, we obtain overall good consistency in the model predictions of the positions, magnifications, and time delays of the multiple images of SN Encore and SN Requiem -- especially from the teams whose models fit the observed image positions with $\chiimsq \leq 25$. 
We predict the next images of SNe Encore and Requiem to reappear with time delays $\gtrsim$$3000$\,days and $\sim$$3700$ to $4000$\,days, respectively, based on a fiducial cosmological model with $H_0=70\,\kmsMpc$ and $\Om = 1-\OL = 0.3$. 
By considering a range of hypothetical time-delay values with the same $\Om = 1-\OL = 0.3$, we obtain relations between $H_0$ and the time delays of SN Encore and SN Requiem. In particular, 
for $H_0=73\,\kmsMpc$, the four lowest $\chiimsq$ models forecast the next image of SN Requiem to appear approximately April-December 2026; for $H_0=67\,\kmsMpc$, they predict it to appear approximately March-November 2027 ($1\sigma$ uncertainties). Using the newly measured time delay between the two detected multiple images of SN Encore by Pierel et al.~(2026) and our mass modeling, we infer $H_0 = \hoEncore\,\kmsMpc$, where the uncertainty is dominated by that of the short time delay between the existing pair of images. The long time delays of the next-appearing SN Requiem and SN Encore images provide excellent opportunities to measure $H_0$ with 2-3\% uncertainty. Our mass models form the basis for cosmological inference from this unique lens cluster with two strongly lensed SNe.
}


\maketitle

\section{Introduction}
\label{sec:intro}

Strong gravitational lensing by massive galaxy clusters provides a powerful approach to address key astrophysical and cosmological questions. As nature’s cosmic telescopes, strong lensing clusters magnify and distort the light of distant background galaxies, enabling studies of high-redshift sources that are otherwise too faint to be detected at the same observational depth \citep[e.g.,][]{Lotz+2017, Coe2019, Salmon2020, Treu2022b, Bezanson2024}. Accurate mass modeling of these systems is critical, not only for reconstructing the statistical and individual properties of magnified background sources \citep[e.g.,][]{Diego2022, Bouwens+2022, Welch+2022, Atek+2023, Bergamini+2023, Vanzella2023, Vanzella2024, Claeyssens2025}, but also for leveraging lensing clusters as probes of their total mass distribution, both to study the nature of dark matter \citep[DM; e.g.,][]{Bradac+2008, Bartalucci2024} and to test theoretical models of structure formation \citep[e.g.,][]{Grillo+2015, Meneghetti+2020}. Strong lensing clusters with multiple, strongly lensed background galaxies at different redshifts allow us to infer the ratio of angular-diameter distances to the lensed background sources and thus probe the geometry and the constituents of the Universe \citep[e.g.,][]{Jullo2010, Acebron2017, Caminha+2022, Grillo+2024}.

Accurate mass models are especially crucial in strong lensing clusters used for time-delay cosmography, a method that exploits time delays between multiple images of variable background sources, such as quasars and supernovae (SNe), to measure the Hubble constant \citep[$H_0$; e.g.,][]{Refsdal1964, Treu+2022, Kelly+2023, Liu+2023, Martinez+2023, Napier+2023, Acebron+2024, Grillo+2024, Suyu+2024, Pascale+2025}.  This method is independent of various approaches to infer $H_0$, including local distance ladders using Type Ia SNe \citep[e.g.,][]{Riess+2022, Riess+2024, Freedman+2020, Freedman+2025}, the cosmic microwave background \citep[CMB;][]{Planck+2020}, expanding photospheres of Type IIP SNe \citep[e.g.,][]{Vogl+2025}, megamasers \citep[e.g.,][]{Pesce+2020}, and standard sirens from gravitational waves \citep[e.g.,][]{Abbott+2017}.  Time-delay cosmography therefore provides crucial insight into the Hubble tension reported between some of the late-Universe and early-Universe measurements \citep[see, e.g., recent reviews by][]{diValentino+2021, Verde+2024}. For a robust determination of $H_0$ through time-delay cosmography with strong lensing clusters, a rigorous quantification of systematic uncertainties in mass modeling is paramount.

In this paper, we present comprehensive mass modeling of the galaxy cluster \macs\ discovered as part of the MAssive Cluster Survey \citep[MACS;][]{Ebeling2001,Repp2018}. The cluster strongly lenses and highly magnifies a background galaxy at $z=1.949$ into giant arcs.  This lensing cluster is unique in having two SNe occurring in this host galaxy: SN Requiem was serendipitously discovered in Hubble Space Telescope (\hst) observations \citep{Rodney+2021}, and SN Encore in James Webb Space Telescope (\jwst) observations \citep{Pierel2024}. These two strongly lensed SNe therefore provide an excellent opportunity to measure the value of $H_0$ through time-delay cosmography, especially since future (time-delayed) images of both SN Encore and SN Requiem will appear and their detections will provide time-delay measurements with percent-level precision. Supernova Encore now has a first time-delay measurement from existing data \citep{Pierel+2026} and is the third strongly lensed SN that enables an $H_0$ measurement, after SN Refsdal \citep{Kelly+2015, Kelly+2023, Grillo+2024, LiuOguri+2025} and SN H0pe \citep{Frye+2024, Pascale+2025}. The new \jwst\ observations of \macs\ also provide a detailed and magnified view of the host galaxy of the SNe, enabling the study of its central supermassive black hole \citep{Newman+2025}.

To quantify potential systematic uncertainties stemming from mass modeling assumptions and choices, we constructed seven independent mass models using six different lens modeling software. We designed a controlled experiment and blind analysis to prevent potential experimenter bias in modeling results and uncertainties. Starting with the same photometric and spectroscopic input data vetted and presented in \citet{Ertl+2025} and \citet{Granata2025}, respectively, the seven modeling teams constructed their mass models completely independently without any exchanges on their results.  Only after all teams finalized their mass models did the teams proceed to unblind their results from one another.  In particular, we imposed a strict submission process requiring each team to submit their modeling results, especially the model-predicted positions, magnifications, and time delays of SN Encore and SN Requiem. The teams further provided additional auxiliary data such as Markov chain Monte Carlo (MCMC) samples of their mass models. The submissions of all teams' results were sent to coauthor A.B.N., who was not involved in any modeling teams and verified that the submissions conformed to the predefined format (e.g., coordinate system) before unblinding. 

We present here the results of the experiment and blind analysis, and we report the model predictions from each team. At the time of writing, the time delay(s) between the multiple images of SN Encore were being measured for the first time \citep{Pierel+2026} and kept secret from the modeling teams. After the time delay was measured, we used the mass models without modifications to measure the value of $H_0$. The seven independent lens mass-model predictions were unblinded only within the modeling teams on December 10, 2024 (and not shared with coauthors involved in the time-delay measurements). The subsequent cosmographic unblinding of both the time delays and the lens mass-model results took place on July 28, 2025. Sects.~\ref{sec:obs} to \ref{sec:H0-td} were written before cosmographic unblinding, and Sects.~\ref{sec:H0} and \ref{sec:SNeReappear} were written after unblinding. Our careful blind analysis enables us to provide a value of $H_0$ without experimenter bias, which is crucial for assessing the Hubble tension. Furthermore, our predictions of the future reappearance of SN Encore and SN Requiem instill in us with prescient knowledge to strategize future observations and catch their reappearance.  

The outline of the paper is as follows.  In Sect.~\ref{sec:obs}, we summarize the observational data.  In Sect.~\ref{sec:data}, we describe the data products used by the modeling teams to construct their mass models. In Sect.~\ref{sec:mass_model}, we present each of the seven independent mass models. In Sect.~\ref{sec:comparison}, we compare the model predictions of the properties of SN Encore and SN Requiem.  Based on these mass models, we obtain relations between $H_0$ and hypothetical values of the time delays of SN Encore and SN Requiem in Sect.~\ref{sec:H0-td}. In Sect.~\ref{sec:H0}, we present our measurement of $H_0$ using the new time-delay measurement by \citet{Pierel+2026}. In Sect.~\ref{sec:SNeReappear}, we present our forecast for the reappearance of SN Encore and SN Requiem. Sect.~\ref{sec:summary} provides a summary of our work.  Throughout this paper, right ascension and declination (RA, Dec) coordinate values are in the ICRS system. The values of $H_0$ for SN Encore are reported as the median with $1\sigma$ uncertainties given by the 16th and 84th percentiles of the probability distribution.

\section{Summary of observations}
\label{sec:obs}

\macs\ has been observed with multiple facilities \citep[e.g.,][]{Newman2018a, Newman2018b, Pierel2024}. In this section,  we summarize the imaging and spectroscopic observations used to construct our lens mass models.

\subsection{\hst\ and \jwst\ imaging}
\label{sec:obs:imaging}

\macs\ has been observed with the \hst\ in 2016 (Proposal ID 14496, PI: Newman), 2019 (Proposal ID 15663, PI: Akhshik), December 2023, and January 2024 (Proposal ID 16264, PI: Pierel). Supernova Requiem was discovered by \citet{Rodney+2021} in 2016 HST imaging after they noticed the absence of the SN multiple images in the 2019 imaging. When SN Encore was discovered in November 2023, we triggered the HST observations in 2023 and 2024 \citep{Pierel2024}, to obtain the light curves of the three multiple images of SN Encore.  

The first \jwst\ observations were taken in November 2023 (Proposal ID 2345, PI: Newman) from which SN Encore was discovered \citep{Pierel2024}. Subsequently, we acquired \jwst\ follow-up observations through Director's Discretionary Time (Proposal ID 6549, PI: Pierel), obtaining three additional epochs of \jwst\ images in six filters.

\begin{figure*}
        \centering
        \includegraphics[width=0.8\linewidth, trim=0cm 0cm 0cm 0cm, clip]{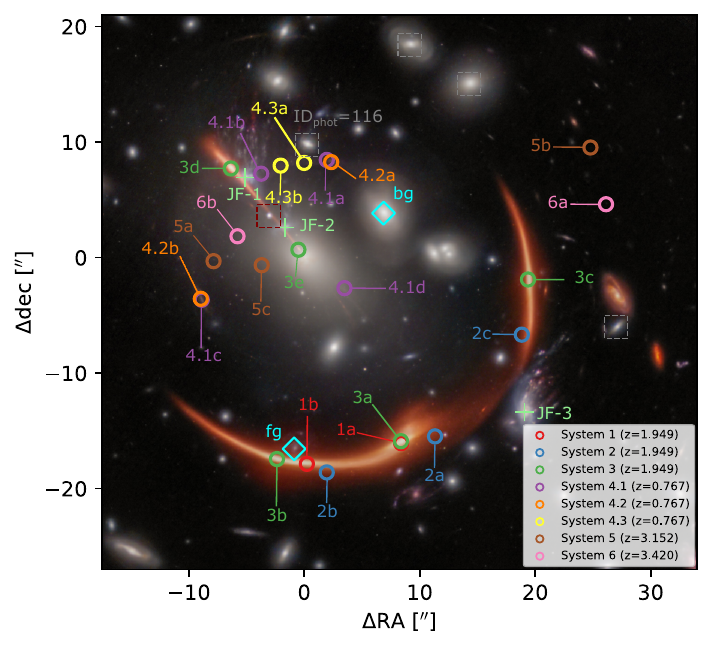}
        \caption{\macs\ as observed through \hst\ and \jwst\ with the following combinations of filters for the color image:
        F105W+F115W+F125W (blue), F150W+F160W+F200W (green), and
F277W+F356W+F444W (red). 
          The positions of the ``gold'' multiple-image systems are shown with circles. SN Encore is System 1, and SN Requiem is System 2. The foreground (fg) and background (bg) galaxies are marked by cyan diamonds. The dashed squares identify some of the freely optimized cluster members in the lens model, i.e., not constrained by the scaling relations for the cluster members (gray for the \zitrin\ lens model, maroon for the \lenstool\ I, and that labeled with $\rm ID_{\rm phot}=116$ for the \GLEE\ and \zitrin\ models). The three jellyfish galaxies are labeled with light-green crosses as JF-1, JF-2, and JF-3. }
        \label{fig:jwst_image}
\end{figure*}

For our lens mass modeling, we used the full-depth combined mosaics of all the data in the five \hst\ and six \jwst\ filters listed in Table 3 of \citet{Ertl+2025}. These mosaics were produced according to the techniques first described by \citet{Koekemoer2011}, updated for the JWST pipeline, with further details about these data presented in \citet{Pierel2024} and \citet{Acebron2025}.  In Fig.~\ref{fig:jwst_image}, we show a color image of \macs, with the multiple image positions of SN Encore and SN Requiem marked by systems 1 and 2, respectively, along with additional multiple-image systems.

\subsection{\jwst\ and MUSE spectroscopy}
\label{sec:obs:spec}

Supernova Encore and its host galaxy were observed with \jwst\ NIRSpec integral field unit (IFU) for one of the multiple images, namely, 1a and 3a, marked in Fig.~\ref{fig:jwst_image} (Proposal ID 2345, PI: Newman).  With the spectra taken at two epochs, \citet{Dhawan2024} showed that SN Encore, at $\zs = 1.949$, is a Type Ia SN similar to local Type Ia SNe in terms of its spectral features, despite its high redshift.

\macs\ was also observed at the Very Large Telescope of the European Southern Observatory using the integral-field spectrograph Multi Unit Spectroscopic Explorer  \citep[MUSE;][]{Bacon2010} through two programs.  Program ID 0103.A-0777 (September 2019, PI: Edge) targeted the central $\sim$1'$\times$1' field of view (FoV) of the galaxy cluster for nearly 1 hour.  Upon the discovery of SN Encore, our team acquired additionally 2.9 hours of observations with the MUSE IFU in December 2023 (Program ID 110.23PS, PI: Suyu). The data from both programs were reduced and analyzed by \citet{Granata2025}, who measured 107 reliable spectroscopic redshifts of galaxies in the cluster FoV, including lensed background galaxies.

\section{Data products}
\label{sec:data}

Building a mass model requires the identification of (1) galaxies that are members of the cluster, (2) line-of-sight (LOS) galaxies that are not part of the cluster but are sufficiently massive to produce significant lensing effects, and (3) background sources that are strongly lensed into multiple images by the foreground cluster and LOS galaxies. Spectroscopic redshifts are also needed for accurate mass models. We summarize the work of \citet{Granata2025} and \citet{Ertl+2025} to obtain the ingredients (1) and (2) in Sect.~\ref{sec:data:cluster_members} and \ref{sec:data:LOS}.  We describe our process of defining the sample of multiple images in Sect.~\ref{sec:data:multiple_images}, which provides crucial constraints on the lens mass models.  

\subsection{Galaxy cluster members}
\label{sec:data:cluster_members}

Through the spectroscopic analysis in \citet{Granata2025} and the photometric analysis in \citet{Ertl+2025} of galaxies in the cluster FoV, we identified 84 galaxy cluster members, including 50 that are spectroscopically confirmed (providing a cluster redshift value of $z_{\rm d}=0.336$). Three of the cluster members are jellyfish galaxies \citep[e.g.,][]{Ebeling+2014, Poggianti+2025, Gibson+2025}, which are highlighted in light green in Fig.~\ref{fig:jwst_image} and labeled as JF-1, JF-2, and JF-3. \citet{Granata2025} measured the stellar velocity dispersions for 14 early-type cluster members, which have sufficiently high spectral signal-to-noise ratio, including the brightest cluster galaxy (BCG).  These velocity dispersion measurements can then be used to calibrate the Faber-Jackson relation for the cluster members, relating their observed magnitudes to velocity dispersions (and thus total masses).  The photometric and velocity-dispersion catalogs of the cluster members are released by \citet{Ertl+2025} and \citet{Granata2025}, respectively.

\subsection{Significant line-of-sight galaxies}
\label{sec:data:LOS}

Two LOS galaxies, both with spectroscopic redshift measurements from \citet{Granata2025}, were explicitly included in the lens mass models. One is a foreground galaxy, labeled as `fg' hereafter, located angularly close to the southern giant arc with (RA, Dec) = (24\fdg5159632, $-$21\fdg9300802) and a redshift of $z_{\rm fg}=0.309$.  The other is a massive background galaxy, labeled as `bg' hereafter, about 8\arcsec\ northwest of the BCG at (RA, Dec) = (24\fdg5136457, $-$21\fdg9244117) and $z_{\rm bg}=0.371$. These two objects are marked in Fig.~\ref{fig:jwst_image}.

\subsection{Strongly lensed multiple images}
\label{sec:data:multiple_images}

Accurate identifications of multiple images are crucial, since wrong identifications introduce biases in the lens mass models. All modeling teams collectively identified the multiple-image systems to ensure both robustness and completeness, given the existing data. We had a two-step procedure: (1) we proposed and reached consensus on the positions of the secure multiple-image systems, and (2) we determined the associated uncertainties for these secure multiple-image systems.  

For step (1), we first distributed the \hst\ and \jwst\ images as well as the photometric and spectroscopic catalogs to all the modeling teams. By building preliminary mass models using these data, each team proposed candidates for strongly lensed systems by marking the candidate positions of multiple images that belong to the same system (of the same background source position). In total, the teams collectively proposed approximately 36 candidate systems with a few more candidate systems whose multiple images overlap partly with some of those in the 36 systems. To determine whether a candidate system is robust, we proceeded by voting. Each team voted on every candidate system to indicate whether it thinks the candidate is a secure strongly lensed system.  If a team considered a candidate system secure, it then voted on the proposed positions of that system's multiple images (various teams sometimes proposed slightly different positions for the same candidate-lensed image, and in the voting round, each team voted for one among the proposals). At the time of voting, there were eight independent modeling teams (two of the teams later merged into a single team). Therefore, the maximum number of positive votes a candidate lensed-image position could have was eight, since each team cast one vote.

We defined the following criteria for ``gold'' and ``silver'' multiple images, where we consider ``gold'' as secure and ``silver'' as likely.  For a multiple image and its proposed position to be classified as ``gold,'' it required seven or eight votes in total and had to belong to a multiple-image system with at least one spectroscopically confirmed multiple image. For a multiple image to be classified as ``silver,'' it needed to have six to eight votes, without requiring spectroscopic redshift.  

Based on the votes, we identified eight gold systems with a total of 23 multiple images, as shown in Fig.~\ref{fig:jwst_image}.  We refer to \citet{Ertl+2025} for more details about these gold systems. We further identified one silver system corresponding to the spectroscopic ID 1755 \citep{Granata2025}, with a ``likely'' rather than ``secure'' redshift measurement from two blended background sources. The silver system has three multiple images, and we refer to Appendix \ref{app:silver_data} for more details.  

After establishing the multiple image positions of the gold and silver systems, we proceeded to step (2) to estimate their positional uncertainties. The teams agreed to adopt elliptical uncertainties based on the shape of the observed lensed images. The uncertainties typically extend $\sim$1-3 pixels, in either the \jwst\ or MUSE data, depending on which dataset the lensed system is detected in. The positions of the spatially extended and highly magnified multiple images (e.g., multiple images 3a and 3b) are harder to pinpoint and thus have larger uncertainties (in terms of the number of pixels) compared to those of point-like multiple images such as SN Encore (images 1a and 1b).  

In summary, the 23 gold multiple-image positions from the eight systems are listed in Table 4 of \citet{Ertl+2025} and shown in Fig.~\ref{fig:jwst_image}. The three multiple images of the single silver system are presented in Table \ref{tab:silver_pos} and Fig.~\ref{fig:silver_sys}.  The lensed background sources span a redshift range between 0.767 to 3.420.

\section{Lens mass models}
\label{sec:mass_model}

In this section, we summarize the seven modeling approaches using six different software with the same data as input. The modeling approaches are divided into two generic categories, parametric and free-form total mass distributions.  
The five parametric models from four different software are presented in Sect.~\ref{sec:mass_model:glafic} to Sect.~\ref{sec:mass_model:zitrin}, and the two hybrid free-form models used in this work are described in  Sects.~\ref{sec:mass_model:mars} and \ref{sec:mass_model:wslap}. 
We summarize in Table \ref{tab:software} the requirements and capabilities of each software. All can model single-plane lenses with circular positional uncertainties for the multiple images. This setup consequently forms a \texttt{baseline} model for direct comparison of the software and approaches. All models adopted a flat $\Lambda$CDM cosmology\footnote{The current standard cosmological model that is spatially flat and consists of dark energy described by the cosmological constant $\Lambda$ and cold dark matter (CDM).} with $H_0=70\,\kmsMpc$ and $\Om=1-\OL=0.3$ for predicting the multiple image positions, magnifications, and time delays, unless otherwise stated.

The seven modeling teams were each invited to build two models: (1) a \texttt{baseline} model to facilitate direct comparison between different models, and, optionally, (2) their ultimate model, representing the team's most suitable mass model for cosmographic inference with SN Encore. For the \texttt{baseline} model, all the mass components (cluster-scale halos, sub-halos, and LOS galaxies) were placed at the cluster redshift of $\zd=0.336$, and the positional uncertainties were assumed to be circular. The size of the circular uncertainty ($\sigma^{\rm circ}$) is the geometric mean of the semimajor ($\sigma^{\rm major}$) and semiminor ($\sigma^{\rm minor}$) axes of the elliptical uncertainty tabulated in Table 4 of \citet{Ertl+2025} and in Table~\ref{tab:silver_pos}, i.e., $\sigma^{\rm circ} = \sqrt{\sigma^{\rm major} \sigma^{\rm minor}}$. In the end, six of the seven teams used their \texttt{baseline} model as their ultimate model. Only the \GLEE\ team built an ultimate model in addition to their \texttt{baseline} model.

\paragraph*{Parametric models.}
In parametric models, the cluster total mass distribution is often separated into multiple mass components (such as cluster-scale DM halos and galaxies) with each parameterized by a functional form with several parameters.  
Depending on the parametric model presented in this work, the cluster-scale mass component of \macs\ (mainly in the form of DM) is parameterized with one or two mass distributions where the mass density profile was chosen by each team. The mass density profile of individual cluster members is a 
dual pseudo-isothermal elliptical (dPIE) mass distribution -- a truncated isothermal profile \citep[e.g.,][]{Eliasdottir+2007, SuyuHalkola2010} with velocity dispersion, $\sigma_{{\rm v}, i}$, and truncation radius, $r_{{\rm cut},i}$, as free parameters for the $i$th cluster member (centroids, axis ratios, and position angles were either varied or fixed to those of the galaxy light distribution, as specified in each mass model's subsection). 
Some teams adopted circular (instead of elliptical) convergence profiles for the galaxies in their models, and we use the same nomenclature of dPIE to refer to the mass profiles of the galaxies, irrespective of their ellipticity. 
Scaling relations between the cluster members' total masses, $M_{{\rm tot},i}\propto\sigma^2_{{\rm v},i}r_{{\rm cut},i}$, and their associated luminosity, $L_i$, enable significant reduction in the number of free parameters. Specifically, the following two scaling relations are generally adopted:
\begin{equation}
\label{eq:fj}
       \sigma_{{\rm v},i} = \sigma_{\rm v}^\mathrm{ref} \left(\frac{L_i}{L_\mathrm{ref}}\right)^{\alpha}, \,
       r_{{\rm cut},i} = r_{\rm cut}^\mathrm{ref} \left(\frac{L_i}{L_\mathrm{ref}}\right)^{\beta},
\end{equation}
where $L_\mathrm{ref}$ refers to the reference luminosity of a galaxy at the cluster redshift. The quantities $\sigma_{\rm v}^\mathrm{ref}$ and $r_{\rm cut}^\mathrm{ref}$ were optimized in the lens modeling, and $\alpha$ and $\beta$ represent the slopes of these scaling relations, respectively. The slope of the total mass-to-light ratio was then obtained as $\beta +2\alpha -1$. The values chosen for the two slopes then determine whether the cluster members have total mass-to-light ratios that are constant or that increase with their luminosity.

\paragraph*{Free-form and hybrid models.}
In contrast to parametric methods, where the mass distribution of the lens is constructed as a sum of analytic halos, whose positions and properties are usually guided by the visible components, the free-form method models the lens by dividing the lens plane into $N_{\rm c}$ small cells without such informed priors. The deflection field $\vec{\alpha}(\vec{\theta})$ is evaluated at the position $\vec{\theta}$ by summing the contributions from all $N_{\rm c}$ cells as follows:
\begin{equation}
\vec{\alpha}(\vec{\theta}) = \frac{4 G}{c^2} \frac{D_{\rm ds}}{D_{\rm s} D_{\rm d}} \sum_{i}^{N_{\rm c}} m_i(\vec{\theta}_i)\frac{\vec{\theta} - \vec{\theta}_i}{| \vec{\theta} - \vec{\theta}_i |^2},
\label{Eq_alpha}
\end{equation}
where $m_i(\vec{\theta}_i)$ is the mass of the $i^{th}$ cell at $\vec{\theta}_i$.

The quantity $m_i(\vec{\theta}_i)$ is obtained by minimizing the distance between the model-predicted and observed positions of multiple images. Since the number of free parameters, $N_{\rm c}$, exceeds the number of observables, the lens model cannot uniquely be determined. To overcome this degeneracy, regularization schemes, such as entropy, can be applied. 

Hybrid models introduce an extra layer of flexibility by incorporating analytic halos, similar to those used in the parametric methods. This hybrid approach is particularly useful when the lensing data are too sparse or when a sharp variation in the deflection field is required, which cannot be adequately represented by free-form mass cells alone. 

\paragraph*{Model optimization and statistical estimators.}
The values of all the model parameters, $\modpar$, were varied to fit the observables, which are the 23 multiple image positions of the eight background source positions for the gold sample. Specifically, the posterior probability distribution of the model parameters, $P(\modpar|{\rm data})$, was either optimized or sampled, where, through the Bayes theorem,
\begin{equation}
\label{eq:posterior}
P(\modpar|{\rm data}) = \frac{P({\rm data}|\modpar) P(\modpar)}{P({\rm data})}.
\end{equation}
The likelihood, $P({\rm data}|\modpar)$, of the multiple image positions, given a set of model parameters $\modpar$, is
\begin{equation}
\label{eq:likelihood}
P({\rm data}|\modpar) = A \exp\left(-\chiimsq/2\right),
\end{equation}
where $A$ is a normalization constant. The $\chi^2_{\rm im}$ function is given by
\begin{equation}
\label{eq:chi2im}
        \chi^2_{\rm im} = \sum_{j=1}^{N_{\rm sys}}\sum_{i=1}^{N_{\rm im}^{j}}\frac{ \left|\Delta \vec{\theta}_{ij} \right|^2}{\sigma_{ij}^2},
\end{equation}
where $N_{\rm sys}=8$ is the number of gold systems, $N_{\rm im}^{j}$ is the number of multiple images of the $j$th system, and
\begin{equation}
\label{eq:posdiff}
        \Delta \vec{\theta}_{ij} = \vec{\theta}_{ij}^{\rm obs}-\vec{\theta}_{ij}^{\rm pred}(\boldsymbol{\eta}, \vec{\beta}_{j}^{\rm mod}),
\end{equation}
with $\vec{\theta}_{ij}^{\rm obs}$ as the observed position of the $i$th lensed image of system $j$, $\vec{\theta}_{ij}^{\rm pred}$ as the corresponding model-predicted image position, $\vec{\beta}_{j}^{\rm mod}$ as the modeled source position of system $j$, and $\sigma_{ij}$ as the positional uncertainty of the $i$th image of system $j$. The above expression is for the case when circular positional uncertainties are considered, and we refer to \citet{Ertl+2025} for the expression of elliptical uncertainties. The prior $P(\modpar)$ is based on information from other datasets, such as kinematic measurements of member galaxies in the cluster.

The rms separation between the observed and model-predicted positions of the multiple images is a quantity widely used to assess the goodness of a given lens model. It is expressed as 
\begin{equation}
\label{eq:rms}
        {\rm rms}=\sqrt{\frac{1}{N_{\rm im}} \sum_{j=1}^{N_{\rm sys}}\sum_{i=1}^{N_{\rm im}^{j}} \left|\Delta \vec{\theta}_{ij} \right|^2},
\end{equation}
with $N_{\rm im} = 23$ in this paper.

Depending on the complexity of the parametric models, they can have several to dozens of variable parameters. The number of degrees of freedom, $N_{\rm dof}$, is computed as
\begin{equation}
N_{\rm dof}=2 \times N_{\rm im} - N_{\rm param}. 
\end{equation}
where $N_{\rm param}=2\times N_{\rm sys}+N_{\rm mass ~ param}$. The first term is for the source position parameters, and the second term is for the lens mass-distribution parameters. Besides using the values of $\chi^2_{\rm im}$ or of the rms to quantify the goodness of a given model, we also compared the values of the Bayesian information criterion \citep[BIC;][]{Schwarz1978}. Even though all the lens models included the same number of multiple images and observables, the BIC is particularly relevant when comparing different total mass parameterizations, with a different number of free parameters. It can be defined as
\begin{equation}
{\rm BIC}=-2\ln(P({\rm data}|\hat{\modpar})) + N_{\rm param} \times \ln(2 N_{\rm im}),
\end{equation}
where $\hat{\modpar}$ are the maximum-likelihood parameter values.  Therefore, up to an (irrelevant) additive constant, the BIC is 
\begin{equation}
{\rm BIC}=\chiimsqmin + N_{\rm param} \times \ln(2 N_{\rm im}),
\end{equation}
where $\chiimsqmin$ is the minimum $\chiimsq$ value.
For the \mrmartian\ model (see Sect.~\ref{sec:mass_model:mars}), the BIC was computed using the effective number of free parameters for $N_{\rm param}$ to account for the effect of regularization, as detailed in Sect.~\ref{sec:mass_model:mars}.

When $N_{\rm dof}>0$, a good-fitting model has $N_{\rm dof} \sim \chiimsqmin$.  However, mass models might not capture the full complexity of the cluster mass distribution and reproduce the observed image positions within the observational uncertainties, in which case $\chiimsqmin$ could be substantially larger than $N_{\rm dof}$. In these cases, the model-parameter uncertainties estimated by, for example, MCMC, would be underestimated given the artificially large changes in the likelihood due to the misfit to the observed data.  One practical way to mitigate this and obtain realistic model parameter uncertainties is to boost the positional uncertainties of the multiple images such that $N_{\rm dof} \sim \chiimsqboost$, where $\chiimsqboost$ is the $\chi^2$ obtained with the boosted positional uncertainty. For each mass model, we specified the boost factor (1 when no boosting was required to fit to the observed image positions). This boosting applied only to mass-model parameter estimates and predictions. For comparing and weighting the different models, the original $\chiimsqmin$ was used. 

\paragraph*{Overview.}
We provide an overview of the setups of the seven lens modeling approaches in Table \ref{tab:model_overview}.  For the details of each model, we describe them in the following subsections.

\begin{table*}[h!]
    \centering
    \caption{Summary of modeling software }
    \label{tab:software}
    \resizebox{\textwidth}{!}{
    \begin{tabular}{p{6.4cm}ccccccc}
        \hline \addlinespace[2pt]
        Description & \glafic & \GLEE & \lenstool & \zitrin & \mrmartian & \wslap \\ 
        \addlinespace[2pt] \hline \addlinespace[2pt]

        Category of mass models &  &  & & & & \\
        \addlinespace[2pt]
          $\bullet$ parametric mass distribution  & \checkmark & \checkmark & \checkmark & \checkmark & & \\
          $\bullet$ free-form mass distribution  &  &  & & & \checkmark & \checkmark   \\
        \addlinespace[2pt] \hline \addlinespace[2pt]

        Observables that can be used as input to constrain the mass-model parameters            &  &  & & & &\\
        \addlinespace[2pt]
          $\bullet$ lensed image positions \texttt{[baseline]}  & \checkmark & \checkmark & \checkmark & \checkmark & \checkmark &  \checkmark  \\
          $\bullet$ fluxes at lensed image positions  & \checkmark & \checkmark & \checkmark & \checkmark & & \\
          $\bullet$ surface brightness of lensed images  & \checkmark & \checkmark  & & &  &   \\
          $\bullet$ time delays & \checkmark & \checkmark & & \checkmark & \checkmark &  \\ 
        \addlinespace[2pt] \hline \addlinespace[2pt]
        
        Uncertainties on image positions &  & & & & & \\
        \addlinespace[2pt]
          $\bullet$ circular positional uncertainties \texttt{[baseline]}  & \checkmark & \checkmark  & \checkmark & \checkmark & \checkmark &  \checkmark   \\
          $\bullet$ elliptical positional uncertainties  &  & \checkmark & $\sim$ & & &   \\
        \addlinespace[2pt]  \hline \addlinespace[2pt]

        Lens planes &  & & & & & \\
        \addlinespace[2pt]
          $\bullet$ single-lens planes \texttt{[baseline]}  & \checkmark & \checkmark  & \checkmark & \checkmark & \checkmark &   \checkmark  \\
          $\bullet$ multi-lens planes  & \checkmark & \checkmark & $\sim$ & \checkmark & &   \checkmark  \\
        \addlinespace[2pt]  \hline \addlinespace[2pt]

    \end{tabular}
    }
\tablefoot{The first column describes the possible input, setup, and capabilities of the software, while the remaining columns list the six independent software packages employed in this work. The tilde ($\sim$) denotes the implemented capabilities that have not yet been extensively tested. All software can use lensed image positions with circular uncertainties as observables to constrain lens mass distributions on a single lens plane.  This setup, as indicated by ``\texttt{[baseline]}'' in the corresponding rows, forms the basis for constructing the \texttt{baseline} models that enable direct comparison of software and models from the different modeling teams.}
\end{table*}

\subsection{\glafic\ (M.O., N.F., B.F., S.N., Y.F.)}
\label{sec:mass_model:glafic}

\glafic\ \citep{2010PASJ...62.1017O,2021PASP..133g4504O} is a
software for parametric strong lens mass modeling. We modeled
\macs\ assuming a single lens plane and using only the positions of
the multiple images as observables. Since \glafic\ cannot handle
elliptical errors for the multiple images positions, circularized errors were adopted for the multiple-image positions. 

In the \texttt{baseline} model, we assumed a halo component modeled by an elliptical Navarro-Frenk-White (NFW) mass density profile \citep{1997ApJ...490..493N}, and a BCG component modeled by an elliptical Hernquist mass density profile
\citep{1990ApJ...356..359H}. Their lensing properties were calculated by
adopting the fast approximation proposed by \citet{2021PASP..133g4504O}.
For the NFW component, the mass, centroid position, ellipticity, position angle, and concentration were included as free parameters. The centroid position of the Hernquist component was fixed to the observed position of the BCG, and the mass, ellipticity, position angle, and scale radius were included as parameters. We adopted Gaussian priors for the ellipticity ($0.4254\pm0.05$), the position angle ($43.59\pm5.0 \deg$), and the scale radius ($10\farcs 5\pm2\farcs0$), based on the observed light profile of the BCG from \citet{Ertl+2025}. The lensing effect of most of the other cluster member galaxies was included assuming the dPIE profile using the scaling relation of the velocity dispersion and truncation radius with $\alpha=0.25$ and $\beta=0.5$ in Eq.~(\ref{eq:fj}), where $L_i$ is the luminosity of each galaxy derived from the F115W-band
magnitude, and the normalizations of these two scaling
relations are free parameters. The galaxies fg and JF-1 were modeled separately, assuming circular dPIE mass density profiles, and their velocity dispersions and truncation radii were free parameters. We also included an external shear in our mass modeling. The number of model parameters is 15, excluding those with informative priors. The best-fitting model has $\chi^2_{\rm im}=10.8$ with 15 degrees of freedom, and the rms of the difference of image positions between the observation and the model prediction is $0\farcs 20$.

In addition to the \texttt{baseline} model, the mass model using the gold plus silver multiple image sample was constructed. We adopted a Gaussian prior of $2.0\pm 0.5$ for the source redshift of the silver multiple-image system. In addition, the member galaxy at (RA, Dec)=(24\fdg5153289, $-21\fdg9192430$), which affects the multiple images of the silver sample significantly (see Fig.~\ref{fig:silver_sys}), was modeled separately with a dPIE mass density profile with its ellipticity and position angle fixed to observed values from the light distribution and the velocity dispersion and the
truncation radius as free parameters. Thus, the number of model
parameters increased to 17, excluding those with informative priors.
The best-fitting model has $\chiimsq=13.9$, and a positional rms of $0\farcs 20$. The posterior of the source redshift of the silver multiple-image system was dominated by the prior.

The errors of the model parameters were estimated with the MCMC method with a total number of steps of 28871 and 15194 for the \texttt{baseline} and gold plus silver models, respectively. A subsample of 3000 random realizations was used to estimate the distribution of the model-predicted time delays and magnifications for SN Encore and SN Requiem. No rescaling factor was introduced for the positional uncertainty of the multiple images. In our models, $100\%$ of these samples predict five (or more) multiple images both for Encore and Requiem.
We note that the \texttt{baseline} model based on only the gold multiple image sample 
is used for comparison with all the other models in the subsequent analysis, and the results of the gold plus silver images are presented instead in Appendix \ref{app:silver}.

\subsection{GLEE (S.E., S.H.S., S.S., G.B.C., G.G., C.G.)}
\label{sec:mass_model:glee}

We built seven different lens mass models with \GLEE\ \citep{SuyuHalkola2010, Suyu+2012} with gold images as constraints, exploring a range of DM halo profiles, the presence of a mass sheet, the differences between multiplane versus single-plane modeling, and the impact of having elliptical versus circular BCG. Two of the seven \GLEE\ models are single-lens plane at $\zd=0.336$, with one adopting elliptical positional uncertainties, and the other one adopting circular positional uncertainties; the latter is the \texttt{baseline} model. The other five \GLEE\ models employed multi-lens planes for the foreground, cluster, and background deflectors, with elliptical positional uncertainties on the multiple images.  We refer to \citet{Ertl+2025} for details on the \GLEE\ models and summarize here the key aspects and results for the model comparison.

Our \GLEE\ mass models included two DM halos, a primary one centered close to the BCG and a perturbative one that is diffuse with a large core, located $\sim30\arcsec$ southwest of the BCG.  In addition, we included the 84 cluster galaxy members and the two LOS galaxies as truncated isothermal profiles into the mass model.  We used the Faber-Jackson relation from \citet{Granata2025} to relate the Einstein radii and truncation radii of the cluster members to each other, except for the BCG and the galaxy \citep[ID$_{\rm phot}$=116 at (RA, Dec) = (24\fdg51563054, $-21\fdg92276878$);][]{Ertl+2025} that is $\sim$$2\arcsec$ north of the multiple image 4.3a (see Fig.~\ref{fig:jwst_image}), whose Einstein and truncation radii were allowed to vary independently of the scaling relation, given their significant effect on the prediction of the multiple image positions.  The three jellyfish galaxies were also not modeled within the scaling relation since the relation applies to early-type galaxies and not jellyfish galaxies. Our reference cluster galaxy for the scaling relations is that with ID$_{\rm phot}=115$ in \citet{Ertl+2025} with $m_{\rm F160W,\ ID115}=17.99$. 
By calibrating the Faber-Jackson relation using the 13 cluster members with velocity dispersion measurements (excluding the BCG), we used the measured value $\alpha = 0.25$ and a prior on $\sigma_{\rm v}^\mathrm{ref}$ of $206\pm25\, \kms$ for the reference galaxy $L_{\mathrm{ID 115}}$ \citep{Granata2025} with $\beta=0.7$.  We refer to \citet{Ertl+2025} for the kinematic priors on the parameters of galaxies not included in the scaling relations.

For each of the seven \GLEE\ models, we ran MCMC chains of $2\times10^6$ steps.  After exploring four different cluster DM halo profiles for the primary lens -- including one model allowing a mass sheet at the cluster redshift -- we find that cored isothermal DM profiles fit the gold sample of observed multiple image positions well, with a reduced $\chiimsq \sim 1$. This includes our \texttt{baseline} model. Therefore, we do not need to boost the positional uncertainties to quantify our mass-model parameter uncertainties. 

We combined four of our five multi-lens plane mass models, which fit well to the observed image positions and have remarkably consistent results, to form our ultimate lens model. We also obtained the \texttt{baseline} model from single lens-plane modeling, for direct model comparison. We find that the \texttt{baseline} model provides a good approximation for our ultimate model. For the predictions of the multiple image positions, magnifications and time delays of SN Encore and Requiem, we thinned the MCMC chains by a factor of 1000.  Therefore, the \GLEE\ ultimate model has 8000 samples, and the \GLEE\ \texttt{baseline} model has 2000 samples.

The number of free mass parameters (without informative priors)\footnote{There are three informative priors on the Einstein radii of three galaxies based on their velocity dispersion measurements, as detailed in \citet{Ertl+2025}. The number of free parameters is thus 25 (total number of parameters) $-$ 3 = 22.} in the \GLEE\ \texttt{baseline} model is 22.  
From our optimization, we obtain the most-probable $\chiimsq$ of 7.3, which we approximate\footnote{The most probable $\chiimsq$ is obtained by maximizing the posterior, whereas $\chiimsqmin$ corresponds to the maximal likelihood.  Of the four mass parameters with Gaussian priors, only one parameter is not prior-dominated and our most-probable $\chiimsq$ of 7.3 should be comparable or only slightly higher than $\chiimsqmin$.}  as $\chiimsqmin$, with an rms value 
of 0.24\arcsec.  We refer to Table 5 of \citet{Ertl+2025} for a detailed breakdown of the $\chiimsqmin$ and rms for the individual gold image systems.  While the first four images of SN Encore (images 1a-1d) and SN Requiem (images 2a-2d) were predicted 100\% of the time by the mass models in the MCMC chains, images 1e and 2e were only predicted a fraction of the time\footnote{The image configuration and the critical curves near the radial arc is complex, given the presence of JF-1, JF-2, and the BCG.  When they are sufficiently massive, the singular nature of their mass profiles prevents the creation of a central image near the BCG. However, there can be additional multiple images produced by the jellyfish galaxies near the radial arc. Future detections of such images will provide opportunities to study the mass distributions of JF-1 and JF-2 in detail.}, as listed in Tables \ref{tab:snencore} and \ref{tab:snrequiem}. Images 1d, 1e, 2d, and 2e are predicted to appear in the future.

\subsection{\lenstool\ I (P.K.)}
\label{sec:mass_model:kamieneski}

The \lenstool\ I model made use of the well-tested \textsc{Lenstool} software \citep{Kneib:1993aa,Kneib:1996aa, Jullo2007, Jullo:2009aa}, adhering to the general methodology followed in \citet{Kamieneski:2024ac, Kamieneski:2024aa} (see also \citealt{Pascale+2025}).
Only a \texttt{baseline} model was contributed in this analysis. The \macs\ cluster was modeled under the assumption of a single lens plane at $z_{\rm d}=0.336$, consisting of an underlying DM halo parameterized as a non-truncated dPIE.
An ensemble of 80 cluster member galaxies (out of a total of 84) were included as small-scale, truncated dPIE mass components held at $z=0.336$, with small core radii values fixed to 0.15 kpc, and each with their ellipticities and position angles fixed to the morphological fitting by \citet{Ertl+2025}.
Together, they were scaled in velocity dispersion based on their measured luminosities in the F200W band, with $\alpha=0.25$ while their truncation radii were scaled with $\beta=0.5$ (which yields a constant total mass-to-light ratio). 
The normalizations of these relations were optimized as free parameters.
Also, the background LOS galaxy, bg, was included in the scaling relations, as it is not in very close proximity to any arcs of the gold image families.

The four cluster members excluded from these scaling relations were instead independently optimized as singular isothermal sphere (SIS) or singular isothermal ellipsoid (SIE) profiles \citep{Kormann:1994aa}, due to a need for greater flexibility in the model (typically as a result of close proximity to one of the gold images).
The first of these is at $({\rm RA, ~Dec}) = (24\fdg51662331, -21\fdg92448273)$, 
optimized as an SIS with fixed centroid (chosen given its proximity to the radial arc of System 3; see the maroon square in Fig.~\ref{fig:jwst_image}). 
The three other members are apparent jellyfish galaxies requiring some extra care.
Specifically, the jellyfish galaxies JF-1 and JF-3
were optimized as an SIS profile (again with fixed centroid). The jellyfish galaxy JF-2 
was instead parameterized as an SIE profile (with free parameters of velocity dispersion, ellipticity, and position angle).

The fg near the multiple image 3b was independently modeled as an SIS 
with its centroid and velocity dispersion as free parameters.
Finally, an external shear component was included in the model, primarily to account for deficiencies in this parameterization.
In total, the model included 20 free parameters.
Uniform priors were used for all the parameters, except for (i) the centroid of the primary cluster halo component, (ii) the centroid of the LOS fg, and (iii) the truncation radius of the cluster halo.
These parameters were optimized with Gaussian priors instead, with $\sigma = 2.0\arcsec$,  $0.5\arcsec$, and $100$ kpc (centered at 1000 kpc), respectively.

The \texttt{baseline} model was constrained using only the observed positions of the gold set of multiple images and optimized on the image plane.
The multiple image positions in the highest-likelihood model were recovered with an image-plane rms offset of $0\farcs 48$.
The positional uncertainties were then uniformly rescaled by a factor of 2.2 to ensure a reduced $\chiimsq < 2$, and a $\chiimsq$ of 19.2 is achieved for 10 degrees of freedom.

The posteriors were estimated from 300 realizations randomly drawn from the final MCMC chain, which contained 10000 iterations after burn-in. In our \texttt{baseline} model, $100\%$ of these samples predict a fourth image of SN Encore (1d), but only $\sim 4\%$ of the realizations predict a fifth image (1e). For SN Requiem, $\sim 16\%$ of iterations predict the fourth image (2d). None of the realizations predict a fifth image (2e).

After the lens modeling unblinding, an alternative model was tested to examine the cause of the large uncertainties for the time delays between different pairs of images (see Sect.~\ref{sec:comparison:SNEncore}). We found significant covariance between the free parameters for JF-2 (namely, ellipticity and velocity dispersion) and the time delays, perhaps due to its proximity to the BCG. For the alternative model, JF-2 was instead parameterized as an SIS potential, thus with only one free parameter instead of three. This resulted in minimal change to the $\chiimsq$ value, 
but relative time-delay uncertainties that were much more consistent with those from the other lens models (see Appendix \ref{app:LenstoolI_alt}).

\subsection{\lenstool\ II (A.A., P.B., P.R.)}
\label{sec:mass_model:lenstoolII}
A second team also used the parametric \lenstool\ software \citep{Jullo2007} to model the total mass distribution of \macs\ following the methodology outlined by, for example, \citet{Caminha2019, Bergamini+2023, Acebron2022}. The strong-lensing analysis is detailed in \citet{Acebron2025}, where several total mass models were explored to assess the impact of systematic uncertainties on the model-predicted positions, magnifications, and time delays of the SN~Requiem and SN~Encore multiple-image systems. We summarize here the characteristics of our best-fitting lens model, labeled as \texttt{reference} in \citet{Acebron2025}, and presented to this comparison challenge as our \texttt{baseline} model. The six lens models explored by \citet{Acebron2025} enable one to assess the impact of modeling choices on the model-predicted time delays of both SNe systems. We found that the different mass parameterizations yield model-predicted values of their magnifications and time delays that are consistent within the $1\sigma$ uncertainties, provided that fg is taken into account.

The total mass distribution of the lens was modeled with a single cluster-scale halo, parameterized with a non-truncated elliptical dPIE mass density profile and an external shear component. The sub-halo component, as well as the three jellyfish members and the two LOS galaxies, were described instead with circular, truncated dPIE mass density profiles, with vanishing cores. 
The values of the normalization and slope of the kinematic scaling relation were anchored on the calibrated Faber-Jackson relation including the BCG \citep[see][using the measurements from \citet{Granata2025}]{Acebron2025}. Specifically, we adopted a Gaussian distribution centered on the measured value of $\sigma_{\rm v}^\mathrm{ref}=329\pm26~\rm km~s^{-1}$ (associated with the BCG with $m_{\rm F160W} = 15.30$), as a prior for the normalization of the kinematic scaling relation. Its slope was instead fixed to the best-fit value of $\alpha=0.21$, while the value of $\beta$ is inferred to be equal to 0.71. Thus, the 81 cluster member galaxies were scaled with total mass-to-light ratios that increase with their F160W luminosities, as $M/L\propto L^{0.2}$, consistent with the observed tilt of the Fundamental Plane.
The three jellyfish member galaxies and the two LOS galaxies were modeled outside of the scaling relations, with their mass parameters free to vary.
The values of all the model parameters except that of the scaling relation normalization, $\sigma_{\rm v}^\mathrm{ref}$, were optimized within large flat priors \citep[see Table 2 in][]{Acebron2025}. 

Our \texttt{baseline} model of the \macs\ total mass distribution includes 20 free mass parameters and reproduces the observed positions of the multiple images with a rms offset of $0\farcs36$. The final model was obtained after rescaling the multiple images positional uncertainty by a factor of 1.5. The median values and the associated $1\sigma$ uncertainties of the model-predicted quantities were obtained from 1000 different models randomly extracted from the MCMC chain, with a total number of $5\times 10^5$ samples, excluding the burn-in phase. Our \texttt{baseline} model predicts the reappearance of both SN~Encore and SN~Requiem in the future, with two and one additional multiple image(s), respectively, in the radial arc of \macs. We note that our model predicts the fifth and central multiple image of SN~Encore, 1e, in only $\sim15\%$ of the chains.

\subsection{\zitrin\ (A.K.M, A.Z.)}
\label{sec:mass_model:zitrin}
This lens model was constructed using a revised version of the parametric approach described in~\citet{2015ApJ...801...44Z}, which is sometimes also referred to as ``Zitrin-analytic.'' The underlying modeling approach is similar to that from other parametric lens modeling techniques and has been successfully applied to many other galaxy clusters \citep[e.g.,][]{2023MNRAS.523.4568F, 2024MNRAS.533.2242F, 2023ApJ...944L...6M, 2022ApJ...938L...6P, Pascale+2025}. The model primarily contains two parametric mass components as isothermal halos to describe the DM, and dPIE for galaxy-scale components. The method calculates the various lensing quantities directly at the positions of multiple images.

For the \macs\ \texttt{baseline} model, we used two DM halos placed around the BCG and bg positions, 
and the DM positions were optimized in the minimization. Each dark-matter halo thus had six free parameters: the velocity dispersion, core radius, ellipticity, position angle, and the two-dimensional position. We also left free to vary the core radius, ellipticity, and position angle of the BCG and of bg. The rest of the cluster galaxies were assumed to be circular and modeled using the scaling relation given by Eq.~\ref{eq:fj}, but with a core. For the scaling relations, we used $(\alpha, \beta, \gamma) = (0.25, 0.50, 0.50)$ where $\gamma$ represents a similar scaling for the core radius. We used a core size value of $r_{\rm c}^{\rm ref}=0.2$ kpc for the reference (roughly $\sim L^{*}$) galaxy, where $\sigma^{\rm ref}$ and $r_{\rm cut}^{\rm ref}$ were optimized in the minimization.  In our model, which does not include an external shear, we also left free to vary the velocity dispersions of four of the circular cluster galaxies with (RA, Dec) = $(24\fdg5114258, -21\fdg9213075), (24\fdg5129665, -21\fdg9203688)$, $(24\fdg5156305, -21\fdg9227688)$, and $(24\fdg5076146, -21\fdg9271504)$. As illustrated with the dashed gray squares in Fig.~\ref{fig:jwst_image}, these four galaxies are located west of the BCG, and their model parameters were free to vary to improve the reproduction of nearby images. We obtained a total of 24 free parameters. 

The lens model was optimized on the source plane using an MCMC sampling (where the $\chi^2$ is defined on the source plane based on the separation of the mapped source positions corresponding to the observed image positions, instead of $\chiimsq$ in Eq.~\ref{eq:chi2im})\footnote{The \zitrin\ team tested and found that their source-plane optimization recovers the input cluster mass models parameters in their cluster simulation. The \mrmartian\ team in the Section \ref{sec:mass_model:mars} also showed in \citet{ChaJee+2026} that \mrmartian's source plane optimization recovers the convergence within $\sim$10\% without significant bias, as well as the magnifications and time delays within $\sim$30\% on the Ares and Hera clusters simulated by \citet{Meneghetti+2017}.}, with a couple of hundred thousand steps after the burn-in phase. For more details about the modeling scheme and minimization procedure, we refer to \citet{2023MNRAS.523.4568F}. With the above setup, we got ($\chiimsq$, $N_{\rm dof}$, image rms) = (143.2, 8, $0\farcs30$), where we boosted the nominal positional uncertainties as described in Sect.~\ref{sec:data:multiple_images} by a factor of 3 (such that the output~$\chiimsq$ was 15.91 in the minimization). To generate the errors, we drew 1000 random points from the final MCMC chain. In our sample, 90\% and 94\% of solutions led to the formation of a fifth image (1e and 2e) close to the center of the lens cluster for Encore and Requiem, respectively.

\begin{table*}[]
\renewcommand\arraystretch{1.2}
	\centering          
\caption{Summary of the characteristics of the seven lens models of \macs.}
\begin{tabular}{|c|c|c|c|c|c|c|c|}
\hline
\textbf{Model} & $\mathbf{N_{\rm param}}$ & $\mathbf{\Ndof}$  & $\mathbf{\chiimsqmin}$ & \textbf{rms} [$\arcsec$] & $\mathbf{\ln{\mathcal{L}}}$ & \textbf{BIC} & weight $\mathbf{{\mathcal{L}}}$ \\
\hline
\noalign{\vskip 1pt}  
 \glafic\ & 31 & 15  &10.8 & 0.20 & $-$5.4 & 129.5 & $1.47\times10^{-1}$ \\
\hline
\noalign{\vskip 1pt}  
\GLEE-baseline & 38 & 8 & 7.3 & 0.24 & $-$3.7 & 152.8 &  $8.45\times10^{-1}$ \\
\hline
\noalign{\vskip 1pt}  
\lenstoolone\ & 36 & 10 & 92.9 & 0.48 & $-$46.5 & 230.7 & $2.18\times10^{-19}$ \\
\hline
\noalign{\vskip 1pt}  
\lenstooltwo\  & 36 & 10 & 24.2  & 0.36 & $-12.1$ & 162.1 & $1.72\times10^{-4}$\\
\hline 
\noalign{\vskip 1pt}  
\mrmartian\ postblind & 2241 (80.2) & $-$ & 16.7$\dagger$ & 0.28 & $-$8.4 & 323.8 & $7.69\times10^{-3}$\\
 & 2241 (27.9) & $-$ & 16.7$\dagger$  & 0.28  & $-$8.4 & 123.5 & $7.69\times10^{-3}$ \\
\hline
\noalign{\vskip 1pt}  
\wslap\ postblind & $272$ & $-$ &  2136.5$\dagger$ & $1.02$ & $-$1068.3 & $-$ &  $3.78\times10^{-463}$ \\
\hline
\noalign{\vskip 1pt}  
\zitrin & 40 & 6 & 143.2$\dagger$ & 0.30 & $-$71.6  & 296.3  &   $2.61\times10^{-30}$ \\
\hline
\end{tabular}
\label{tab:model_stat}
\tablefoot{We report the total number of model free parameters ($N_{\rm param}$), which includes the 16 source-position parameters ($x$ and $y$ for the eight gold systems); the resulting degrees of freedom ($N_{\rm dof}$); and the values of the different statistical estimators used (see Sect.~\ref{sec:mass_model}). The reported $\chiimsqmin$ values correspond to the minimum $\chi^{2}_{\rm im}$ before rescaling the positional uncertainties of the multiple images, where the dagger symbol ($\dagger$) denotes the lens models run on the source plane (the reported values $\chiimsqmin$ are consistently computed on the image plane for direct model comparison). The quantity $\ln{\mathcal{L}}$ is $-\chiimsq/2$, i.e., the natural log likelihood up to a constant offset that cancels out when weighting models by $\mathcal{L}$.}
\end{table*}

\subsection{\mrmartian\  (S.C., M.J.J.)}
\label{sec:mass_model:mars}
\mrmartian\ stands for Multi-resolution MAximum-entropy Reconstruction Technique Integrating Analytic Node and augments the free-form capability of MARS \citep{2022ApJ...931..127C, 2023ApJ...951..140C, 2024ApJ...961..186C} by incorporating analytic nodes. This hybrid approach allows us to model lens features smaller than the grid size using analytic halo mass-density profiles where required by the data, while efficiently representing larger scales with a flexible grid regularized by entropy \citep{ChaJee+2026}. The grid component of \mrmartian\ differs from that of MARS in that the grid size is nonuniform across the field, enabling multi-resolution capability by assigning higher-resolution grids to the regions where the mass density changes more rapidly. To predict the lens properties at any arbitrary location, we summed the contributions from both the analytic nodes and the multi-resolution grid.

In our lens modeling of \macs, we used only one truncated pseudo-elliptical NFW node consisting of seven free parameters: $x$, $y$ coordinates, concentration, scale radius, ellipticity, position angle, and truncation radius. We did not incorporate any observed photometric information as informed priors; only the BCG position was used to set the initial location of the node center. Together with the multi-resolution grid, which consists of three resolution levels ($2\farcs4$/pix, $1\farcs2$/pix, and $0\farcs6$/pix), the total number of free parameters amounts to 2225. While this is higher than for the parametric methods, the effective number of free parameters would be lower than 2225, as the regularization term prevents each parameter from varying independently. To quantify the impact of the regularization, we calculated the effective number of free parameters using two different methods. The resulting values are 64.2 and 11.9, based on \citet{NIPS1991_d64a340b} and \citet{Spiegelhalter_2002}, respectively.
The rms of the predicted multiple image positions is $0\farcs28$, and the offset between the BCG and the halo position from our lens model is $0\farcs5\pm0\farcs06$. To estimate the posterior distributions, we generated 200 realizations from randomly selected initial conditions.\footnote{Given the high number (2225) of model parameters and the complex regularization, it is nontrivial to robustly estimate the posterior distributions of these parameters and their impact on positional uncertainties. While the predicted image-positional uncertainties may be underestimated in this approach, the predicted uncertainties in magnification and time delay are of the same order of magnitude as those derived from other approaches (as shown later in Sect.~\ref{sec:comparison}).} All of our realizations predicted the fourth and fifth images of both SN Encore and SN Requiem.
Following the submission of the blinded model by the \mrmartian\ team, a minor issue was discovered involving the fixed prior intervals for the position angles in the analytic halos. While the best-fit results remained unaffected, the parameter uncertainties were overestimated. The postblind model offers more reliable uncertainty estimates.

We also constructed a lens model using the gold plus silver systems. To represent a compact halo located south of the silver images (see Fig.~\ref{fig:silver_sys}), we included an additional analytic node, which is also modeled with seven free parameters. The redshift of the silver multiple-image system was treated as a free parameter in the lens modeling, with a flat prior range of $[0.436, 15]$. The total number of free parameters amounts to 2233, and the rms of the predicted multiple image positions is $0\farcs21$. The model-predicted redshift of the silver system is $z=2.623_{-0.822}^{+2.558}$. Similar to the model using gold samples, all of our realizations predict the fourth and fifth images of both SN Encore and SN Requiem.  We refer to Appendices \ref{app:silver_data} and \ref{app:silver} for more details.

\subsection{\wslap\ (J.M.D.)}
\label{sec:mass_model:wslap}
This code was first described in \cite{Diego2005} and later expanded in \cite{Diego2007} to include weak lensing measurements 
as additional constraints. A further development was presented in \cite{sendra2014} where the code migrated from its native free-form nature to a hybrid type of modeling, with prominent member galaxies added to the lens model using a mass distribution that matches the observed light distribution.  

\wslap\, placed Gaussians in a predetermined grid of positions in the lens plane. 
The mass of each Gaussian was optimized by the algorithm. 
A contribution from member galaxies was also precomputed by adopting a fiducial mass for them, which was later optimized. A joint solution for the masses of the Gaussians, renormalization of the mass of the member galaxies, and unknown position of the lensed galaxies in the source plane was obtained by solving a system of linear equations:
\begin{equation}
\Phi = \mathbf{\Gamma} X,
\label{Eq_SLWL_matrix_simple}
\end{equation}
where $\Phi$ is an array containing the observed positions of the multiple images, $\mathbf{\Gamma}$ is a known matrix, and $X$ is the vector with all the unknowns: total masses in the Gaussian decomposition ($M$), multiplicative factors for the fiducial mass of the member galaxies ($C$), and source positions ($\beta_x$ and $\beta_y$). The solution was obtained using a quadratic programming optimization algorithm with the constraint $X_i>0$, where $i$ denotes the $i$th component of vector $X$.\footnote{The origin of the image coordinates used by \wslap\ is set at the southeast corner of the cluster image FoV, so by construction the multiple image positions and their corresponding source positions, $\beta_x$ and $\beta_y$, should have positive coordinate values, being located near the center of the FoV. This choice of source coordinates is purely for computational convenience, ensuring the condition $X_i>0$ imposed on the masses also applies to the source positions.}

For the particular case of \macs, we used 252 Gaussians with varying widths, where smaller Gaussians were placed near the BCG, thus increasing the resolution. We also superimposed additional layers of mass components associated with the luminous galaxies, where each layer has one single mass-to-light scaling parameter that is free to scale the observed light to mass. The luminous galaxies expected to have different mass-to-light scaling were thus placed in different layers. For \macs, we adopted four layers (i.e., four additional free parameters) for the total mass associated with the member galaxies, which was assumed to trace the observed light in the F356W filter. Layer 1 contained the BCG, layer 2, all the remaining non-jellyfish member galaxies, layer 3, the jellyfish galaxies, and layer 4, fg and bg. 
All galaxies were assumed to be at the same redshift. Since \wslap\ needed to optimize also for the unknown source coordinates ($\beta_x$ and $\beta_y$) of the eight gold systems, the total number of free parameters amounted to $N_\text{dof}=252+4+2\times8=272$.   
The derived lens model has an rms varying between $0\farcs3$ arcsec for some of the positions of the SN Requiem and up to $1\farcs82$ arcsec for system 4. 
Considering the full sample of multiple images, the rms is $1\farcs02$. The $\chiimsq$ value from Eq.~\ref{eq:chi2im} is very high, 2136.5, as reported in Table~\ref{tab:model_stat}. This high value is mostly due to the relatively large offset between some of the observed and predicted positions of the SNe and host galaxy. More precisely, $99.898\%$ of the $\chiimsq$ (that is, 2134.32 out of 2136.5) comes from the constraints on the host galaxy. This is not unusual in \wslap\ models that sometimes predict larger offsets (along the arcs) in images close to forming Einstein rings, as it is the case for the host galaxy. Nine separate modeling runs were used to estimate the uncertainties of the predicted image positions, magnifications and time delays of the SNe by taking the mean and standard deviation of the results from the nine runs. All models predict five multiple images for SN Encore and SN Requiem. After unblinding, it was discovered that the computation of the time delay (from the deflection field and potential) was affected by incorrectly rescaling the lensing potential from the fiducial redshift ($z_{\rm s}=3$) to the redshift of the two SNe. Hence, we labeled these corrected predictions for the time delay as ``postblind,'' although both the fiducial deflection field and potential were computed before unblinding.

\section{Comparison of the lens mass models}
\label{sec:comparison}

The key characteristics of the best-fitting lens models from each team are summarized in Table~\ref{tab:model_stat}. For each of them, we report the number of free model parameters, degrees of freedom, as well as the values of the statistical estimators introduced in Sect.~\ref{sec:mass_model}. 
We compare in Table~\ref{tab:model_stat} the \texttt{baseline} models that are constructed by all the seven teams using the same circular positional uncertainties (which enable direct model comparison). Only the \GLEE\ team constructed a separate ultimate model with multiplane lensing and elliptical positional uncertainties; we report the \GLEE\ predictions of SN properties but cannot compare the values of the statistical metrics to the other models given the difference in the data (elliptical versus circular positional uncertainties). 
For a direct comparison of $\chiimsqmin$ from the various models, we use the multiple image positional uncertainties for the \texttt{baseline} models without boosting, even though some teams boosted the positional uncertainties to obtain their model parameter constraints and model predictions (see Sect.~\ref{sec:mass_model}).
The goodness of fit to the observables varies across the seven lens models, where four teams obtain a $\chiimsqmin$ of $<$25, with two teams reaching $\chiimsqmin<11$. The corresponding values of the rms vary between $1\farcs02$ to $0\farcs20$. We note that the relationship between the $\chiimsqmin$ and rms is nonlinear, since image positions identified from JWST have uncertainties that are approximately ten times smaller than those from MUSE \citep[see][]{Ertl+2025}.
The \GLEE\ and the \glafic\ total mass models achieve the lowest values of $\chiimsqmin$ and the BIC, respectively.

\subsection{Surface mass density of the cluster mass distribution}
\label{sec:comparison:kappa}
\begin{figure}
\includegraphics[width = \columnwidth]{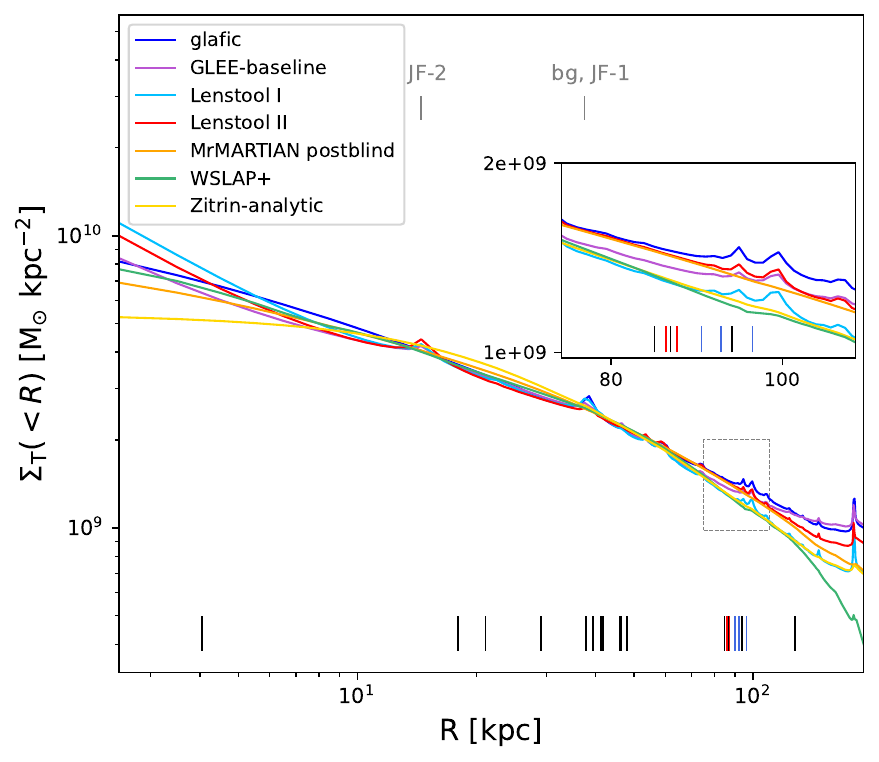}
\caption{Total average surface-mass-density profiles of \macs\ as a function of projected distance from the BCG center for the different best-fit strong-lensing models. The vertical lines (at the bottom of the panel) show the observed positions of the 23 multiple images in the gold sample, where the positions of the multiple images of SN~Encore and SN~Requiem are highlighted in red and blue, respectively. The vertical gray lines (at the top of the panel) mark the positions of the JF-1, JF-2, and bg galaxies, which contribute significantly to the surface mass density at their locations in some models. The inset shows a zoom-in of the region delimited by the dashed rectangle, where the observed multiple images of the two SNe are located. }
\label{fig:Sigmatot}
\end{figure}

In Fig.~\ref{fig:Sigmatot}, we show the best-fit total surface-mass-density profiles of the seven lens models presented in Sect.~\ref{sec:mass_model}. We find that the profiles are in fairly good agreement, despite the different modeling approaches and assumptions. The agreement is especially noteworthy at projected distances from the BCG between 20 and 70 kpc, where most of our observables are located (marked with vertical lines in the figure).
The largest discrepancies arise both at small ($\lesssim 10$ kpc) and large ($\gtrsim 100$ kpc) projected distances from the BCG, where only a few multiple images are identified and model extrapolation become relevant. In particular, in the innermost region of \macs, where our models predict the reappearance of multiple images (d and e) of SNe Requiem and Encore, the inferred slope varies considerably between lens models, from relatively cuspy (\lenstool\ I and II) to flat (\zitrin) total surface-mass-density profiles.

\subsection{SN Encore image positions, magnifications and time delays}
\label{sec:comparison:SNEncore}

In Fig.~\ref{fig:SN_Encore_pos_mag_td} and Table~\ref{tab:snencore}, we show the model predictions from the different teams for each of the multiple images of SN Encore, based on the assumed flat $\Lambda$CDM cosmology with $H_0=70\,\kmsMpc$.
We start noticing that the majority of the blindly model-predicted quantities from the different teams are consistent, given the estimated statistical uncertainties. In particular, three parametric models (\glafic, \GLEE, and \lenstool\ II) can reproduce very well the observed positions of images 1a and 1b (labeled in Fig.~\ref{fig:jwst_image}), and one of the free-form models (\wslap) cannot reproduce the same images within their errors. Images 1a and 1b are, respectively, predicted to be magnified with factors ranging from approximately $-30$ to $-20$ (the sign is negative since image 1a is a negative-parity lensed image) and from $30$ to $40$. Image 1b is predicted by most of the models to have a time delay of about $-$40 days (with a relative error of $\approx 10\%$) with respect to image 1a (i.e., $\tbaEncore \equiv t_{\rm 1b} - t_{\rm 1a} \sim -40\,{\rm days}$).  

All models predict three additional multiple images of SN Encore, although two of them (1c and 1e) are not present in all the iterations of the final chains of the teams. In cases where an image is not predicted in all the iterations of a team's model, the properties of the predicted image are computed based only on the iterations that have the image predicted. Interestingly, images 1d and 1e are expected to appear in the future, more than 3000 days after image 1a, while image 1c should have appeared about a year before image 1a. Given the long time-delay values of images 1d and 1e, their relative errors can be predicted with statistical uncertainties as low as 2$\%$ by some of the models and 1\% by the \mrmartian\ model. The predicted image positions of 1d and 1e show scatter among the models due to the presence of nearby jellyfish galaxies JF-1 and JF-2 (see Fig.~\ref{fig:Encore_Requiem_pos_radial_arc}); the predicted positions of 1d and 1e are thus sensitive to the modeled mass distributions of JF-1 and JF-2, which are currently not well constrained by the existing multiple-image systems. The magnification factors of images 1c, 1d, and 1e are significantly smaller than those of images 1a and 1b, going from slightly more than 10 for image 1c, to approximately $-4$ for image 1d, to order of unity for image 1e.

\begin{figure*}
\includegraphics[scale=0.45]{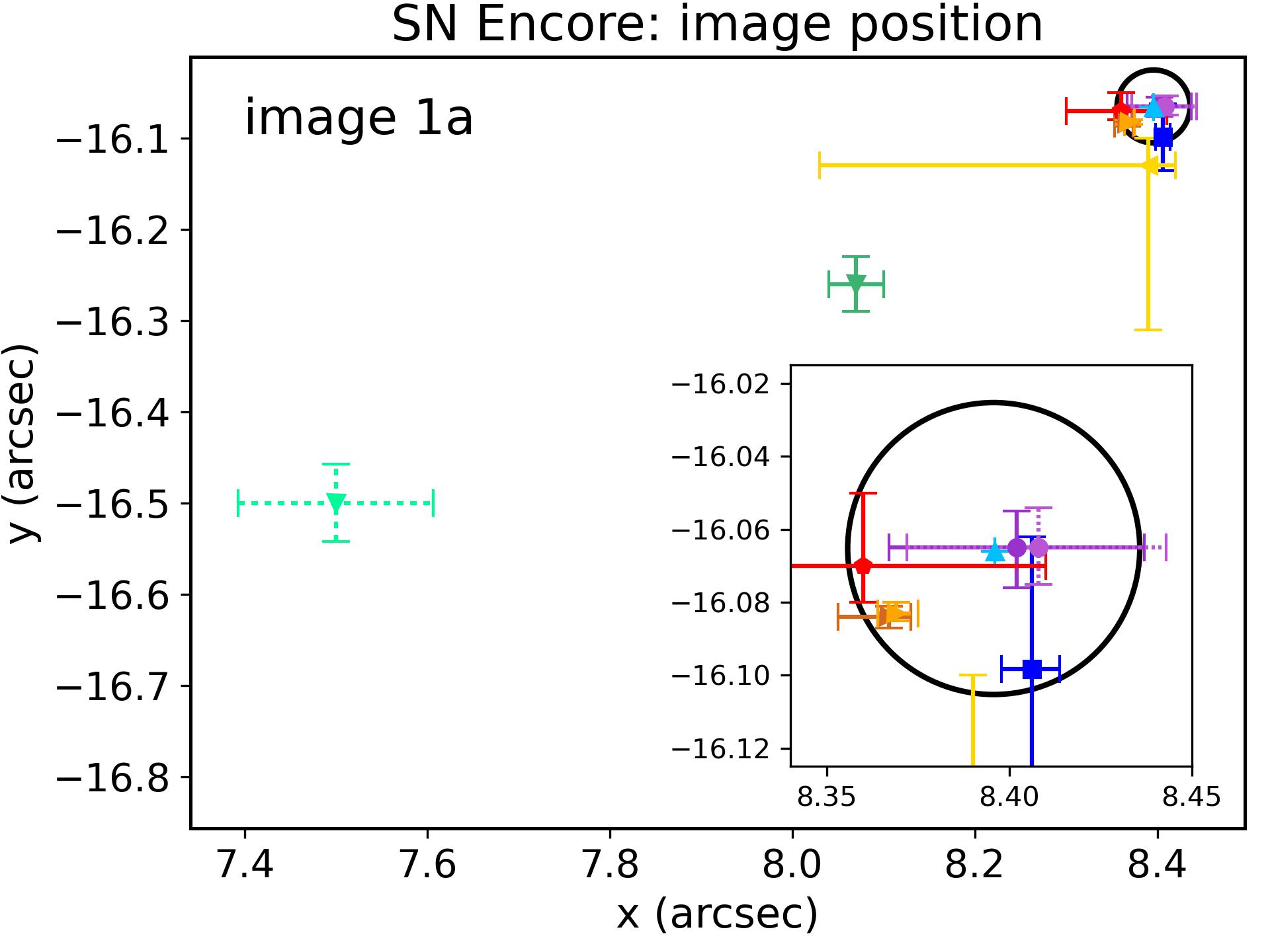}
\includegraphics[scale=0.55]{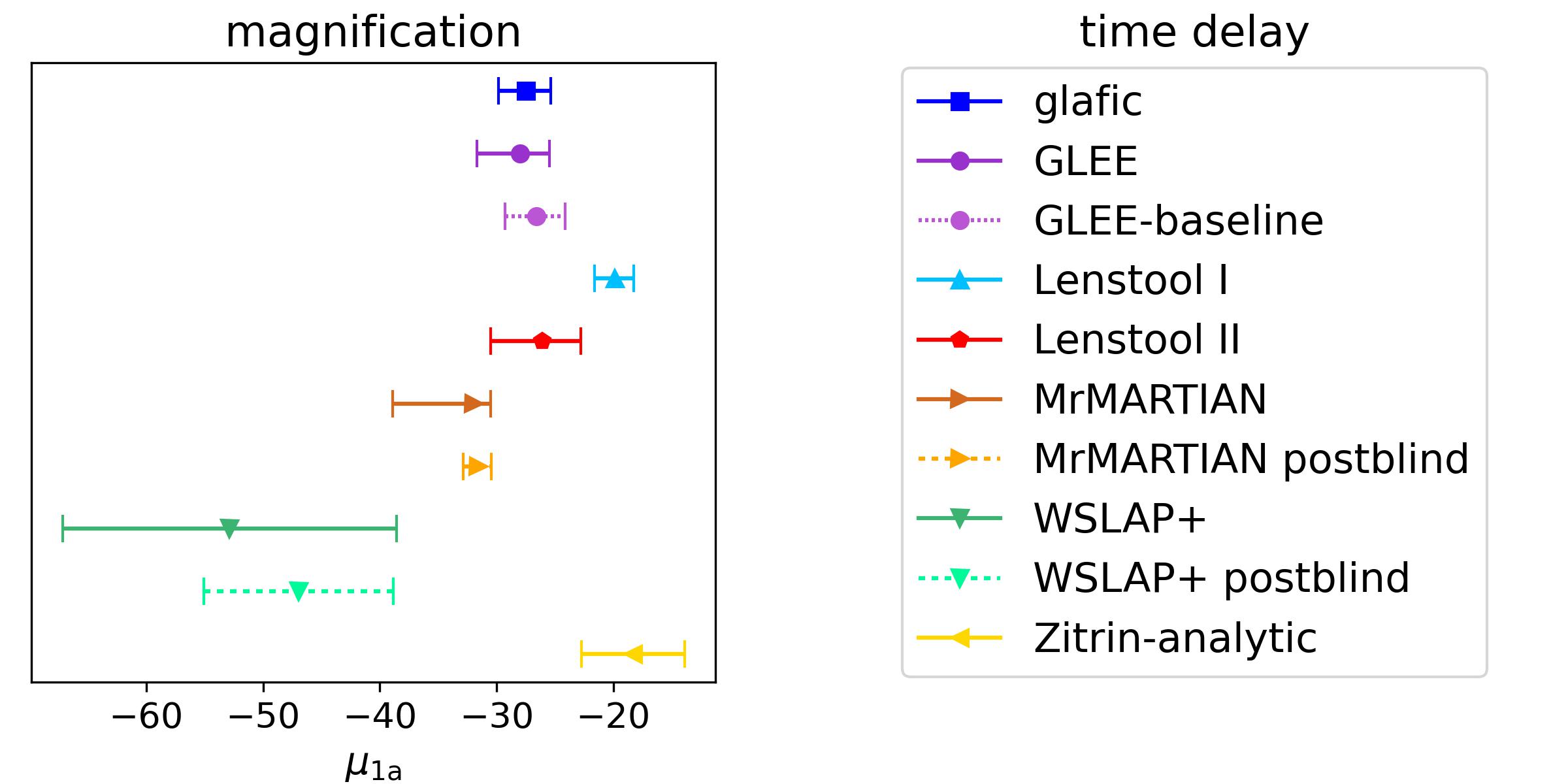}
\includegraphics[scale=0.45]{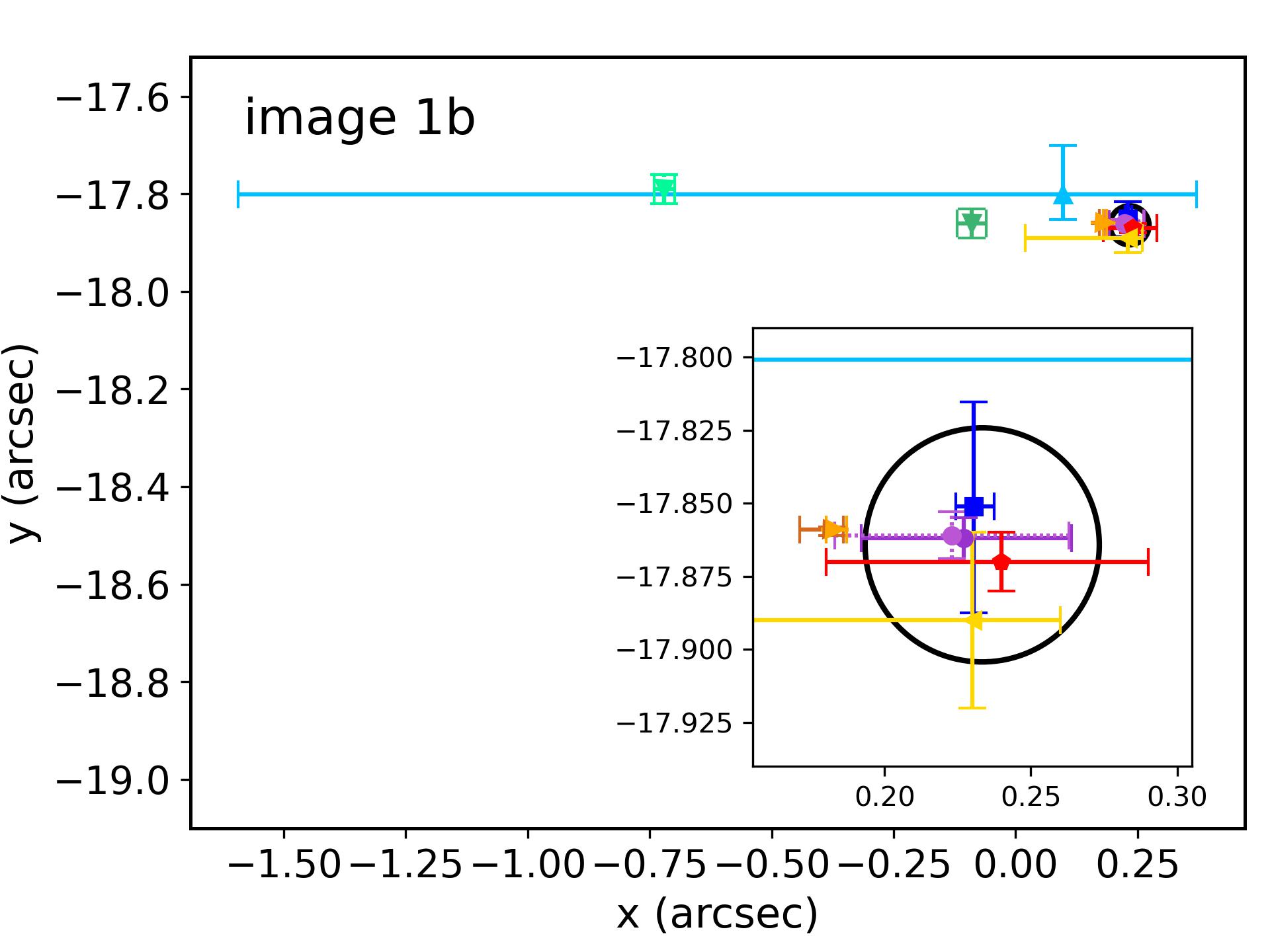}
\includegraphics[scale=0.55]{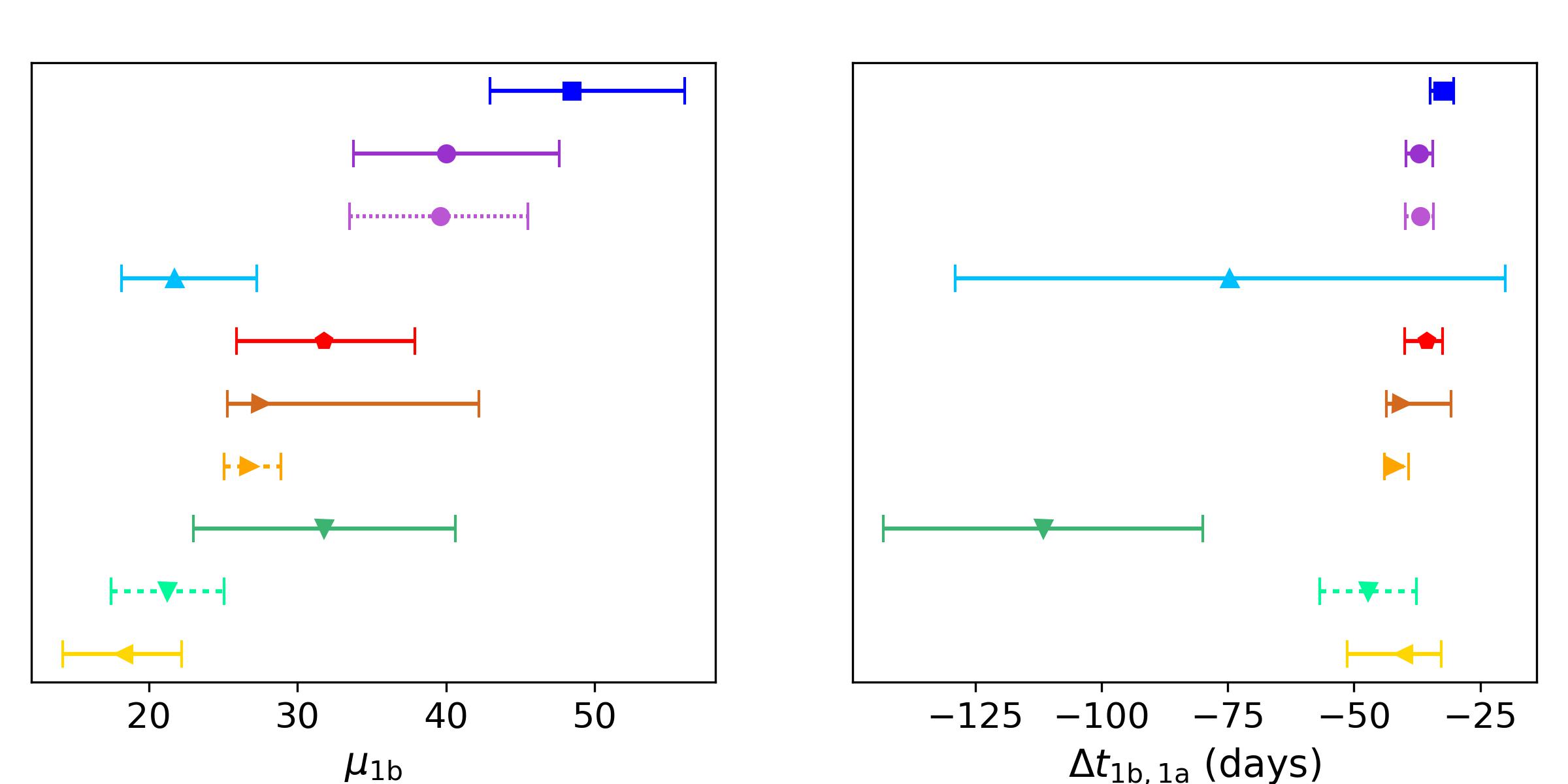}
\includegraphics[scale=0.45]{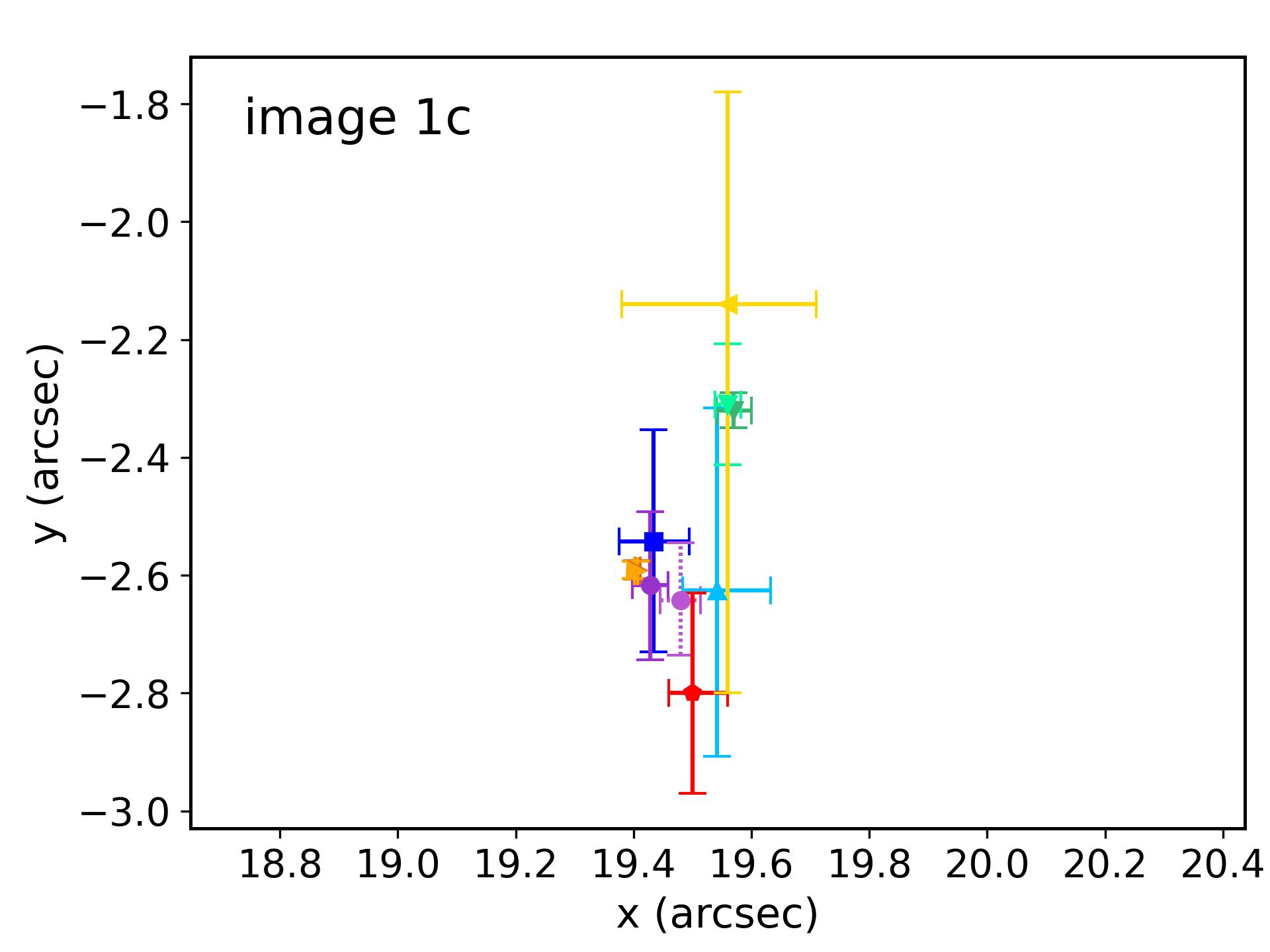}
\includegraphics[scale=0.55]{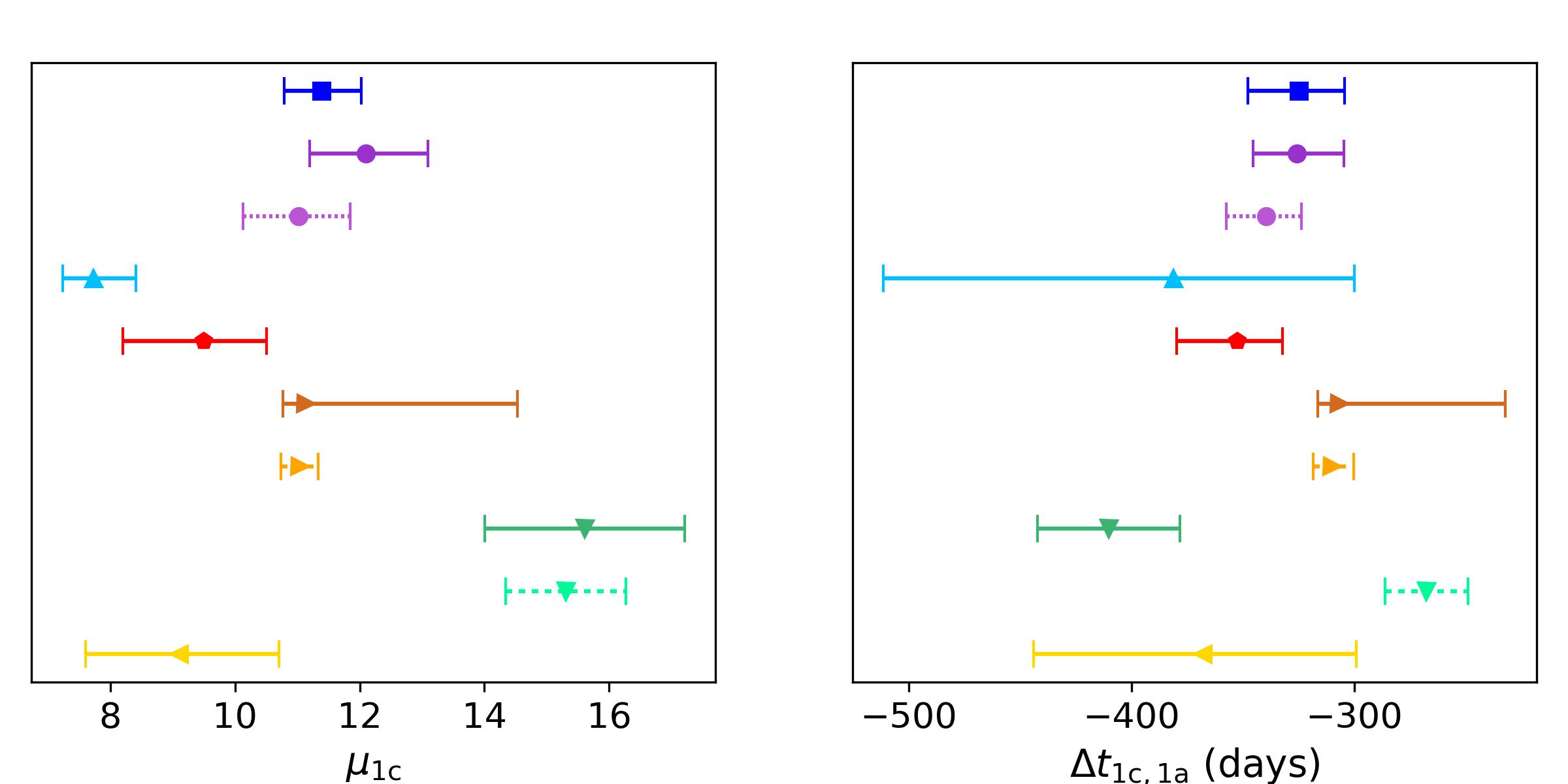}
\includegraphics[scale=0.45]{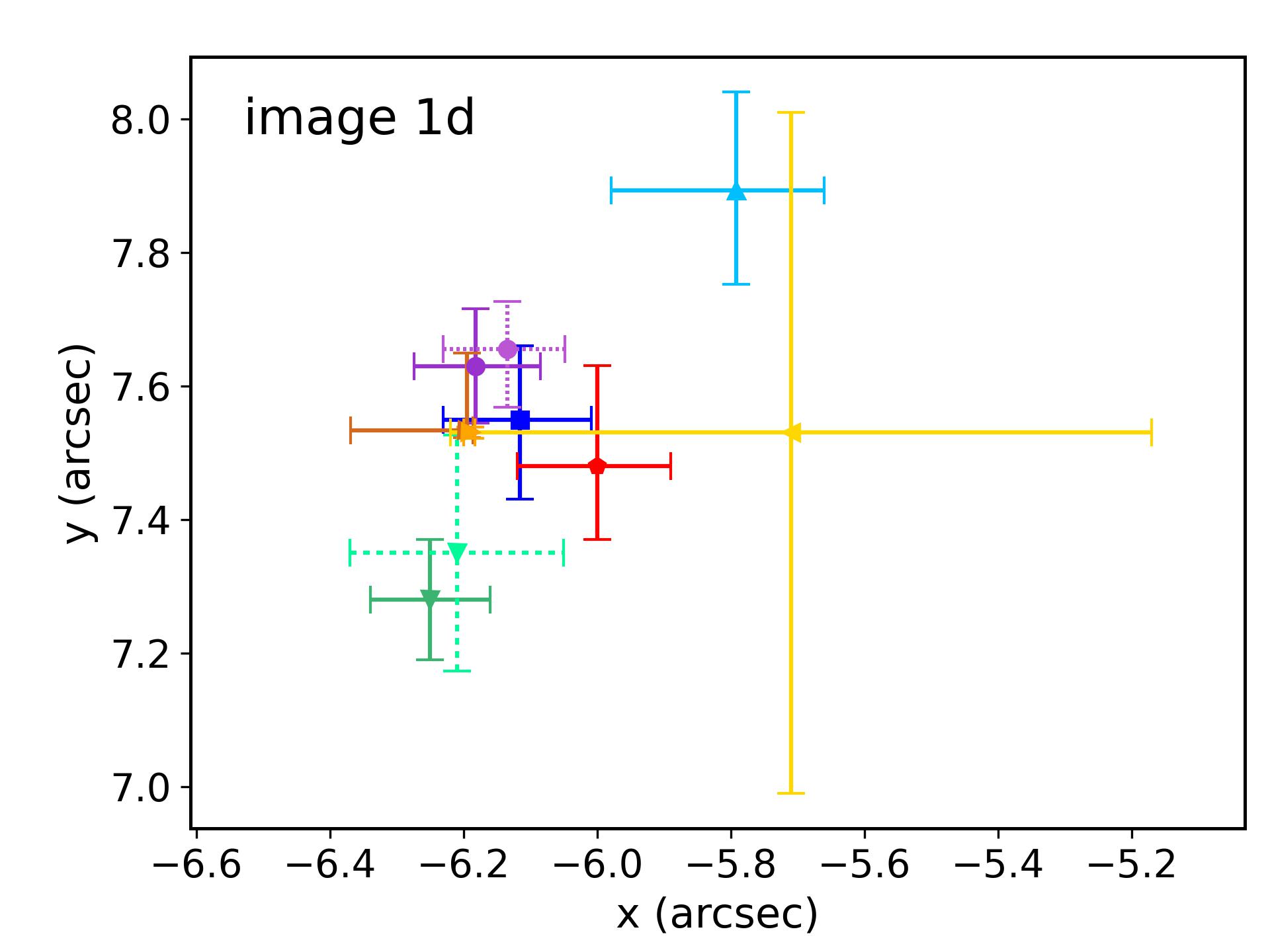}
\includegraphics[scale=0.55]{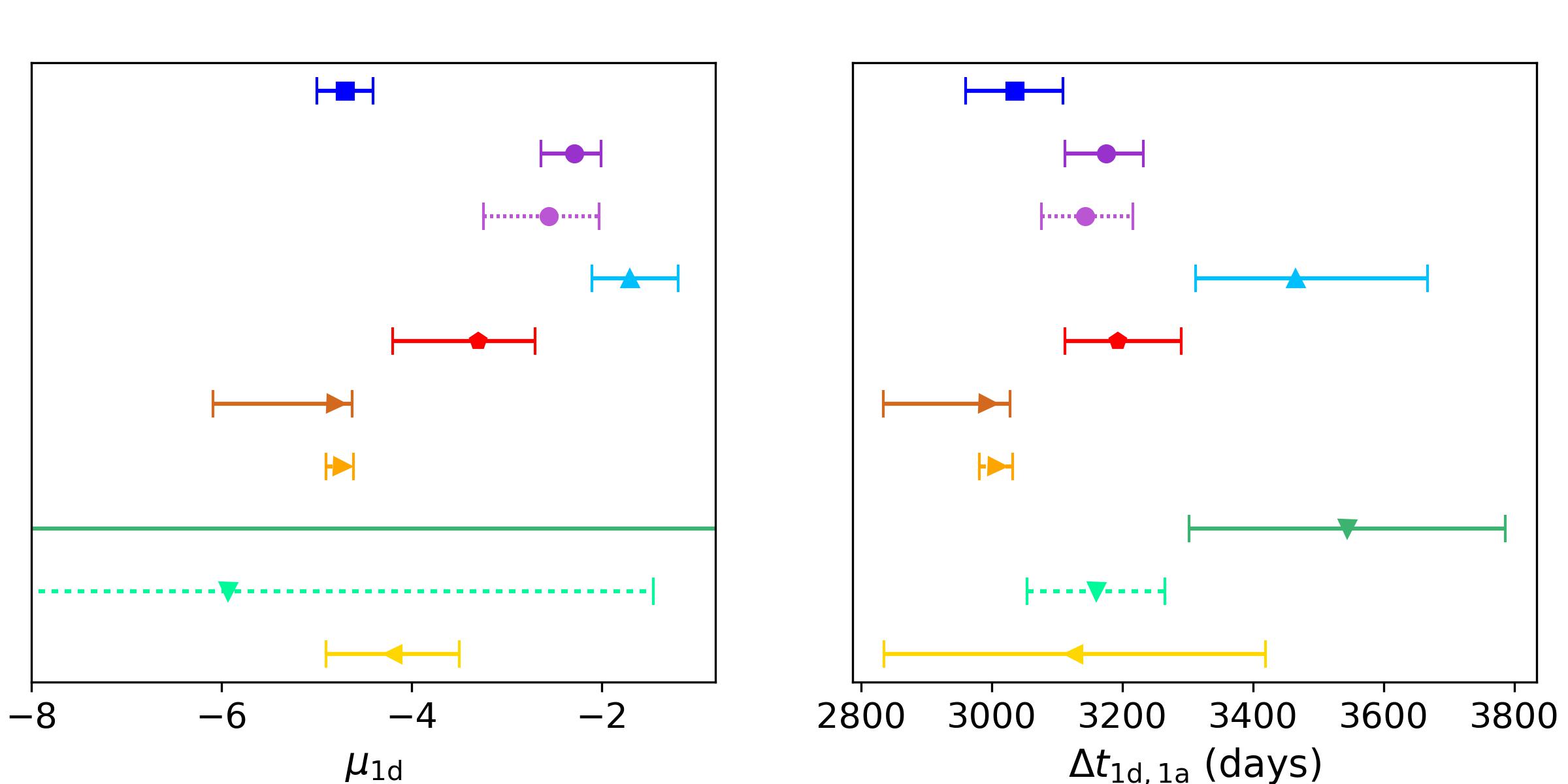}
\caption{Model predictions for SN Encore's image positions (left), magnifications (middle), and time delays (right) for a fixed cosmological model (flat $\Lambda$CDM with $\Hc = 70\,\kmsMpc$ and $\Om = 0.3 = 1-\OL$). Each row corresponds to one lensed image, as labeled in the top-left corner of each panel in the left column. The coordinate positions are relative to the BCG in arc seconds. The black circle marks the observed image positions of 1a and 1b (left panel in first and second rows, respectively) with the circular radius corresponding to the 1$\sigma$ positional uncertainty. The insets for the image positions 1a and 1b (left panels) show a zoom-in of the region near the observed image position. The time delays (in days) are relative to image 1a of SN Encore, discovered in November 2023, i.e., $\Delta t_{i,{\rm 1a}} \equiv t_{i}-t_{\rm 1a}$. }
\label{fig:SN_Encore_pos_mag_td}
\end{figure*}

\addtocounter{figure}{-1}
\begin{figure*}
\includegraphics[scale=0.45]{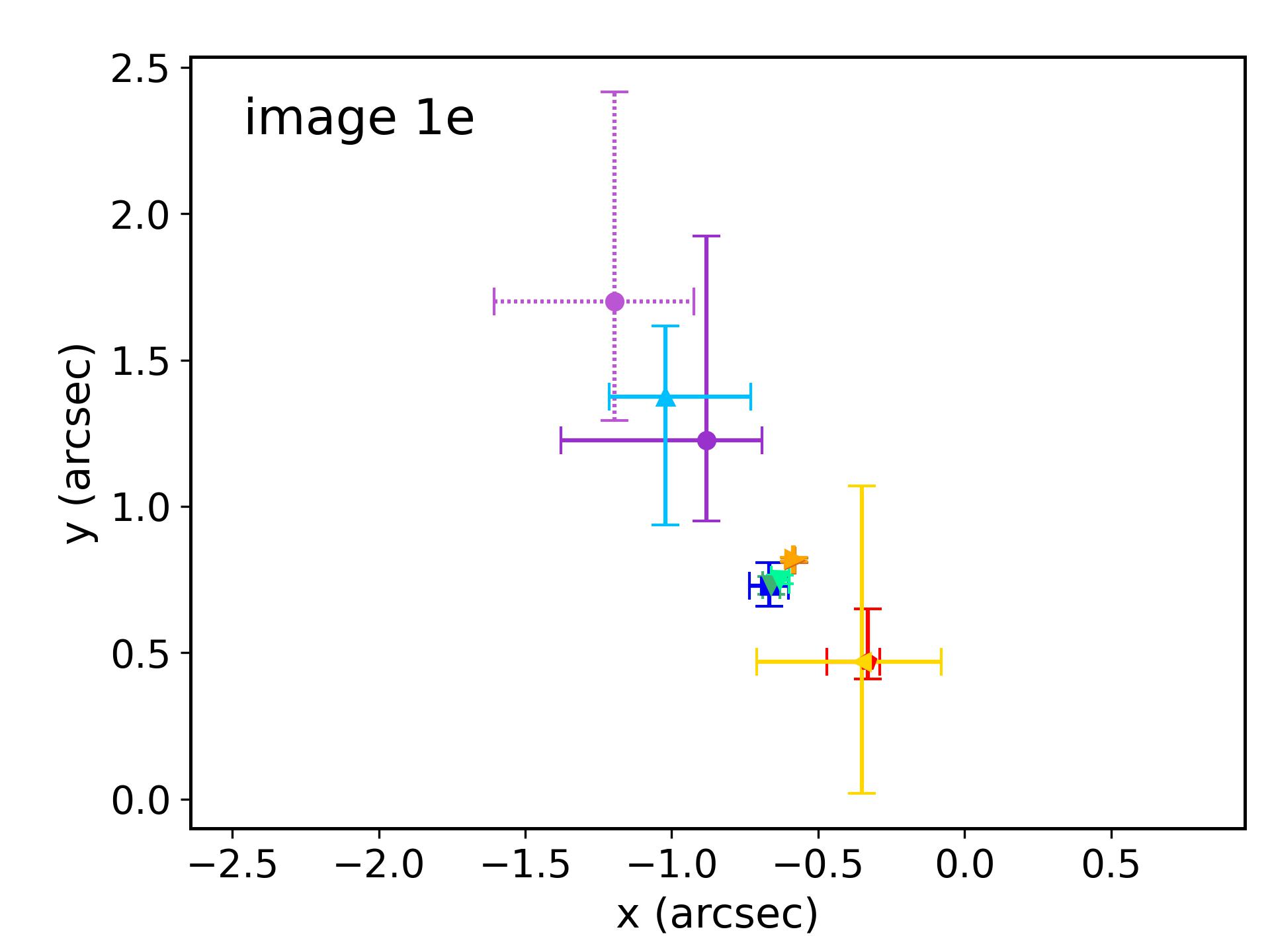}
\includegraphics[scale=0.55]{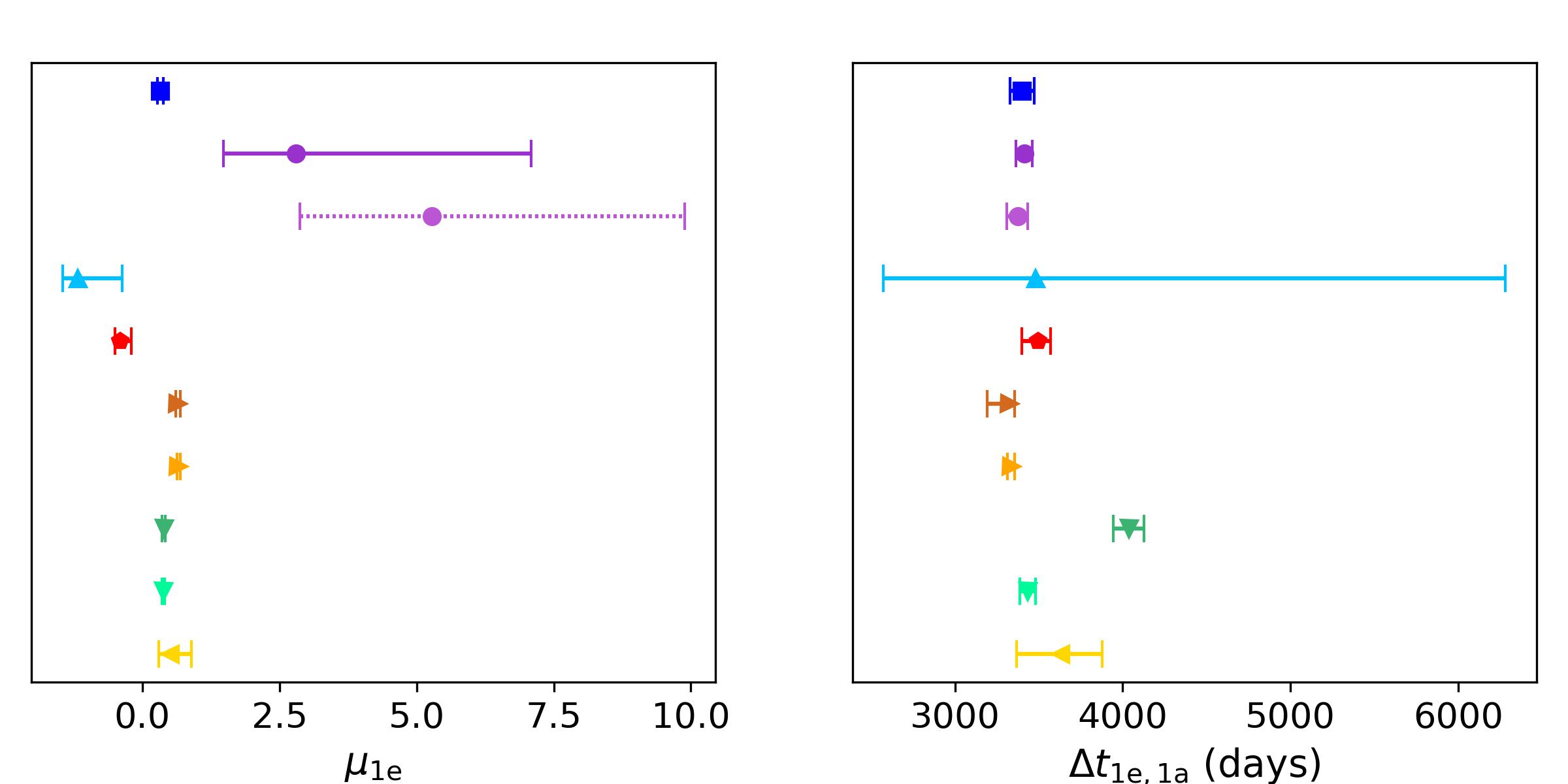}
\caption{Continued.}
\end{figure*}

\subsection{SN Requiem image positions, magnifications and time delays}
\label{sec:comparison:SNRequiem}

Similarly to Sect.~\ref{sec:comparison:SNEncore}, we compare here the model predictions from the different teams for each of the multiple images of SN Requiem, as shown in Fig.~\ref{fig:SN_Requiem_pos_mag_td} and Table~\ref{tab:snrequiem}. The general remarks about the overall consistency of the model-predicted quantities for SN Encore are valid also for SN Requiem. It is confirmed that the same three parametric models (i.e., \glafic, \GLEE, and \lenstool\ II) and the free-form \mrmartian\ model can reproduce more accurately than the other models the observed positions of images 2a, 2b, and 2c. These three images of SN Requiem are always predicted by all the models, with magnification factors mostly between $-40$ and $-30$, $20$ and $30$, and $10$ and $20$, respectively. The magnitudes of the magnification values are several times higher than those predicted by the models of \citet{Newman2018a} and \citet{Rodney+2021}, likely due to the new \jwst\ and MUSE datasets that enable more accurate mass models \citep[see also the discussion in][]{Acebron2025}. 
The predicted positions of images 2d and 2e show larger scatter, due to the presence of nearby jellyfish galaxies JF-1 and JF-2 (see Appendix \ref{app:SN_Encore_Requiem_de_pos} and Fig.~\ref{fig:Encore_Requiem_pos_radial_arc}).

The time delays of images 2b and 2c with respect to image 2a were estimated in \citet{Rodney+2021} by using color curves of SN Ia templates (see Extended Data Fig.~5 of \citet{Rodney+2021} that is based only on SN colors and not on lens mass models). Both values are close to $-120$ days, with $1\sigma$ relative errors of about 25$\%$. These measurements were not used by the teams in their model optimizations. While most of the model-predicted time delays of $\tcaRequiem$ (with $H_0$ fixed to $70\,\kmsMpc$) are consistent with the measured value within $1\sigma$, the model-predicted time delays of $\tbaRequiem$ are approximately a factor of 2 shorter than the measured ones. This is not particularly concerning, given that the measured delay of SN Requiem is based on only a single epoch of photometry and the assumption that it is of Type Ia, both of which introduce significant uncertainties that may not be fully captured in the Requiem delay measurement. In contrast, SN Encore is spectroscopically classified to be Type Ia and is observed in multiple epochs, leading to reliable time-delay measurement uncertainties for cosmography \citep{Pierel2024, Pierel+2026}.

Two additional multiple images of SN Requiem are predicted: image 2d by all teams, image 2e only by some teams. Images 2d and 2e are expected to be less magnified than the other three images, with magnification factors of approximately $-3$ and $1.5$, respectively. Both images have model-predicted time delays of about 4000 days relative to image 1a, so they should be visible in the future.

\begin{figure*}
\includegraphics[scale=0.45]{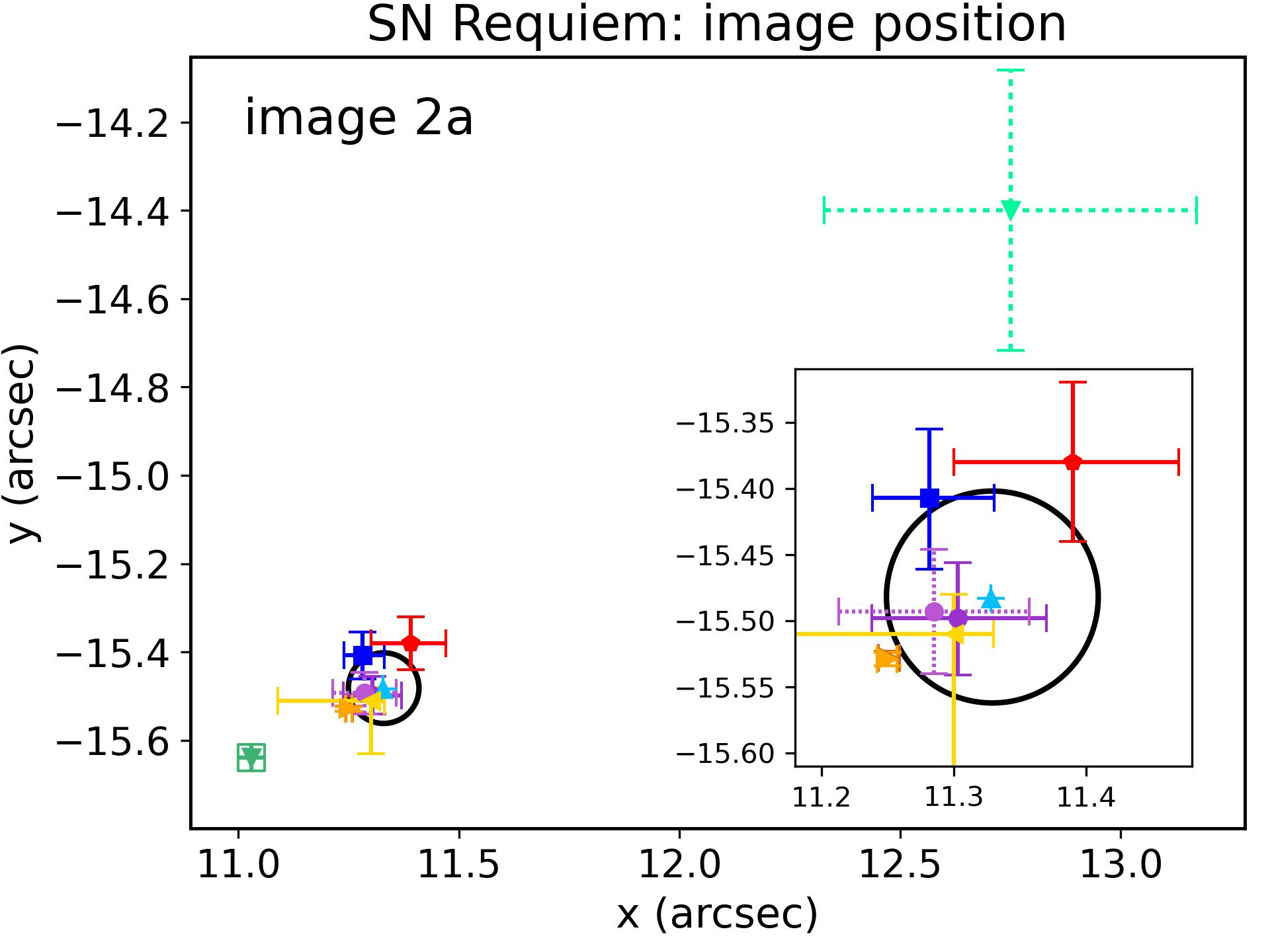}
\includegraphics[scale=0.55]{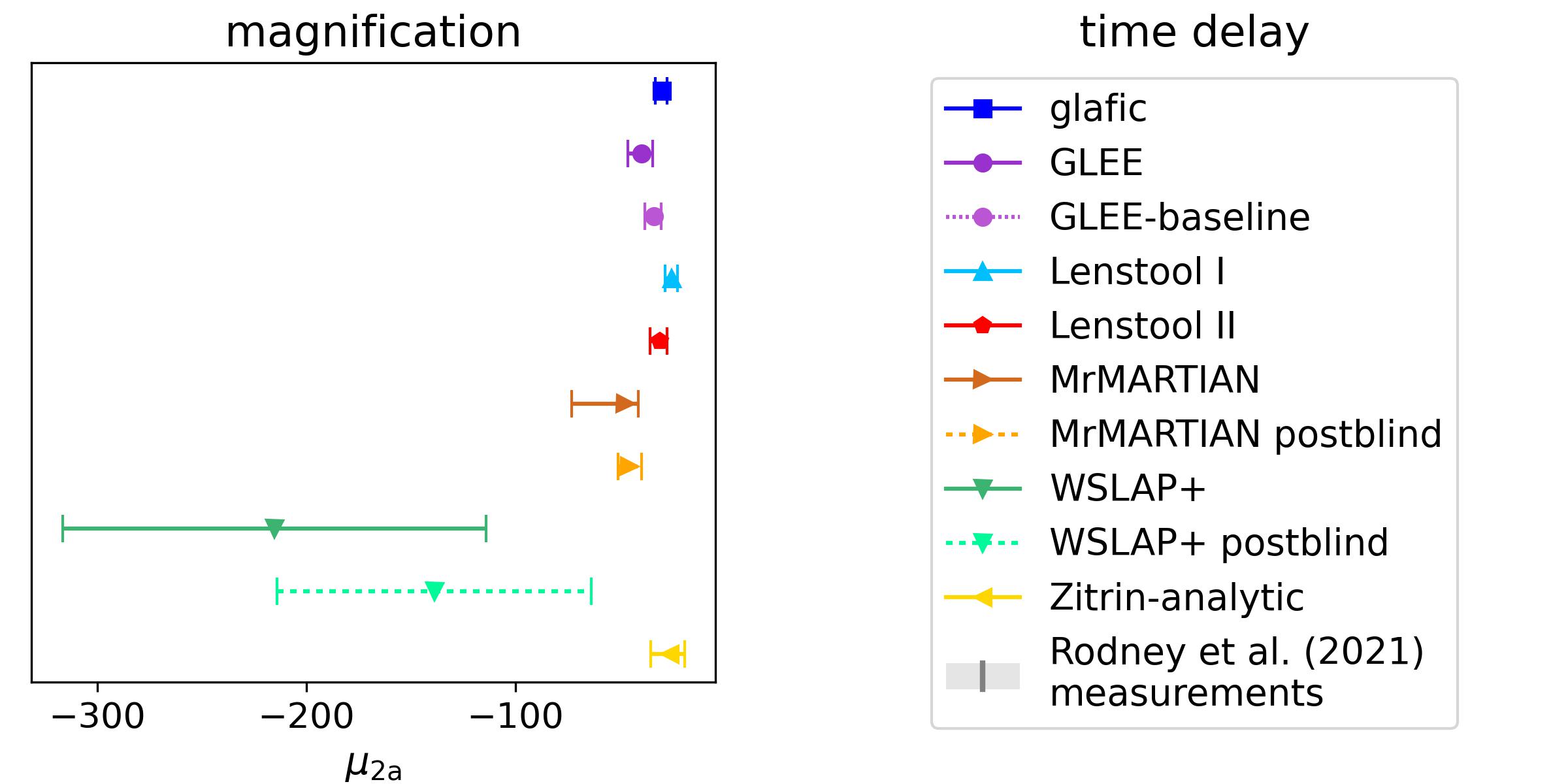}
\includegraphics[scale=0.45]{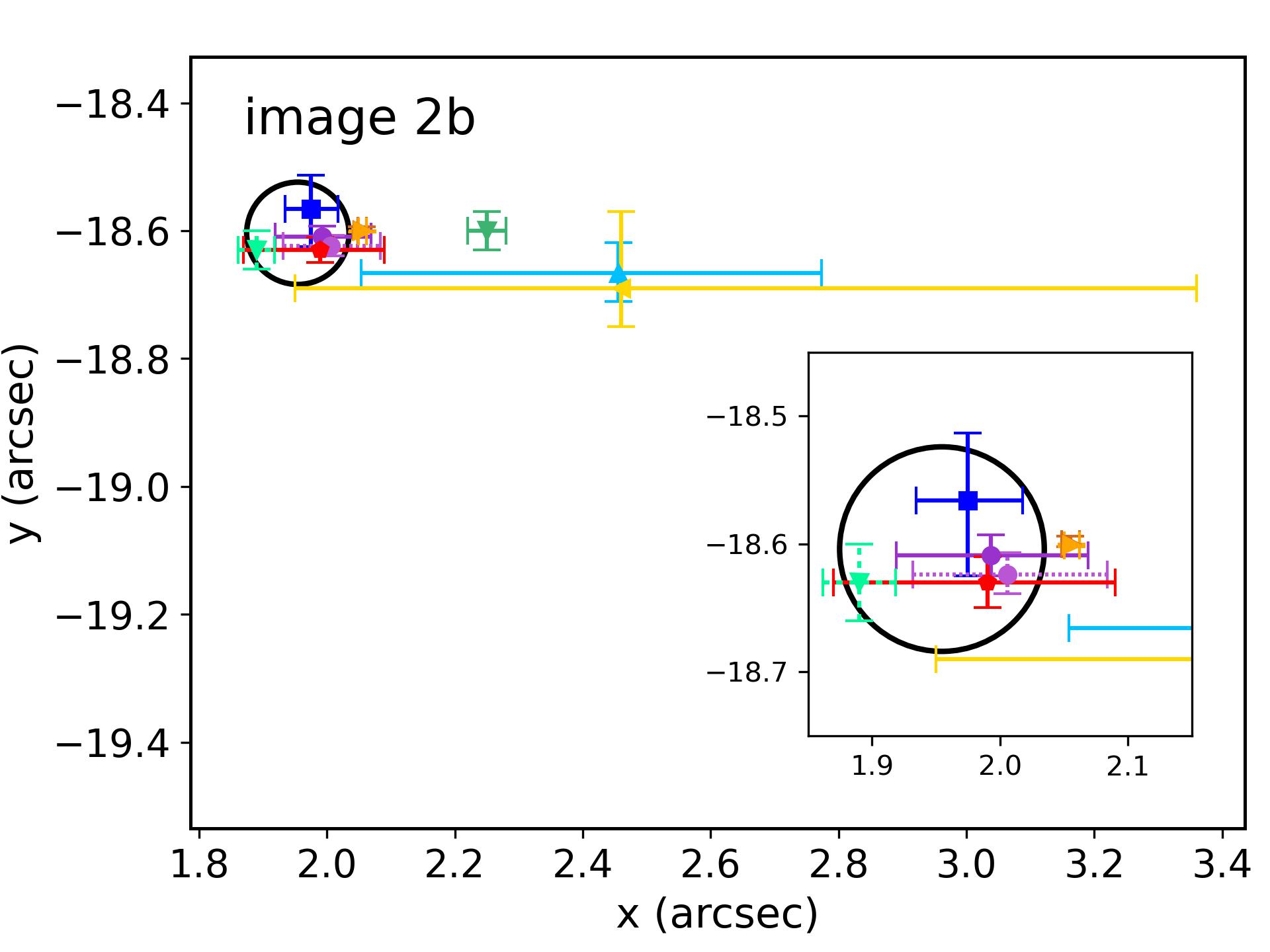}
\includegraphics[scale=0.55]{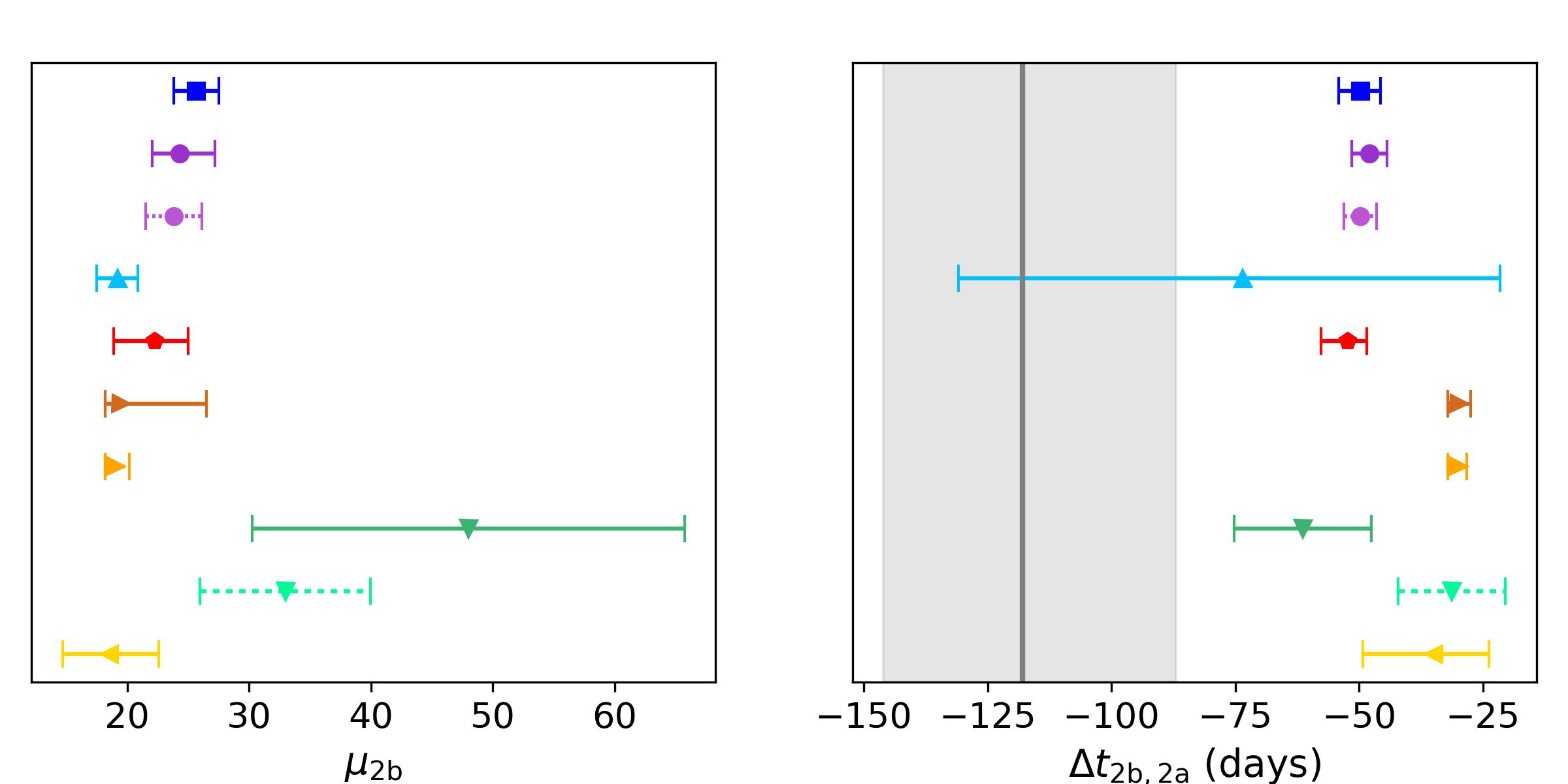}
\includegraphics[scale=0.45]{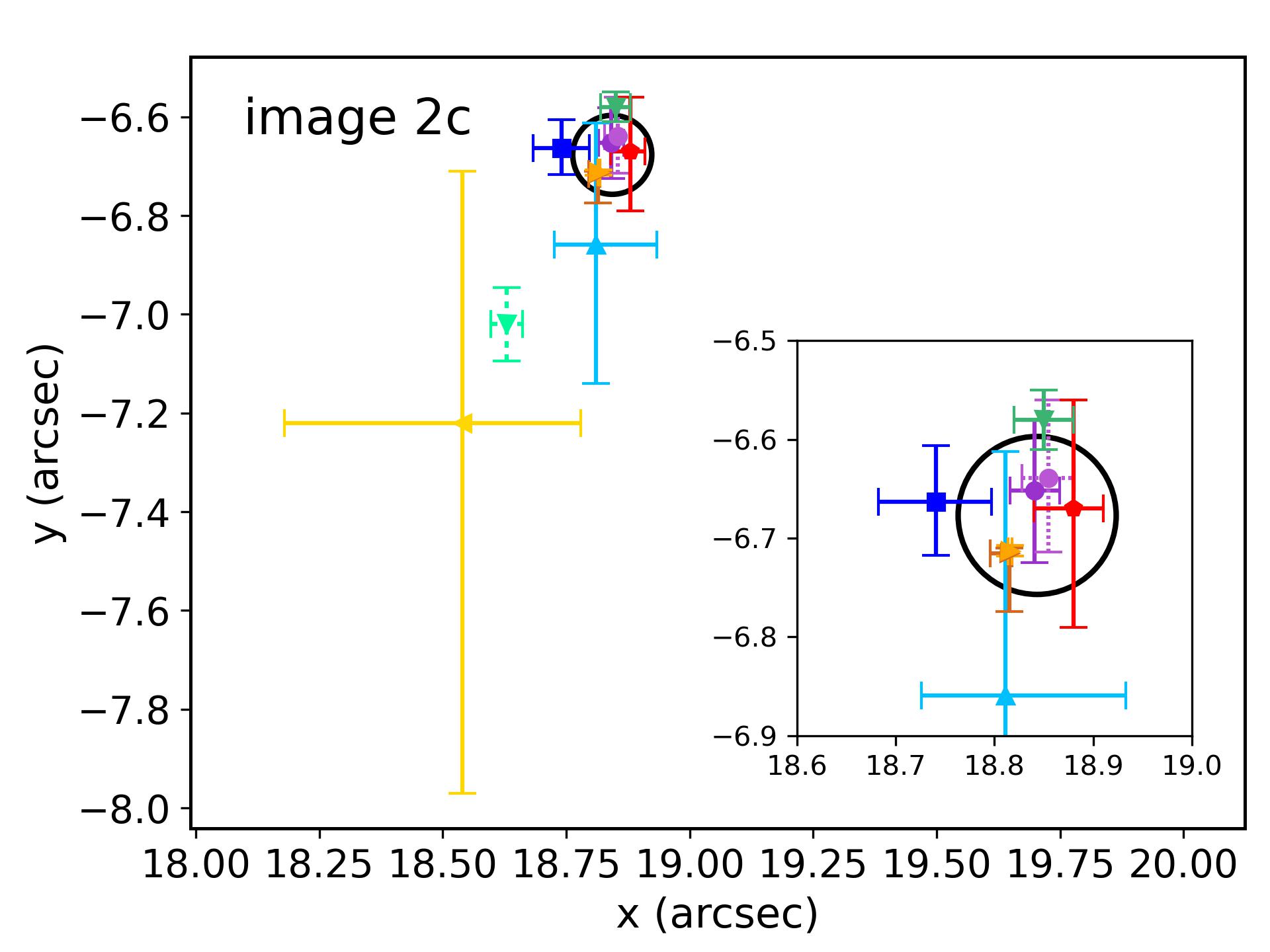}
\includegraphics[scale=0.55]{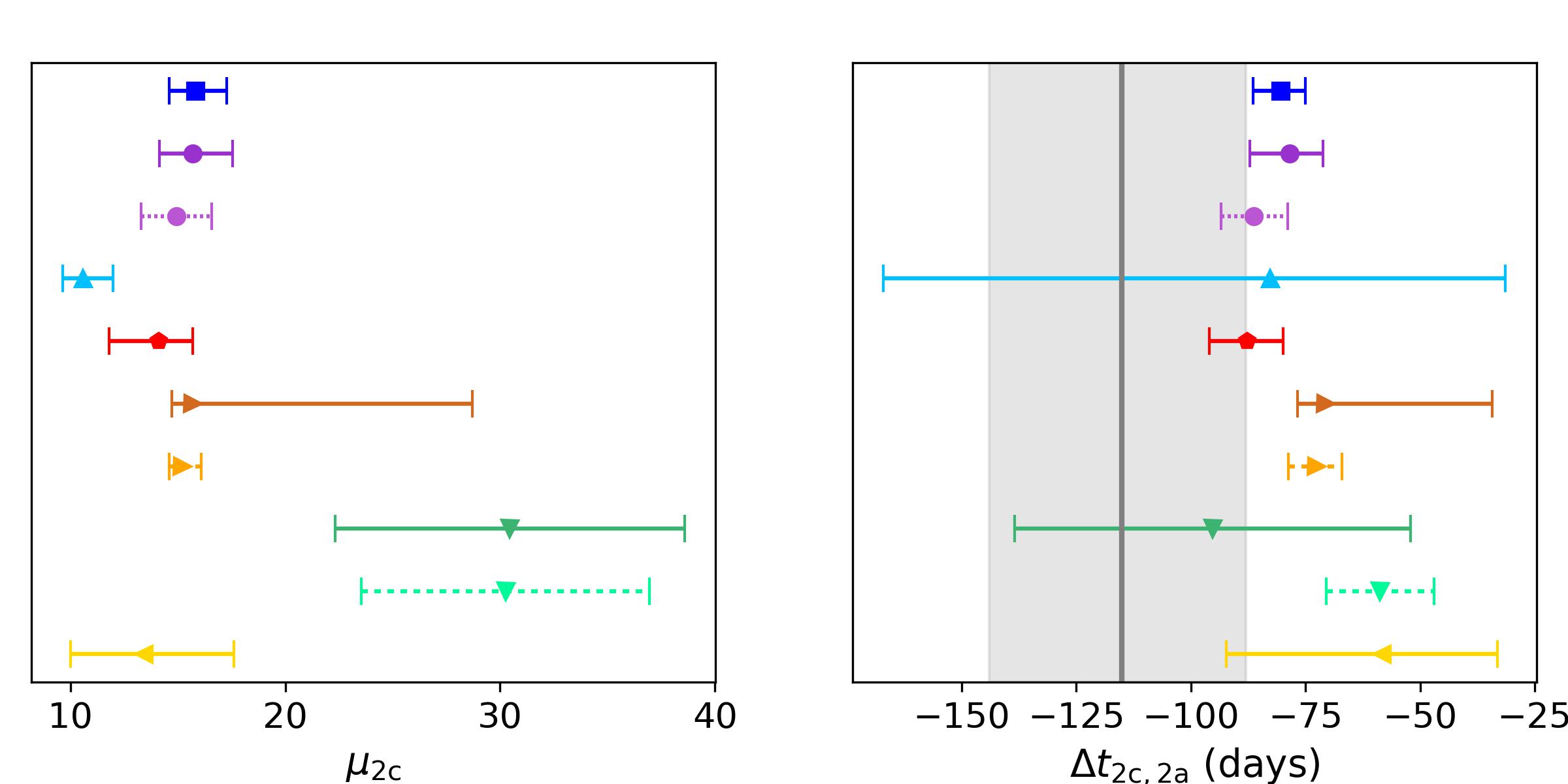}
\includegraphics[scale=0.45]{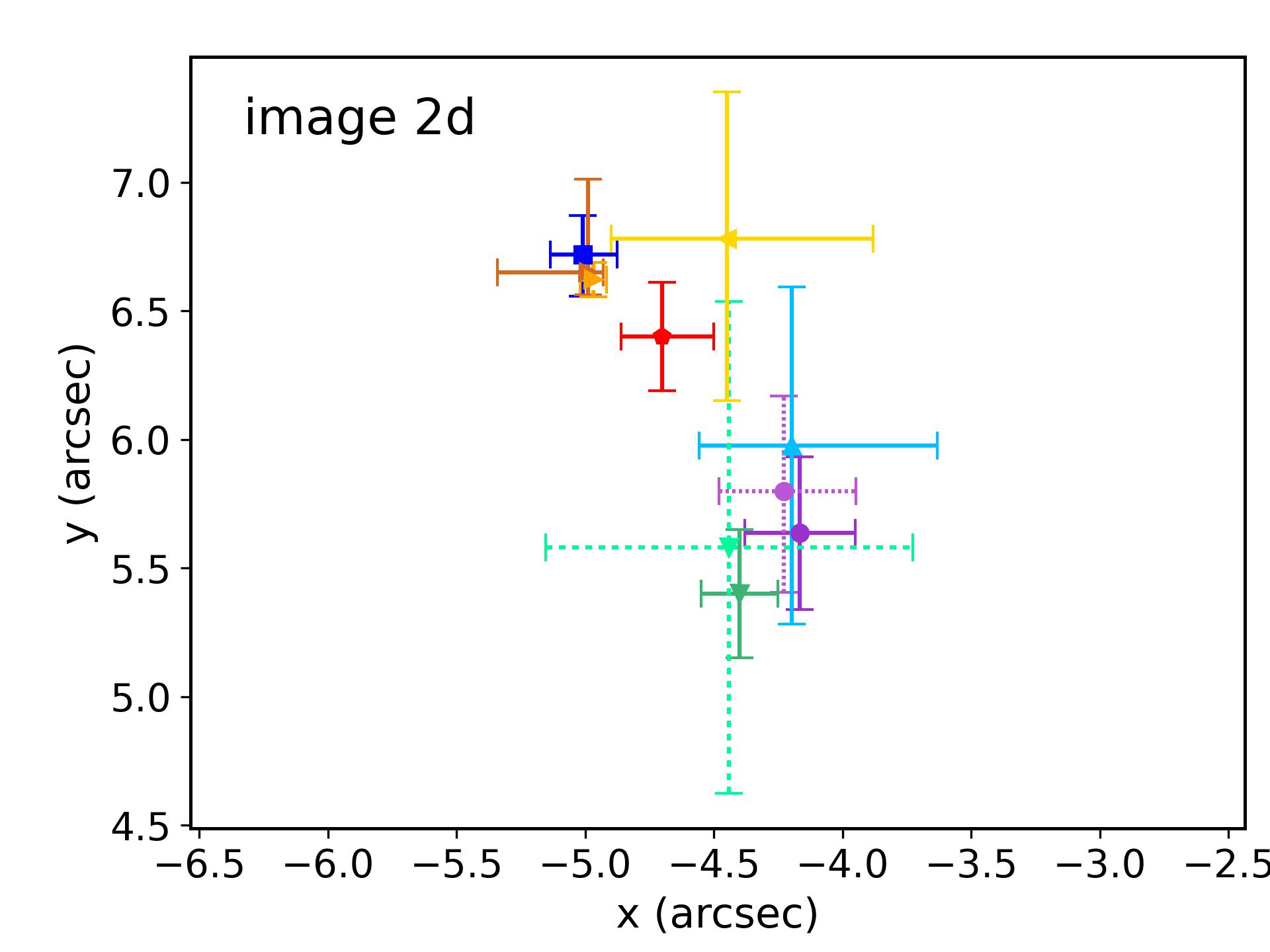}
\includegraphics[scale=0.55]{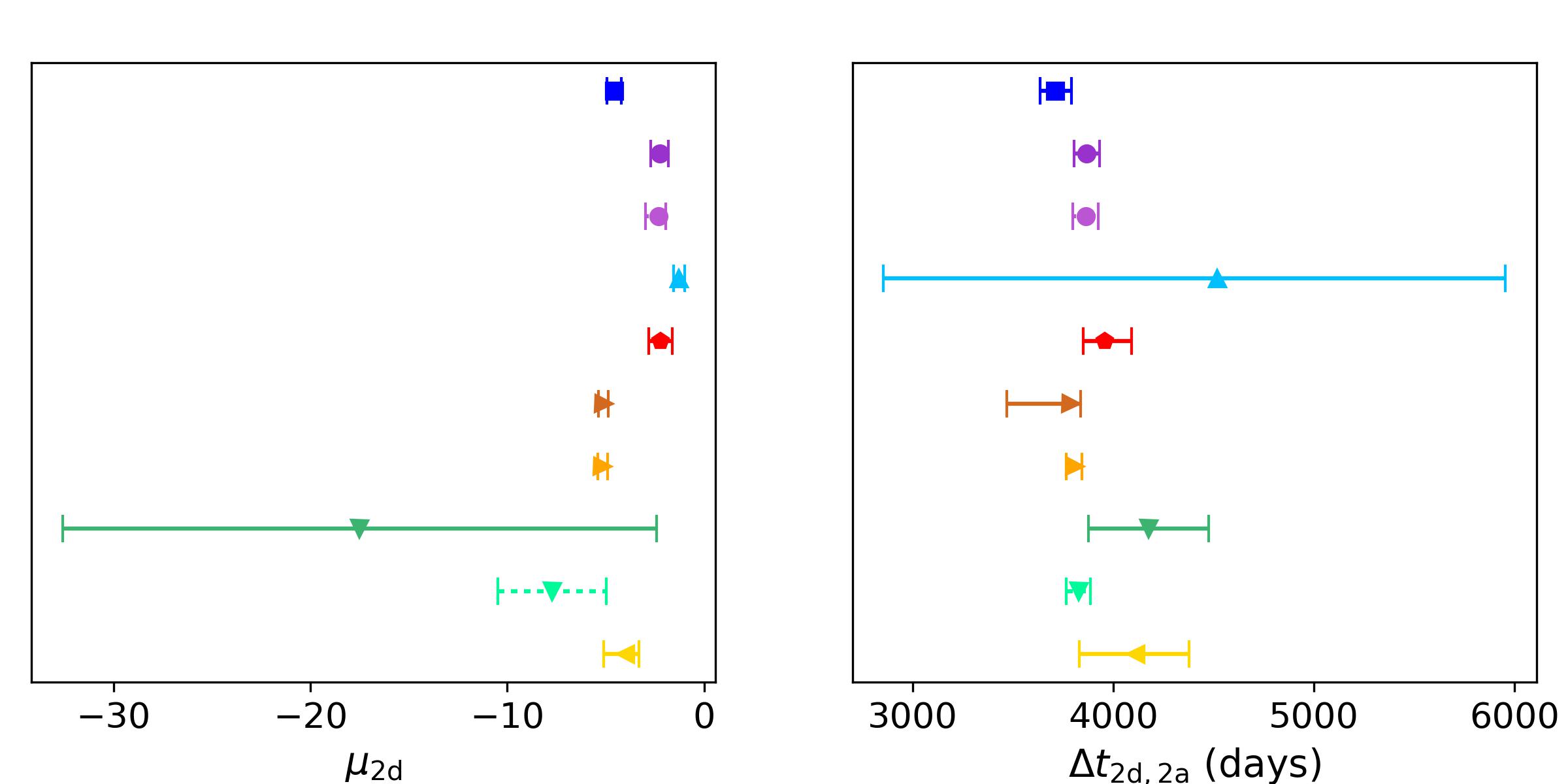}
\caption{Model predictions for SN Requiem's image positions (left), magnifications (middle), and time delays (right) for a fixed cosmological model (flat $\Lambda$CDM with $\Hc = 70\,\kmsMpc$ and $\Om = 0.3 = 1-\OL$), in the same format as Fig.~\ref{fig:SN_Encore_pos_mag_td}.  The black circle marks the observed image positions of 2a, 2b, and 2c with the circular radius corresponding to the 1$\sigma$ positional uncertainty. The time delays (in days) are relative to image 2a of SN Requiem. The measurements of $\tbaRequiem$ and $\tcaRequiem$ from \citet{Rodney+2021} based on the color curves of the SN Ia template are shown by the vertical line, with the shaded interval marking the 1$\sigma$ uncertainty.}
\label{fig:SN_Requiem_pos_mag_td}
\end{figure*}

\addtocounter{figure}{-1}
\begin{figure*}
\includegraphics[scale=0.45]{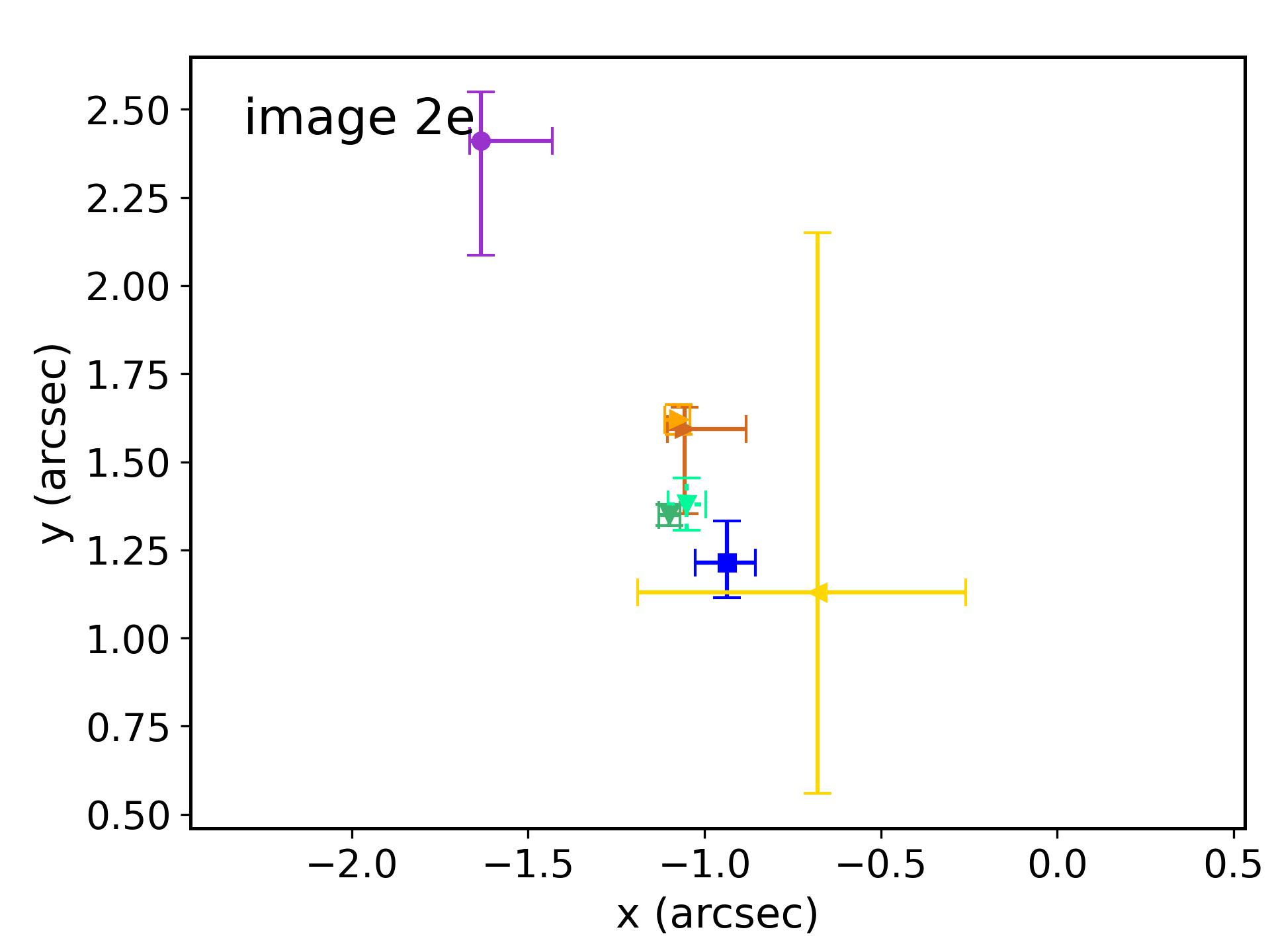}
\includegraphics[scale=0.55]{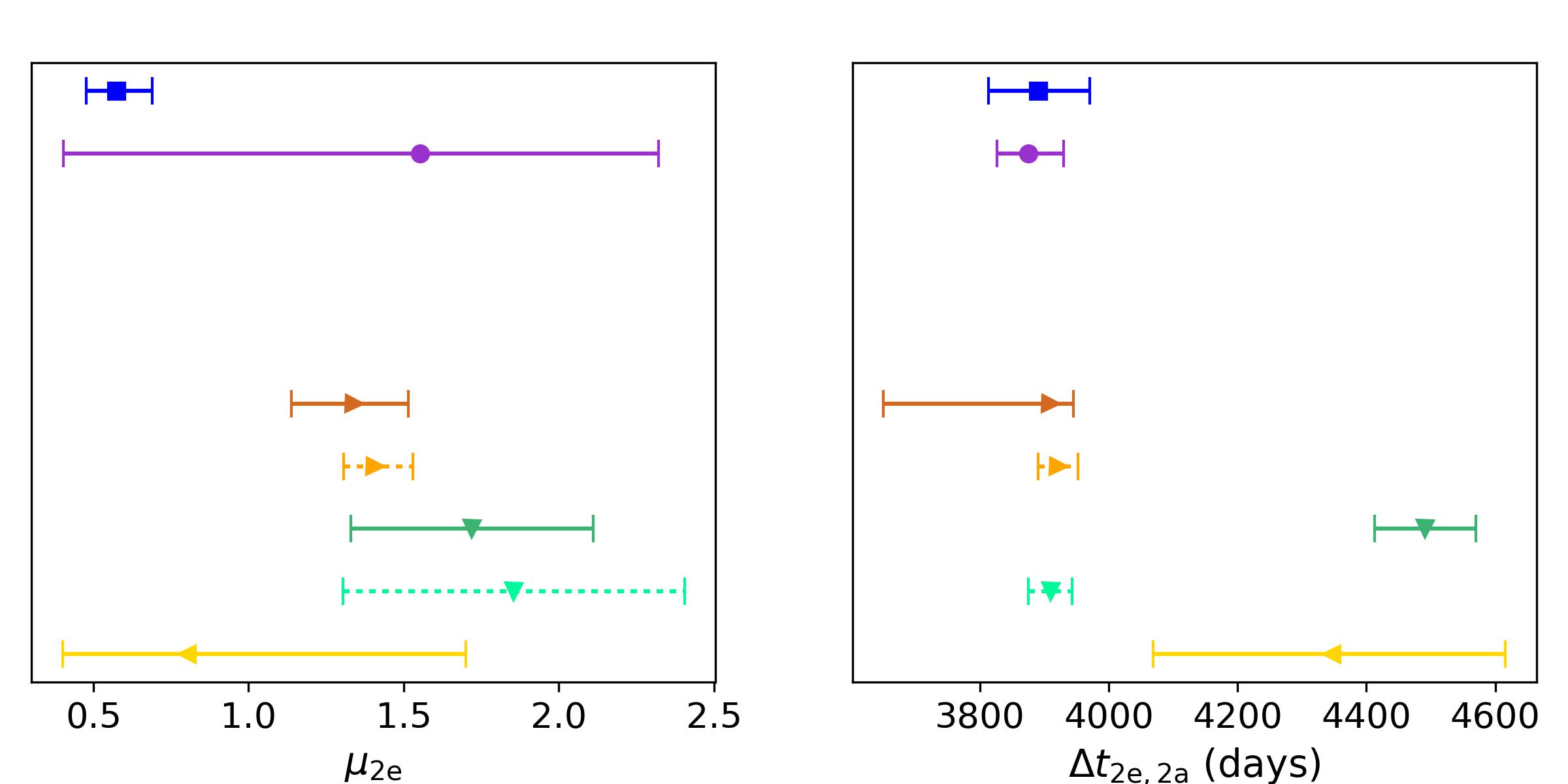}
\caption{Continued.}
\end{figure*}

\section{Relation between $H_0$ and time delays}
\label{sec:H0-td}

For a given mass model, we can predict the time delays, $\boldsymbol{\Delta t}_{\rm fc}$, for fixed cosmological parameters, where $\boldsymbol{\Delta t}_{\rm fc}$ is a vector with a length equal to the number of multiple images that will have time-delay measurements. From Sect.~\ref{sec:mass_model}, we obtain from each team the following predicted time delay for a given set of mass-model parameter values $\boldsymbol{\eta}$ and fixed cosmological parameters:
\begin{equation}
    \boldsymbol{\Delta t}_{{\rm fc},i}(\boldsymbol{\eta}, \Hc=70\,\kmsMpc, \Om = 0.3, \OL = 0.7), 
\end{equation}
where the subscript `fc' stands for fixed cosmology, and $i$ denotes the $i$th sample in the distribution.

The scaled deflection angles depend on the ratios of angular diameter distances for different lens and source redshifts; therefore, they are independent of $\Hc$ since $\Hc$ cancels in the ratio of distances. However, the ray tracing does depend on non-$\Hc$ cosmological parameters such as $\Om$ for systems with multiple-lens or source redshifts. Therefore, by fixing all cosmological parameters except for $\Hc$, the ray tracing and thus the predicted image positions remain invariant. This is true for both single-lens plane and multi-lens plane modeling. In this case, we can simply predict the time delays for a given $\Hc$ value using
\begin{equation}
\label{eq:td_h0_scale}
    \boldsymbol{\Delta t}_i (\boldsymbol{\eta}, H_0, \Om=0.3, \OL=0.7) =  \frac{\Hcfc}{\Hc} \boldsymbol{\Delta t}_{{\rm fc},i},
\end{equation}
where $\Hcfc=70\,\kmsMpc$.

Armed with Eq.~(\ref{eq:td_h0_scale}), we can derive the constraints on $H_0$ for hypothetical time-delay measurements in a flat $\Lambda$CDM cosmology with $\Om=0.3$. Suppose we have hypothetical time-delay measurements of $\boldsymbol{\Delta t}_{\rm data}$ with the covariance matrix between the delay measurements denoted by $C_{\rm \Delta t}$. The time-delay likelihood of the measurements ($\boldsymbol{\Delta t}_{\rm data}$) given model-predicted delays ($\boldsymbol{\Delta t}_{\rm model}$) is 
\begin{equation}
\label{eq:td_likelihood}
\begin{split}
    \mathcal{L}_{\Delta t} = & \frac{1}{Z} \exp \Bigl[-\frac{1}{2} \left(\boldsymbol{\Delta t}_{\rm data}-\boldsymbol{\Delta t}_{\rm model}\right)^{\rm t} C_{\rm \Delta t}^{-1} \Bigr. \\& 
    \Bigl. \left(\boldsymbol{\Delta t}_{\rm data}-\boldsymbol{\Delta t}_{\rm model}\right)\Bigr],
\end{split}
\end{equation}
where $^{\rm t}$ denotes transpose and $Z$ is the normalization constant. We can importance sample the original distribution of time delays $\boldsymbol{\Delta t}_{{\rm fc},i}$ to obtain the resulting $\Hc$ constraint as follows. For each sample $i$ in the distribution, we
(i) draw a random $H_0$ value, (ii) compute the predicted time-delay $\boldsymbol{\Delta t}_i$ via Eq.~(\ref{eq:td_h0_scale}), (iii) compute the likelihood $\mathcal{L}_{\Delta t,i}$ of the hypothetical time-delay measurement $\boldsymbol{\Delta t}_{\rm data}$ and predicted delay $\boldsymbol{\Delta t}_i$ (i.e., $\boldsymbol{\Delta t}_{\rm model}=\boldsymbol{\Delta t}_i$ in Eq.~(\ref{eq:td_likelihood})), and use this as weight $w_i$ for the sample $i$.  The weighted distribution of $H_0$ is then the probability distribution of $H_0$.

For background cosmological models where non-$H_0$ parameters are not the flat $\Lambda$CDM model with $\Om=0.3=1-\OL$, then the ray tracing changes in the modeling and the corresponding image position $\chi_{\rm im}^2$  changes as well. To get cosmological constraints in these cases, we would need to sample the joint image-position and time-delay likelihoods with our mass model and cosmological parameters \citep[see, e.g.,][]{Grillo+2024}. Since we completed the lens mass modeling before the time-delay measurements of SN Encore as the core experimental design of our blind analysis, we therefore defer such $H_0$ inference in more general cosmological models to future work. 

For SN Encore, where images 1a and 1b are clearly visible in our JWST data and image 1c is barely visible, it may be possible to derive two time delays with the existing data. The 1b-1a delay, $\tbaEncore$, will likely be substantially more precise than $\tcaEncore$, given that image 1c is faint. In this case, the cosmological inference will be mostly determined by $\tbaEncore$.  We therefore consider here only $\tbaEncore$ in order to derive the relation between $\Hc$ and the hypothetical time-delay values. This will avoid exploring implausible combinations of $\tbaEncore$ and $\tcaEncore$, which we cannot currently foresee since we do not know the time delays, in order to keep our analysis blind. 
\

\begin{figure*}
\includegraphics[scale=0.45]{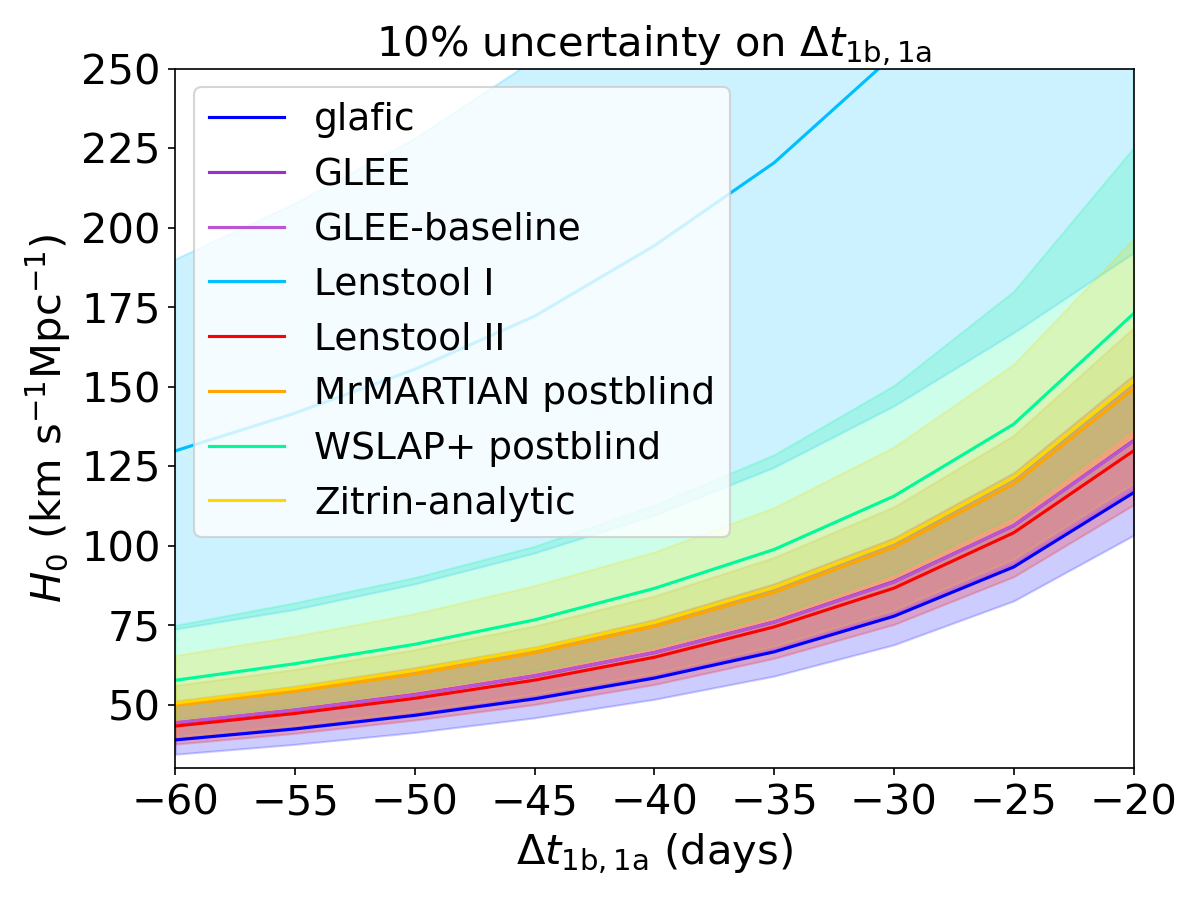}
\includegraphics[scale=0.33]{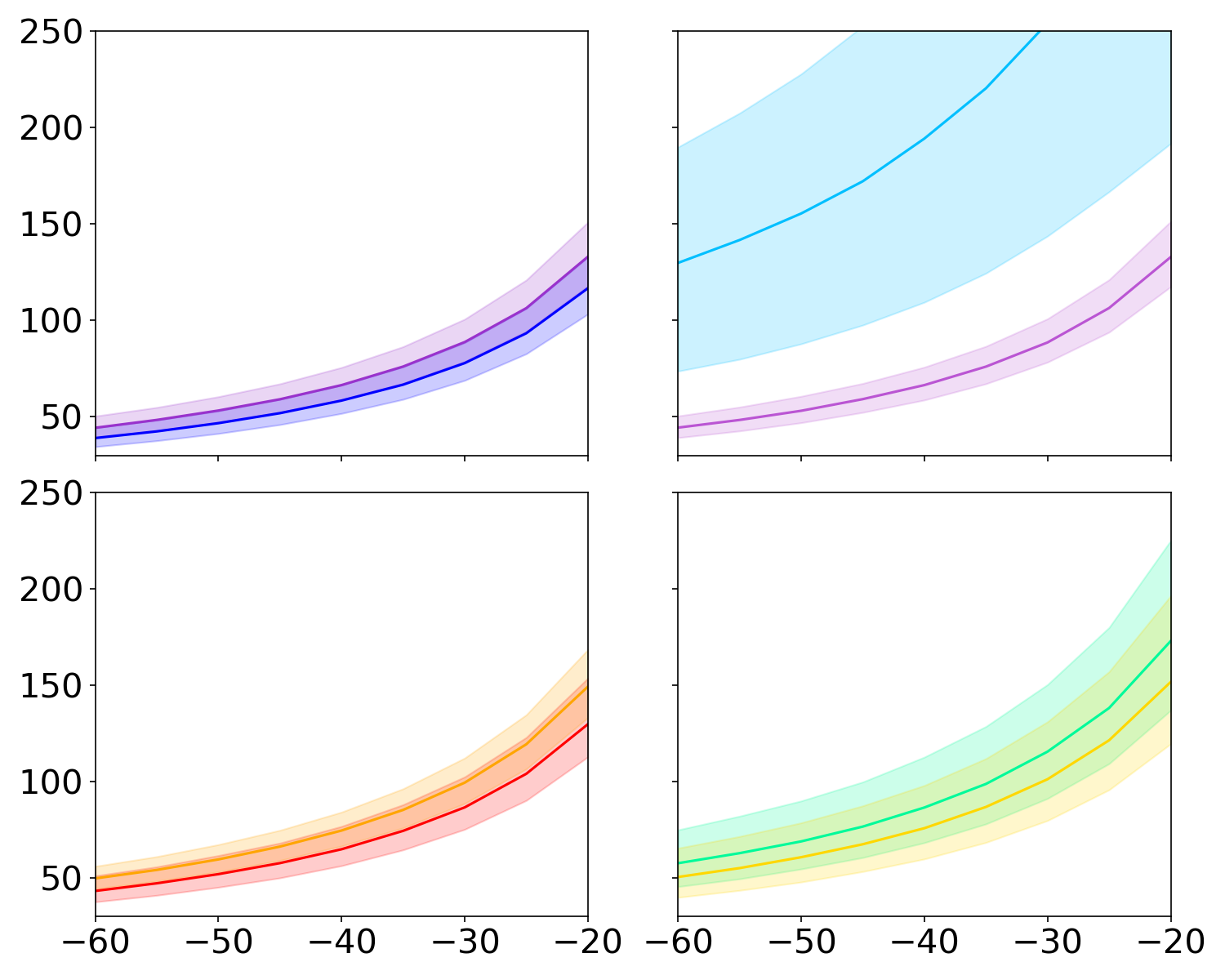}

\caption{Resulting $H_0$ from different mass models for a range of possible $\tbaEncore$ measurements with 10\% uncertainty from SN Encore in a flat $\Lambda$CDM cosmology with $\Om = 1-\OL = 0.3$.  The median values of $H_0$ are shown as solid lines, while the shaded regions indicate the 1$\sigma$ uncertainty. Left panel: All model predictions overlaid, as indicated in the legend. Right four panels: Subset of the model predictions plotted in the same style as in the left panel, for better visibility. }
\label{fig:H0_td_1b1a}
\end{figure*}

We considered 10\% relative uncertainties on $\tbaEncore$ of SN Encore and obtained Fig.~\ref{fig:H0_td_1b1a}, which ranges from $-60$ to $-20$ days, covering the 1$\sigma$ range of predictions from five of the seven teams and lying partly within 2$\sigma$ of the predictions from the remaining two teams. The corresponding $\Hc$ values range approximately from 40 to 170 $\kmsMpc$ for six of the seven models (except for the \lenstoolone\ model that spans to $\sim$330 $\kmsMpc$), with minimum relative errors of about 12\% for the 10\% relative uncertainties on $\tbaEncore$.

\begin{figure*}
\includegraphics[scale=0.45]{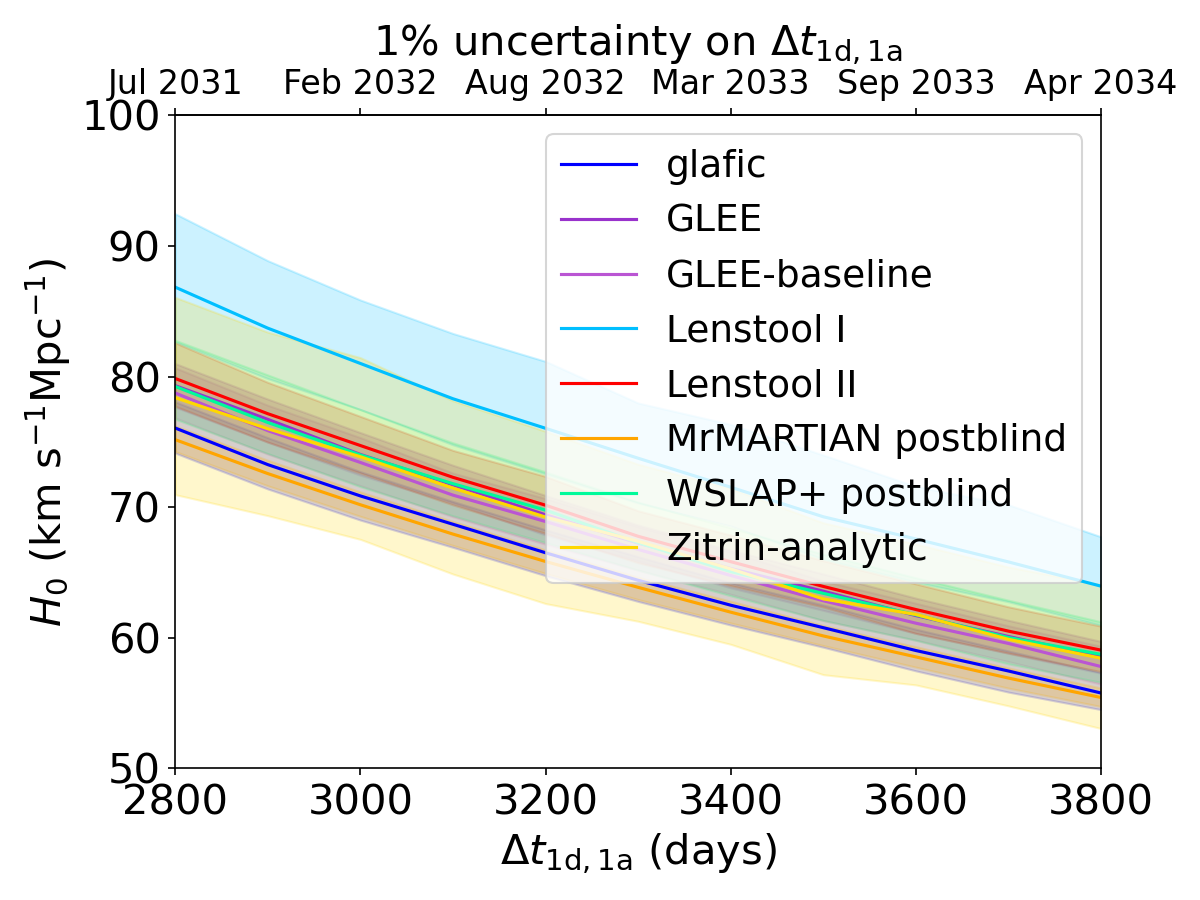}
\includegraphics[scale=0.33]{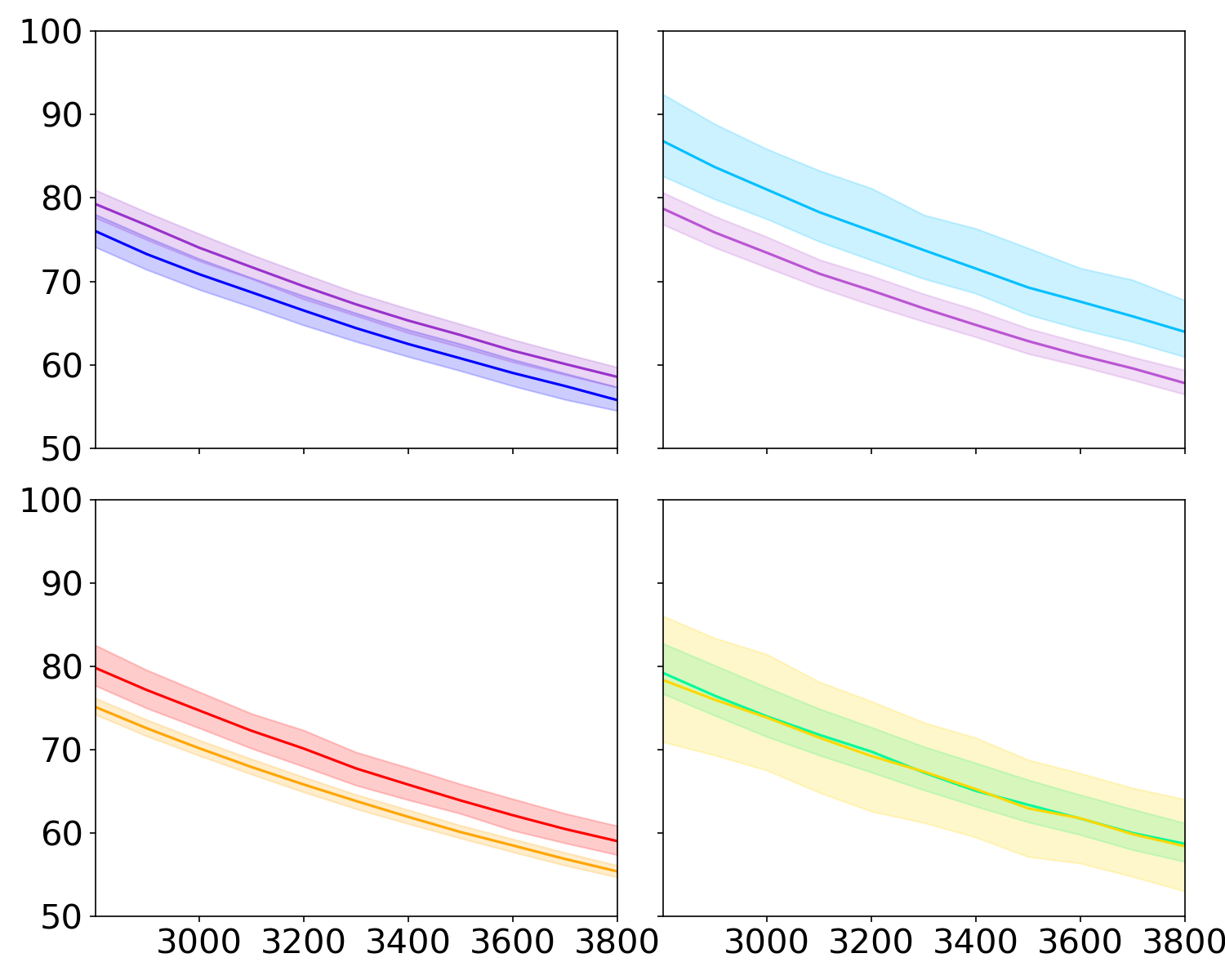}
\caption{Resulting $H_0$ from different mass models for a range of possible $\tdaEncore$ measurements from SN Encore in a flat $\Lambda$CDM cosmology with $\Om = 1-\OL = 0.3$.  The median values of $H_0$ are shown as solid lines, and the shaded regions indicate the 1$\sigma$ uncertainty.  The assumed 1\% uncertainty on $\tdaEncore$ is achievable given the long time delay of $\sim$$3000$ days. }
\label{fig:H0_td_1d1a}
\end{figure*}

\begin{figure*}
\includegraphics[scale=0.45]{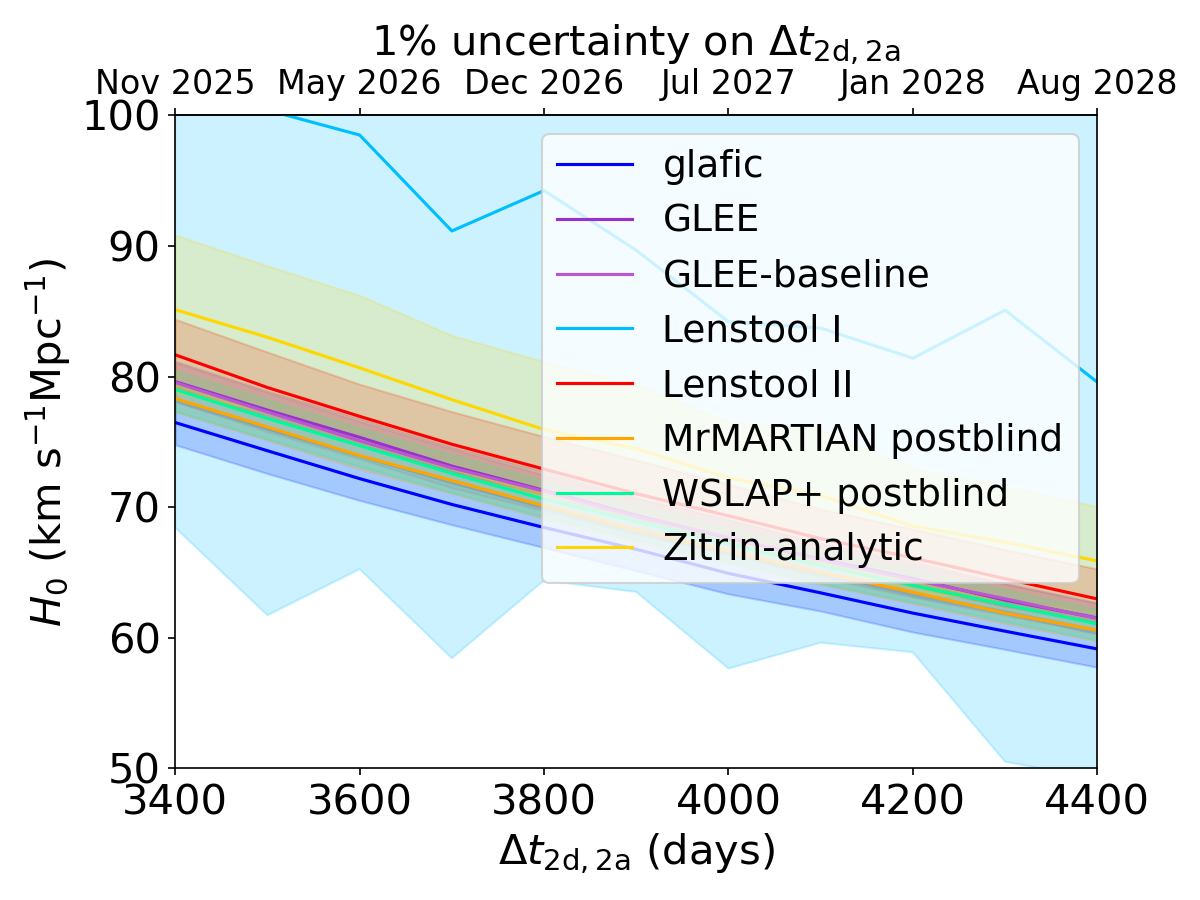}
\includegraphics[scale=0.33]{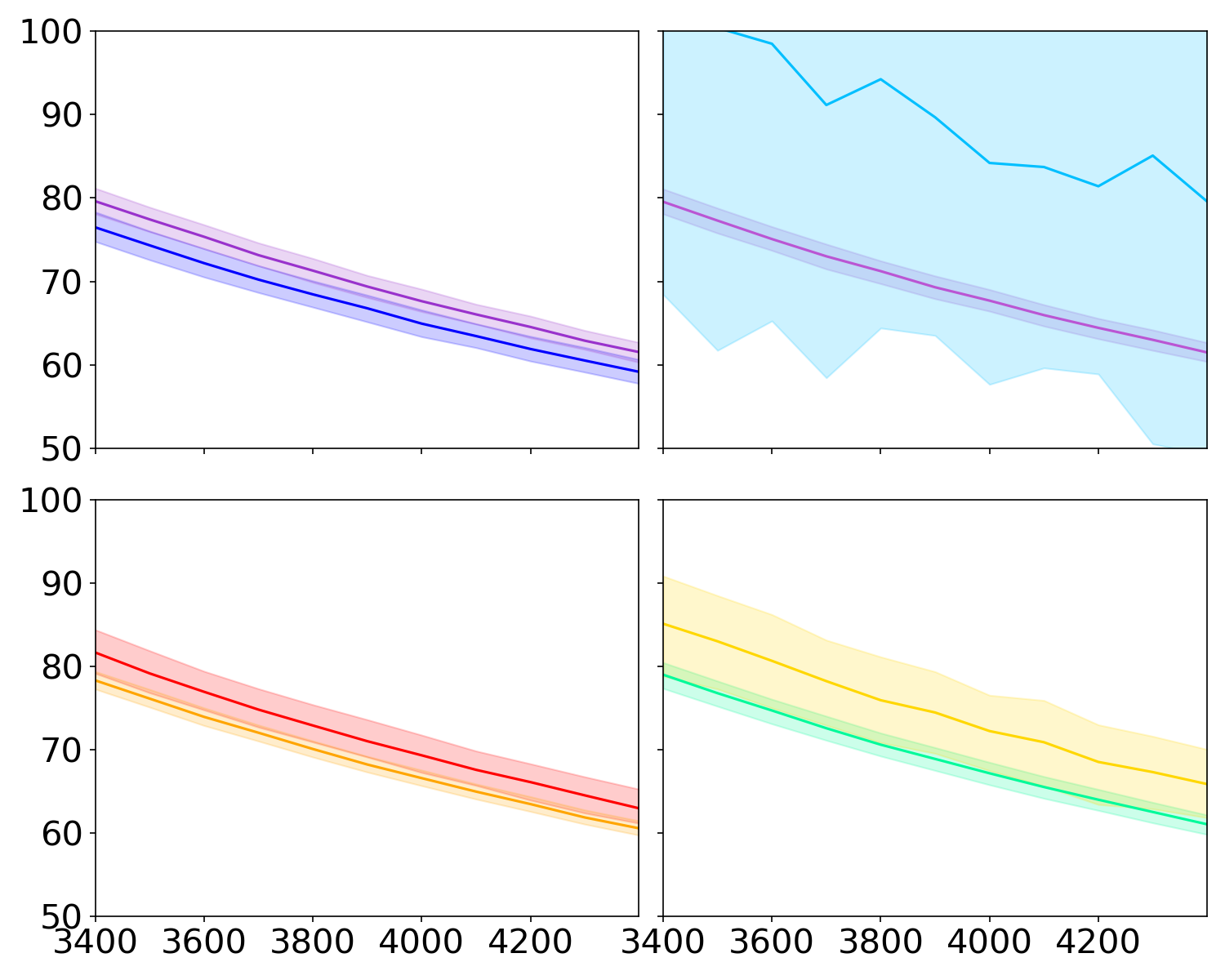}
\caption{Resulting $H_0$ from different mass models for a range of possible $\tdaRequiem$ measurements from SN Requiem in a flat $\Lambda$CDM cosmology with $\Om = 1-\OL = 0.3$.  The median values of $H_0$ are shown as solid lines, and the region delineated by shades indicate the 1$\sigma$ uncertainty.  The relation from the \lenstoolone\ model exhibits larger statistical fluctuations since image 2d is predicted in only 16\% of the samples for that model (see Table \ref{tab:snrequiem}).  The assumed 1\% uncertainty on $\tdaRequiem$ is achievable given the long time delay of $>$$3000$ days. }
\label{fig:H0_td_2d2a}
\end{figure*}

In Figs.~\ref{fig:H0_td_1d1a} and \ref{fig:H0_td_2d2a}, we show potential estimates of the value of $\Hc$ from hypothetical time-delay measurements of the future multiple images 1d and 2d of, respectively, SNe Encore and Requiem. Given the long (between $\approx$3000 and $\approx$4000 days) model-predicted time delays, it might be possible to measure them with a relative uncertainty as low as 1\%. A measurement of one of these time delays would result in a $\Hc$ measurement with a competitive statistical relative error of $\approx$ 2-3\%. Interestingly, according to our models, it is likely that the next reappearance of SN Requiem (image 2d) could be detected already at the end of 2025, while that of SN Encore (image 1d) not earlier than mid-2031. Only HST/JWST imaging might be able to reveal these faint and unique transient events.

\section{$H_0$ inference from the first SN Encore time-delay measurement}
\label{sec:H0}

\subsection{Time-delay measurement $\tbaEncore$}
\label{sec:H0:td}

\citet{Pierel+2026} obtained the photometry of SN Encore images 1a and 1b with two different methods, while the photometry of image 1c cannot be robustly determined with existing data.  \citet{Pierel+2026} measured a time delay of $\tbaEncore = \tbaEncoreMeasured$ days using their photometric measurements, SN Ia template light curves from BayesSN \citep{Mandel+22, Ward+23, Grayling+24} and two different fitting methods, {\sc Glimpse} \citep{Hayes+24, Hayes+2026} and SNTD \citep{PierelRodney2019}.  The measurement uncertainties account for the effect of microlensing using a suite of microlensing simulations based on four different models of SN Ia progenitors, following the microlensing simulations of \citet{Huber+2021}.  The time-delay uncertainties of $\sim$9-10\% are currently dominated by the photometric precision.

\subsection{Combination of lens mass models for cosmography}
\label{sec:H0:combo_model}

The independent mass models built by the seven teams using a variety of software allow us to quantify systematic uncertainties arising from lens modeling software and choices. By combining these models to infer $H_0$, our measurement incorporates both statistical and systematic uncertainties associated with cluster mass modeling.  

One key aspect in the combination of models is the weighting of the models, since the models have different numbers of parameters and goodness of fit to the observables, especially the multiple image positions. With the wide range of software used, including both parametric and free-form approaches, and the different priors adopted by the teams on some of the model parameters, the number of degrees of freedom is highly nontrivial to compute. The BIC values in Table \ref{tab:model_stat} do not fully account for the subtleties associated with the adopted priors on the parameters. Furthermore, in the case of free-form mass models, the methods for computing the effective number of free parameters for \mrmartian\ can alter the BIC values significantly (as shown in Table \ref{tab:model_stat}).  Therefore, for inferring $H_0$, we weighted the models based on their maximum likelihood of the observables, which is the product of the likelihood of the observed image positions and the likelihood of the time delay. Given the single time-delay measurement, $\tbaEncore$, the value of the parameter $H_0$ can adjust to each model and yield the same maximum time-delay likelihood. Thus, the weights of each model was determined solely by the maximum likelihood of the observed image positions.  

We tabulate in Table \ref{tab:model_stat} the value of the multiple image positions likelihood $\mathcal{L}$ in the last column and use these as the weights of the models. We emphasize that the weights of the models were determined before the time delay of SN Encore was measured and before the unblinding of the $H_0$ inference that we present next.

\subsection{$H_0$ inference}
\label{sec:H0:value}

\begin{figure}
\centering\includegraphics[width=\columnwidth]{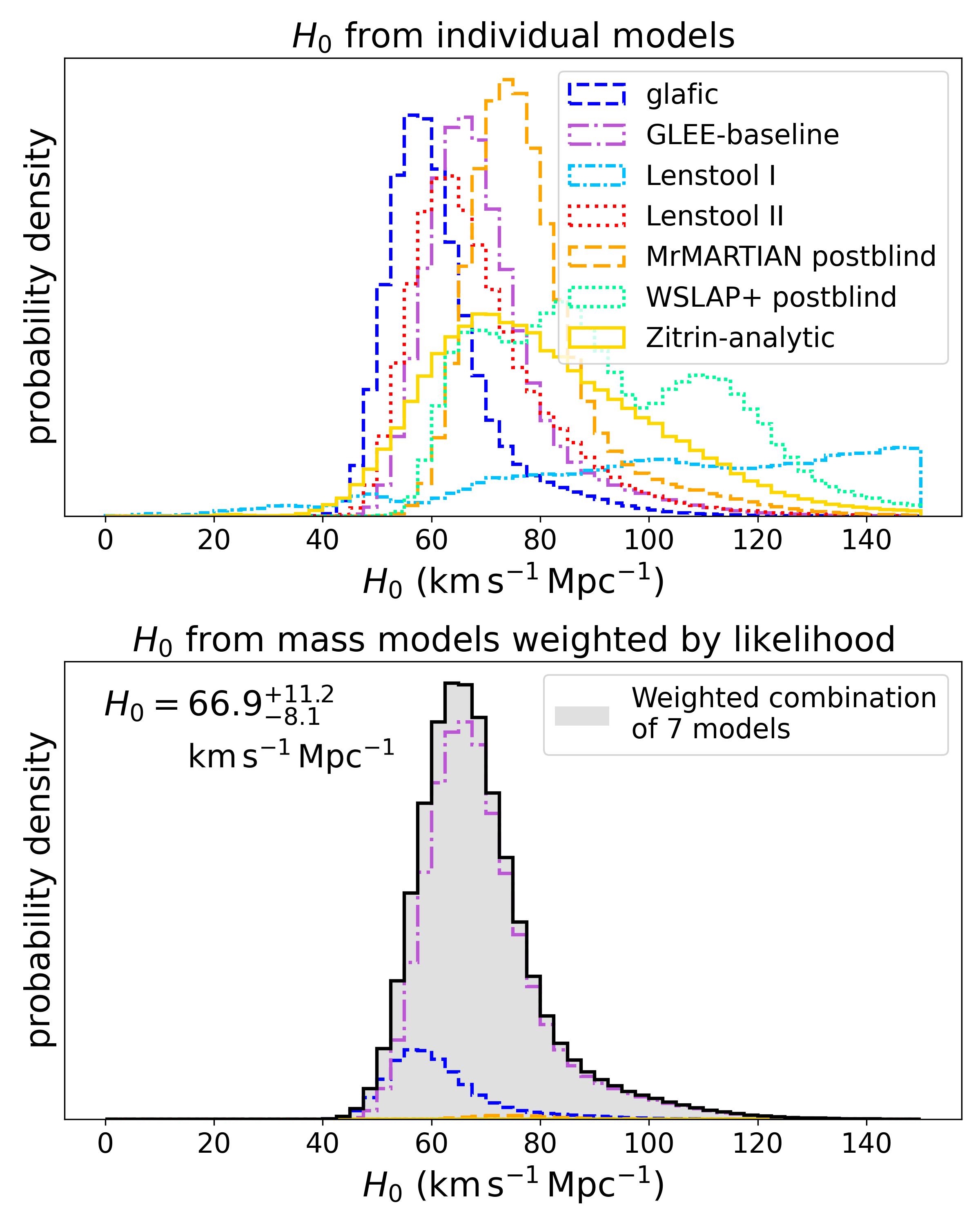}
\caption{$H_0$ inference from SN Encore using the time-delay measurement $\tbaEncore$ by \citet{Pierel+2026}. Top: Inferred $H_0$ distribution from the mass models of the seven modeling teams via blind analysis, where the mass models and the time delay were kept blind to each other throughout and combined after unblinding without modifications. Bottom: $H_0$ from the combined mass model, weighted by each model's likelihood. The weights were determined before unblinding (see Table~\ref{tab:model_stat}). Our $H_0 = \hoEncore\,\kmsMpc$ has most of its uncertainty stemming from the current time-delay measurement, which will be significantly reduced thanks to approved HST and JWST observations of this unique lens system.}
\label{fig:H0final}
\end{figure}

Using the formalism developed in Sect.~\ref{sec:H0-td}, we can compute the $H_0$ distribution for each mass model given the measured time-delay distribution of $\tbaEncore$ by \citet{Pierel+2026}. We adopted a prior range on $H_0$ that is uniform between [0,150]\,$\kmsMpc$, and fix $\Om=1-\OL=0.3$.  In the top panel of Fig.~\ref{fig:H0final}, we show the $H_0$ probability distribution from each of the mass models. In the bottom panel, we show instead the $H_0$ distribution from the combination of the mass models weighted by the likelihood.  The inferred value is $H_0 = \hoEncore\,\kmsMpc$ (68\% CI). Most of the uncertainty on $H_0$ stems from the time-delay uncertainties  of $\sim$9-10\% and the nonlinear relation between $H_0$ and $\tbaEncore$ shown in Fig.~\ref{fig:H0_td_1b1a}. 

\begin{figure*}
\centering\includegraphics[width=0.8\textwidth]{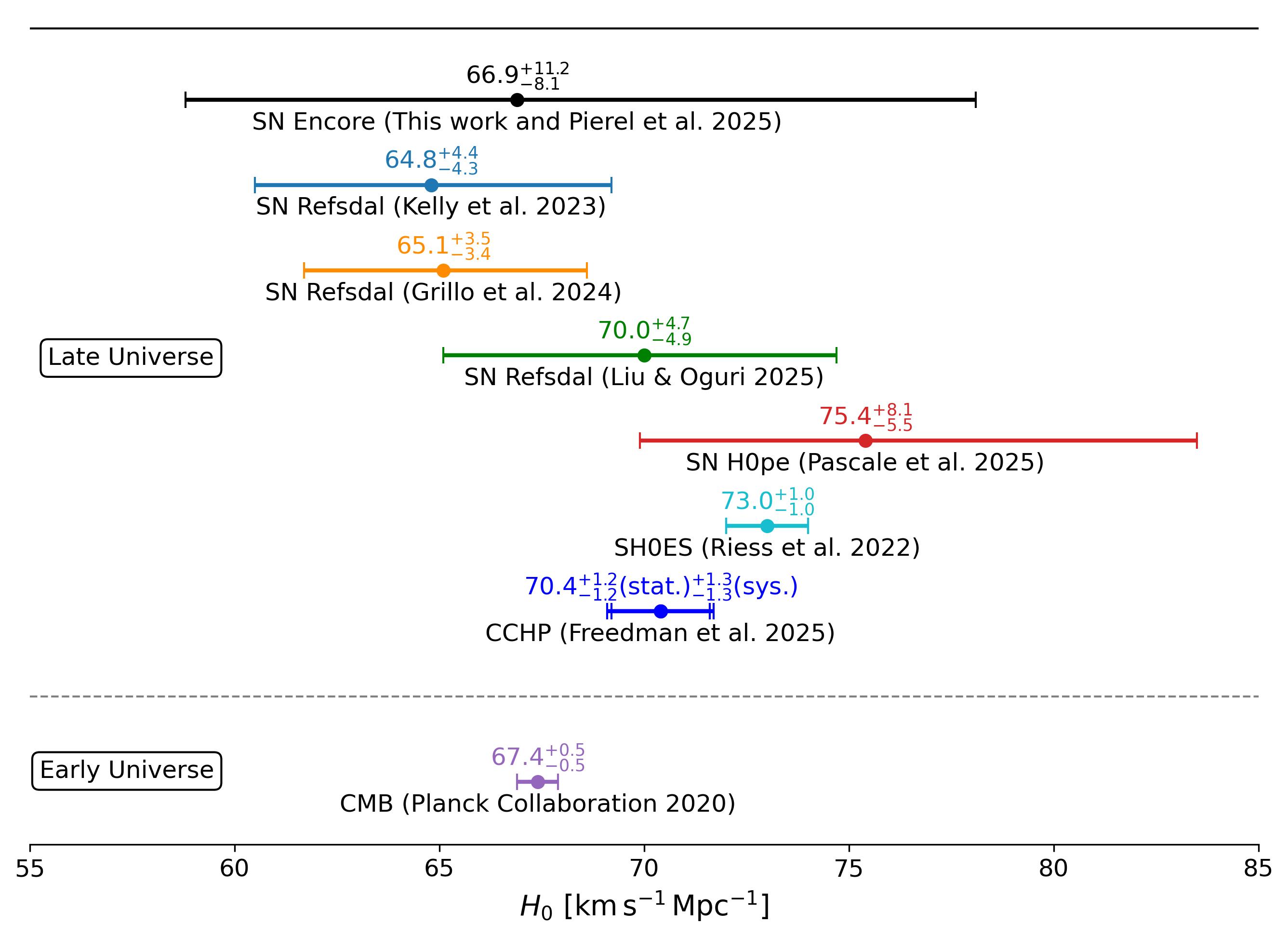}
\caption{Comparison of $H_0$ measurements from lensed SNe, local distance ladders, and the CMB \citep{Planck+2020}.  The $H_0$ values inferred from the lensed SNe shown in the figure assume a flat $\Lambda$CDM cosmological model with $\Om=0.3 = 1-\OL$, except for the measurement by \citet{Grillo+2024} from SN Refsdal, which assumes a more general cosmological model with variable matter density, $\Om$, spatial curvature, and dark energy equation of state, $w$.  With the current uncertainties, the $H_0$ from SN Encore is statistically consistent with measurements from SN Refsdal \citep{Kelly+2023, Grillo+2024, LiuOguri+2025}, SN H0pe \citep{Pascale+2025}, and the two distance-ladder approaches of the SH0ES \citep{Riess+2022} and CCHP \citep{Freedman+2025} programs, and the CMB.}
\label{fig:H0_LSNe_compare}
\end{figure*}

Our $H_0$ measurement is statistically consistent with previous $H_0$ measurements from SN Refsdal \citep{Kelly+2023, Grillo+2024, LiuOguri+2025} and SN H0pe \citep{Pascale+2025}, the only other two lensed SNe from which $H_0$ has been inferred, as shown in Fig.~\ref{fig:H0_LSNe_compare}. Supernova Refsdal is a core-collapse SN \citep{Kelly+2015}, whereas SN H0pe is of type Ia \citep{Frye+2024}.  The measurement of $H_0$ by \citet{Kelly+2023} for SN Refsdal and by \citet{Pascale+2025} for SN H0pe also combined multiple mass models weighted by observables including the image positions, time delays and magnifications of the lensed SN images. In our case with SN Encore, we currently have only a single time-delay measurement $\tbaEncore$; our weighting by the likelihood of the image positions is thus equivalent to weighting by the likelihood of both the image positions and time delay, as explained in Sect.~\ref{sec:H0:combo_model}. Since magnifications can be affected by microlensing and millilensing, especially in the case of SN Encore with the fg near SN image 1b, we did not incorporate the magnification of the SN Encore images into the weighting of the mass models.

Furthermore, our $H_0$ value is also statistically consistent with that of the SH0ES \citep{Riess+2022}, CCHP \citep{Freedman+2025} and \citet{Planck+2020} measurements, as shown in Fig.~\ref{fig:H0_LSNe_compare}. Future observations of SN Encore and Requiem will help significantly reduce our $H_0$ uncertainties for assessing the Hubble tension, which we show next.

\section{Predicted reappearance of SN Requiem and SN Encore}
\label{sec:SNeReappear}

\begin{figure}
        \centering
        
        \includegraphics[width=\columnwidth]{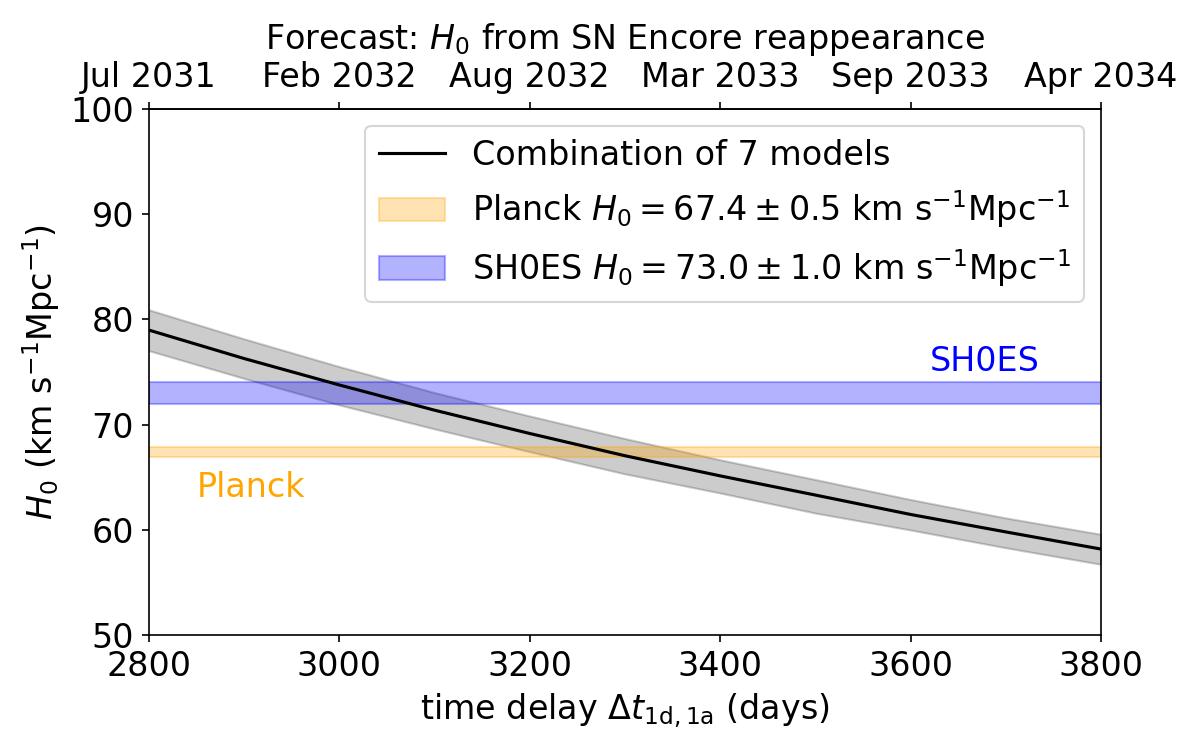}
        \includegraphics[width=\columnwidth]{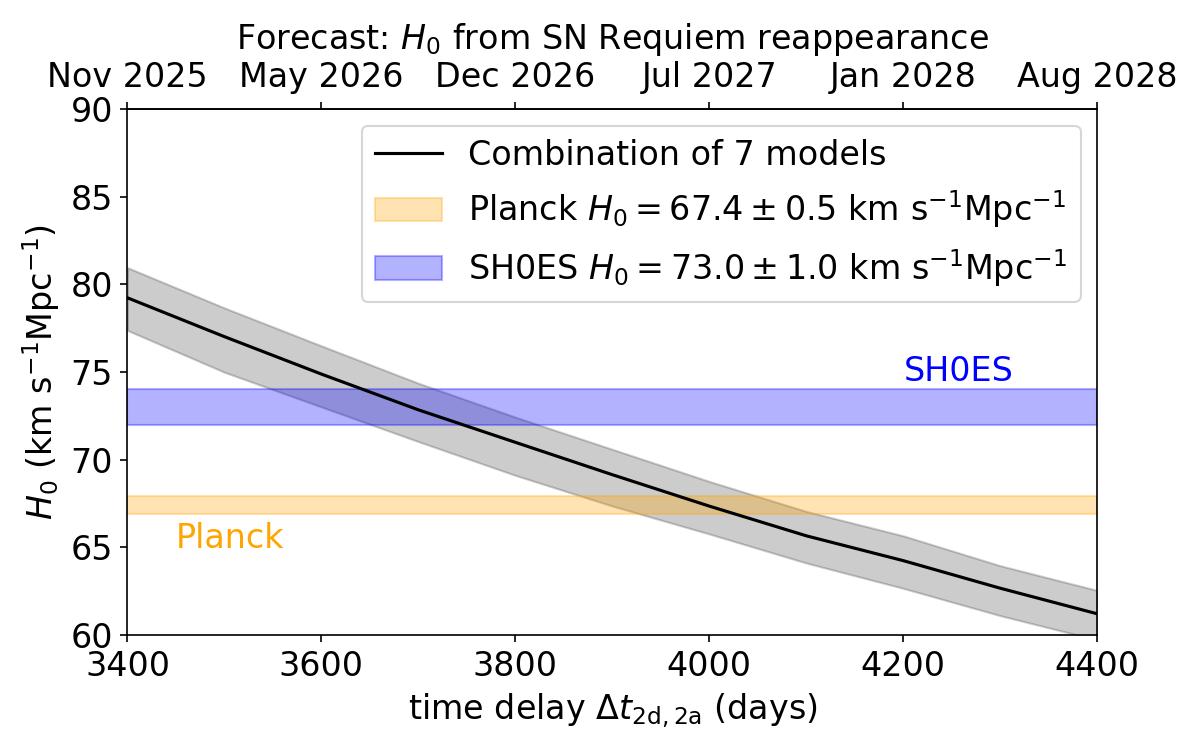}
        \caption{Forecast of the reappearance dates for SN Encore image 1d (top panel) and SN Requiem image 2d (bottom panel). The prediction from the weighted combination of the seven models is shown in black, with shaded 1$\sigma$ uncertainties, assuming a 1\% time-delay uncertainty. The $H_0$ measurements from \citet{Planck+2020} and SH0ES \citep{Riess+2022} with their 1$\sigma$ uncertainties are overlaid. Given the long time delays of images 1d and 2d, detecting the reappearance of each SN enables time-delay measurements with $\sim$1\% uncertainty and $H_0$ with 2-3\% uncertainty. }
        \label{fig:SNeReappear}
\end{figure}

Using the model weighting described in Sect.~\ref{sec:H0:combo_model}, we combined the seven models to predict the reappearance of SN Requiem and SN Encore. Specifically, we weighted the predictions in Tables \ref{tab:snencore} and \ref{tab:snrequiem} (see also Figs.~\ref{fig:H0_td_1d1a} and \ref{fig:H0_td_2d2a}) by the likelihoods $\mathcal{L}$ in Table \ref{tab:model_stat}, to obtain the results shown in Fig.~\ref{fig:SNeReappear}, assuming a time-delay uncertainty of 1\% for each SN. Overlaid on the plots are the measurements of $H_0$ from \citet{Riess+2022} and \citet{Planck+2020}. As noted in Sect.~\ref{sec:H0-td}, a detection of the reappearance of either SN will provide a time-delay measurement with $\sim$1\% uncertainty and allow us to distinguish between the SH0ES and Planck $H_0$ values. In particular, if $H_0=73\,\kmsMpc$ matching SH0ES, then the four best-fitting mass models (with the lowest $\chiimsqmin$ values) predict that the next image (2d) of SN Requiem will appear approximately April-December 2026; if $H_0=67\,\kmsMpc$, then it is predicted to appear  approximately March-November 2027 (based on $1\sigma$ uncertainties).

Recently, \citet{ODonnell+2025} have predicted the appearance of SN Requiem image 2d to be between January 2027 and November 2028, using their mass model and adopting the $H_0$ value of \citet{Planck+2020}. Our prediction for the appearance of image 2d is in the earlier range of the prediction of \citet{ODonnell+2025}. We caution against a direct comparison because the datasets used to constrain the cluster mass model are different between ours and that of \citet{ODonnell+2025}.  In particular, we used 23 multiple images from four distinct background sources, whereas \citet{ODonnell+2025} focused on additional multiple-image systems in the host galaxy of the two SNe and did not include source galaxies at other redshifts. \citet{ODonnell+2025}'s resulting lens model and thus inferred $H_0$ value could be more affected by the mass-sheet degeneracy, as shown in Fig.~A1 of \citet{Grillo2020}.

Using the same model weighting approach, we obtained the magnification of the weighted combination of models for SN Encore image 1a as $\mu_{\rm 1a,mod}=\magEncoreA$ and image 1b as $\mu_{\rm 1b,mod}=\magEncoreB$.  Since SN Encore is of type Ia, \citet{Pierel+2026} determined the absolute magnifications of images 1a and 1b by comparing their observed distance modulii to the distribution of normal non-lensed SNe Ia at the same redshift ($z = 1.95$).  Our model-predicted magnifications 
agree within $1\sigma$ with the measured values from \citet{Pierel+2026} of $| \mu_{\rm 1a,obs}|=21.8^{+9.3}_{-8.6}$ and $|\mu_{\rm 1b,obs}|=32.4^{+11.1}_{-11.0}$. We note that these observed magnifications were not used as constraints in the mass model. This agreement suggests that the effects of microlensing and millilensing for images 1a and 1b are subdominant compared to the effect of photometric uncertainties in measuring the magnification.

\section{Summary}
\label{sec:summary}

We constructed and compared state-of-the-art mass models of the galaxy cluster \macs\ through a blind analysis. This system is unique in featuring two strongly lensed SNe and enabling blind tests of predictions for the reappearance of their future multiple images, with delays of nearly a decade. Throughout our modeling, model comparison, and paper write-up stages (of Sects.~\ref{sec:obs} to \ref{sec:H0-td}), the time-delay measurement of $\tbaEncore$ by \citet{Pierel+2026} was kept blind from the modeling teams. Our key findings are as follows: 

\begin{itemize}
    \item Using HST, JWST, and MUSE data processed by \citet{Pierel2024}, \citet{Ertl+2025}, and \citet{Granata2025}, we identified eight gold systems consisting of 23 multiple images, and one silver system with three multiple images.
    \item Seven teams modeled the high-quality gold image systems using six independent modeling software, with four parametric software and two free-form software.
    \item By comparing the model-predicted image positions to the observed multiple image positions, four teams obtained a $\chiimsqmin$ of $<$25, with two teams obtaining $\chiimsqmin<11$.
    \item Predictions of the positions, magnifications, and time delays of SN Encore and SN Requiem from the different teams using the gold sample of images are in good agreement, mostly within $\sim2\sigma$ from the teams who obtained $\chiimsq<25$.  
    \item Two teams, \glafic\ and \mrmartian, constructed mass models for the gold plus silver samples, finding broad consistency between the results of the gold-only sample and the gold plus silver sample.  
    \item Based on the model predictions and $\Om = 1-\OL=0.3$, we predict the $H_0$ values for a range of hypothetical time delays $\tbaEncore$ of SN Encore with a 10\% time-delay uncertainty as illustration. For 10\% uncertainty on $\tbaEncore$, we expect an $H_0$ uncertainty as low as 12\% from an individual mass model.
    \item All seven mass models predict that SN Encore and SN Requiem will have multiple images appearing in the future, with time delays of $\sim 3000$ days and $\sim 4000$ days, respectively.  These images, if detected, would provide delays with $\sim1\%$ uncertainty, potentially yielding $H_0$ with $\sim$2-3\% uncertainty from this single lens cluster.
    \item By using the new time-delay measurement of $\tbaEncore$ by \citet{Pierel+2026} and weighting our mass models by the likelihood of the observed image positions, we obtain $H_0 = \hoEncore\,\kmsMpc$. The $\sim$$14\%$ (statistical plus systematic) uncertainty is currently dominated by that of the time delay.
    \item The next image of SN Requiem is expected to appear in the very near future. If $H_0=73\,\kmsMpc$, the four best-fitting mass models (with the lowest $\chiimsqmin$ values) predict it will appear approximately April-December 2026; if $H_0=67\,\kmsMpc$, they predict it will appear  approximately March-November 2027 (based on $1\sigma$ uncertainties).

\end{itemize}

Our work establishes a robust framework for future cosmological analyses utilizing strongly lensed SNe, particularly SN Requiem and SN Encore. An approved \jwst\ program (Proposal ID 8799, PIs: Suyu, Pierel) will acquire additional NIRCam imaging of this cluster, providing a template image of the giant arcs after images 1a, 1b, and 1c fade. These images are expected to improve the photometric measurements of SN Encore multiple images through difference imaging. Furthermore, an approved \hst\ program (Proposal ID 18069, PIs: Pierel, Suyu) will monitor this cluster to catch the SN Requiem reappearance, yielding a time delay $\tdaRequiem$ with $\sim1\%$ precision (given the long delay). These future observations will reduce time-delay uncertainties and increase the precision of future $H_0$ measurements from this unique system with two lensed SNe.  

\begin{acknowledgements}
We would like to thank the knowledgeable and meticulous referee whose insightful and constructive comments improved the presentation of our work.
SHS, SE, EM and HW thank the Max Planck Society for support through the Max Planck Fellowship for SHS. This project has received funding from the European Research Council (ERC) under the European Union's Horizon 2020 research and innovation programme (LENSNOVA: grant agreement No 771776).
This work is supported in part by the Deutsche Forschungsgemeinschaft (DFG, German Research Foundation) under Germany's Excellence Strategy -- EXC-2094 -- 390783311. 
AA acknowledges financial support through the Beatriz Galindo programme and the project PID2022-138896NB-C51 (MCIU/AEI/MINECO/FEDER, UE), Ministerio de Ciencia, Investigación y Universidades. 
CG, PB, GG, and PR acknowledge support from the Italian Ministry of University and Research through grant PRIN-MIUR 2020SKSTHZ. This work was supported by JSPS KAKENHI Grant Numbers JP25H00662, JP25H00672, JP22K21349.
SC and MJJ acknowledge support for the current research from the National Research Foundation (NRF) of Korea under the programs 2022R1A2C1003130, RS-2023-00219959, and RS-2024-00413036. 
SS has received funding from the European Union’s Horizon 2022 research and innovation programme under the Marie Skłodowska-Curie grant agreement No 101105167 — FASTIDIoUS.
EEH is supported by a Gates Cambridge Scholarship (\#OPP1144).
CL acknowledges support under DOE award DE-SC0010008 to Rutgers University and support from HST-GO-17474.
MM acknowledges support by the SNSF (Swiss National Science Foundation) through return CH grant P5R5PT\_225598.
This work is based on observations made with the NASA/ESA/CSA James Webb Space Telescope. These observations are associated with programs GO-2345 and DD-6549. Support for programs GO-2345 and DD-6549 was provided by NASA through a grant from the Space Telescope Science Institute, which is operated by the Association of Universities for Research in Astronomy, Inc., under NASA contract NAS 5-03127.
This research is based on observations made with the NASA/ESA {\it Hubble Space Telescope} obtained from the Space Telescope Science Institute, which is operated by the Association of Universities for Research in Astronomy, Inc., under NASA contract NAS 5–26555. These observations are associated with programs 14496, 15663 and 16264.
The data described here may be obtained from the MAST archive at 
\url{https://doi.org/10.17909/snj9-an10}.

\end{acknowledgements}

\bibliographystyle{aa}
\bibliography{main}

\makeatletter
\if@longauth
  \par\kern6pt\hrule\kern6pt
  \institutename
  \@longauthfalse
\fi
\makeatother

\clearpage

\begin{appendix}

\section{Silver multiple-image system}
\label{app:silver_data}

\begin{figure}
        \centering
        \includegraphics[width = 0.9\columnwidth]{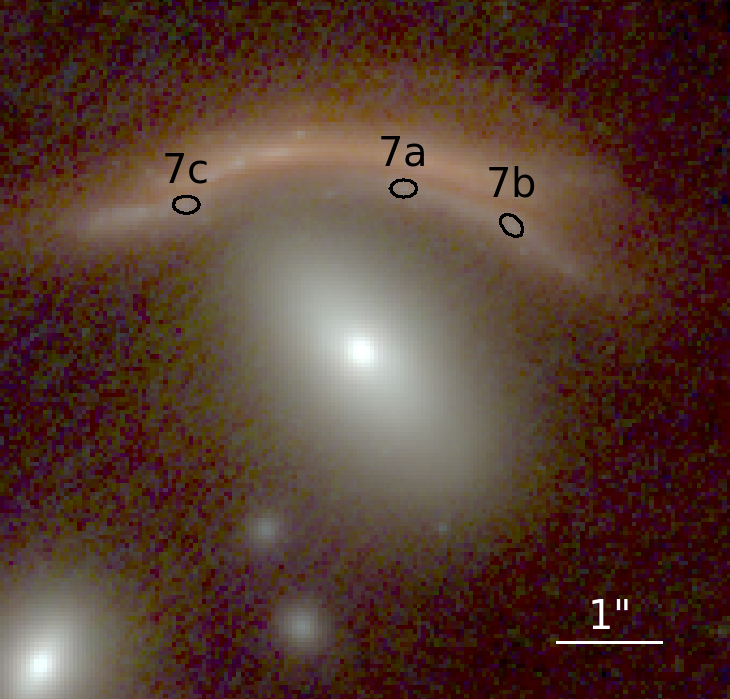}
        \caption{Silver multiple-image system, showing an arc composed of two blended background sources lensed into multiple images around a cluster member. We identified three multiple-image positions associated with one of the background sources. The system is classified as silver due to its ``likely'' rather than ``secure'' spectroscopic redshift and the ambiguous association of the redshift with either of the two blended sources.  }
        \label{fig:silver_sys}
\end{figure}

The silver image system, shown in Fig.~\ref{fig:silver_sys}, has all its multiple images around a single cluster member and therefore mostly depends on the total mass distribution of the specific cluster member rather than the global cluster mass distribution.  By extracting the MUSE spectrum in a mask covering the arc and cross-correlating it with spectral templates, we found the spectrosopic redshift to likely be $z_{\rm s,silver} = 1.945$.

We identified three multiple images of one of the background sources in the silver system, and list their positions and uncertainties in Table \ref{tab:silver_pos}.

\begin{table*}
  \caption{Observed multiple image positions and positional uncertainties of the silver system.}
  \label{tab:silver_pos}
  \begin{center}
  \begin{tabular}{lccc}
    \addlinespace[-13pt]
    \hline
    \addlinespace[2pt]
    Silver image & RA [\degree] & Dec [\degree] & Elliptical positional uncertainty \\
    \addlinespace[2pt]
    \hline
    \addlinespace[2pt]
     7a & 24.5152267  & $-$21.9188283 & 0.12, 0.08, 90.0 \\
     7b & 24.5149262   & $-$21.9189236 &  0.12, 0.08, 47.3 \\
     7c & 24.5158325 & $-$21.9188706 & 0.12, 0.08, 90.0 \\
    \hline
    \addlinespace[-13pt]
  \end{tabular}
  \end{center}
  \label{tab:img_systems}
  \tablefoot{The RA and Dec are in the ICRS system, and the relative elliptical errors are in the format: major axis $['']$, minor axis $['']$, position angle [\degree], with position angle measured counterclockwise from north toward east.}
  \end{table*}

\section{Overview of mass models}

We summarize in Table \ref{tab:model_overview} the setup of the seven independent modeling approaches, which are described in Sect.~\ref{sec:mass_model}.

\begin{sidewaystable*}
\centering
\caption{Overview of the lens mass models from the seven independent modeling approaches using the gold image positions as constraints.}
\resizebox{\columnwidth}{!}{
\begin{tabular}{ccccccccc}
\hline\hline \noalign{\smallskip}
 Description & \glafic & \GLEE & \GLEE \texttt{-baseline} & \texttt{Lenstool} I & \texttt{Lenstool} II & \zitrin & \texttt{MrMARTIAN} & \wslap \\ 
\noalign{\smallskip} \hline \noalign{\smallskip}
 BCG profile & Hernquist & dPIE & dPIE & dPIE & dPIE & dPIE & --- (hybrid) & --- (free-form)   \\ \noalign{\smallskip}
 Number of DM halos & 1 & 2 & 2 & 1 & 1 & 2 & 1 + grid & --- (free-form) \\ \noalign{\smallskip}
 DM halo profile & NFW & various$^{\mathrm{a}}$ & isothermal & dPIE & dPIE & isothermal & NFW + grid & --- (free-form) \\ \noalign{\smallskip}
 External shear & Yes & No & No & Yes & Yes & No & --- & --- \\ \noalign{\smallskip}
 Galaxy scaling relation &  &  &  &  & & & & \\
 $\alpha$ & 0.25 & 0.25$^{\mathrm{b}}$ & 0.25$^{\mathrm{b}}$ & 0.25 & 0.21$^{\mathrm{b}}$ & 0.25 & --- & ---  \\ 
 $\beta$ & 0.5 & 0.7 & 0.7 & 0.5 & 0.71 & 0.5 & --- & ---  \\ 
 Galaxies not in scaling relation & fg, JF-1 & multiple$^{\mathrm{c}}$ & multiple$^{\mathrm{c}}$ & multiple$^{\mathrm{d}}$ & multiple$^{\mathrm{e}}$ & multiple$^{\mathrm{f}}$ & --- & --- \\ \noalign{\smallskip}
 positional error boost factor & 1 & 1 & 1 & 2.2 & 1.5 & 3 & 1 & 1  \\ 
\noalign{\smallskip} \hline
\end{tabular}
}
\tablefoot{
\begin{list}{}{}
\item[$^{\mathrm{a}}$] Various halo profiles were explored, including isothermal, power-law and NFW; one model also included a mass sheet.
\item[$^{\mathrm{b}}$] The parameter $\alpha$ of the scaling relation  (Eq.~\ref{eq:fj}) was optimized using cluster member galaxies with velocity dispersion measurements in the \GLEE\ \citep{Granata2025, Ertl+2025} and \lenstooltwo\ \citep{Acebron2025} models, whereas $\alpha$ was fixed to the tabulated values in the other models.  
\item[$^{\mathrm{c}}$] The multiple objects excluded from the galaxy scaling relation of the \GLEE\ model are: BCG, object ID=116, JF-1, JF-2, JF-3, fg, and bg \citep[see][]{Ertl+2025}.
\item[$^{\mathrm{d}}$] The multiple objects excluded from the galaxy scaling relation of the \lenstoolone\ model are: object at $({\rm RA, ~Dec}) = (24\fdg51662331, -21\fdg92448273)$, JF-1, JF-2, JF-3, and fg.
\item[$^{\mathrm{e}}$] The multiple objects excluded from the galaxy scaling relation of the \lenstooltwo\ model are: JF-1, JF-2, JF-3, fg, and bg.
\item[$^{\mathrm{f}}$] Four objects excluded from the galaxy scaling relation of the \zitrin\ model are located at (RA, Dec) = $(24\fdg5114258, -21\fdg9213075), (24\fdg5129665, -21\fdg9203688)$, $(24\fdg5156305, -21\fdg9227688)$, and $(24\fdg5076146, -21\fdg9271504)$.
\end{list}
}
\label{tab:model_overview}
\end{sidewaystable*}

\section{Comparison of the models using the gold plus silver samples of lensed images}
\label{app:silver}
In addition to the gold samples, both \glafic\ and MrMARTIAN included System 7, classified as a silver sample, in their lens modeling. See Sect.~\ref{sec:mass_model:glafic} and Sect.~\ref{sec:mass_model:mars} for more details on the lens modeling using the combined gold plus silver samples in \glafic\ and MrMARTIAN, respectively. For both models, the results based on the gold plus silver samples are broadly consistent with those derived from the gold-only samples. In the case of \glafic\, the differences in magnification and time delays between the two models are minimal, typically much smaller than the 1$\sigma$ statistical uncertainties. While the MrMARTIAN results also remain consistent within the 1$\sigma$ uncertainties, they exhibit larger differences—on the order of ~10–30\%—in magnification and time delays. We suspect that, because MrMARTIAN relies solely on the positions of multiple images, the limited constraints near the added analytic halo allow its parameters to vary more significantly within the prior range. We therefore conclude that the inclusion of the silver sample does not significantly affect the time-delay predictions or the estimation of the value of $H_0$ in either modeling approach.

\section{$H_0$ forecast with alternative \lenstoolone\ model}
\label{app:LenstoolI_alt}

\begin{figure*}
\centering
\includegraphics[scale=0.4]{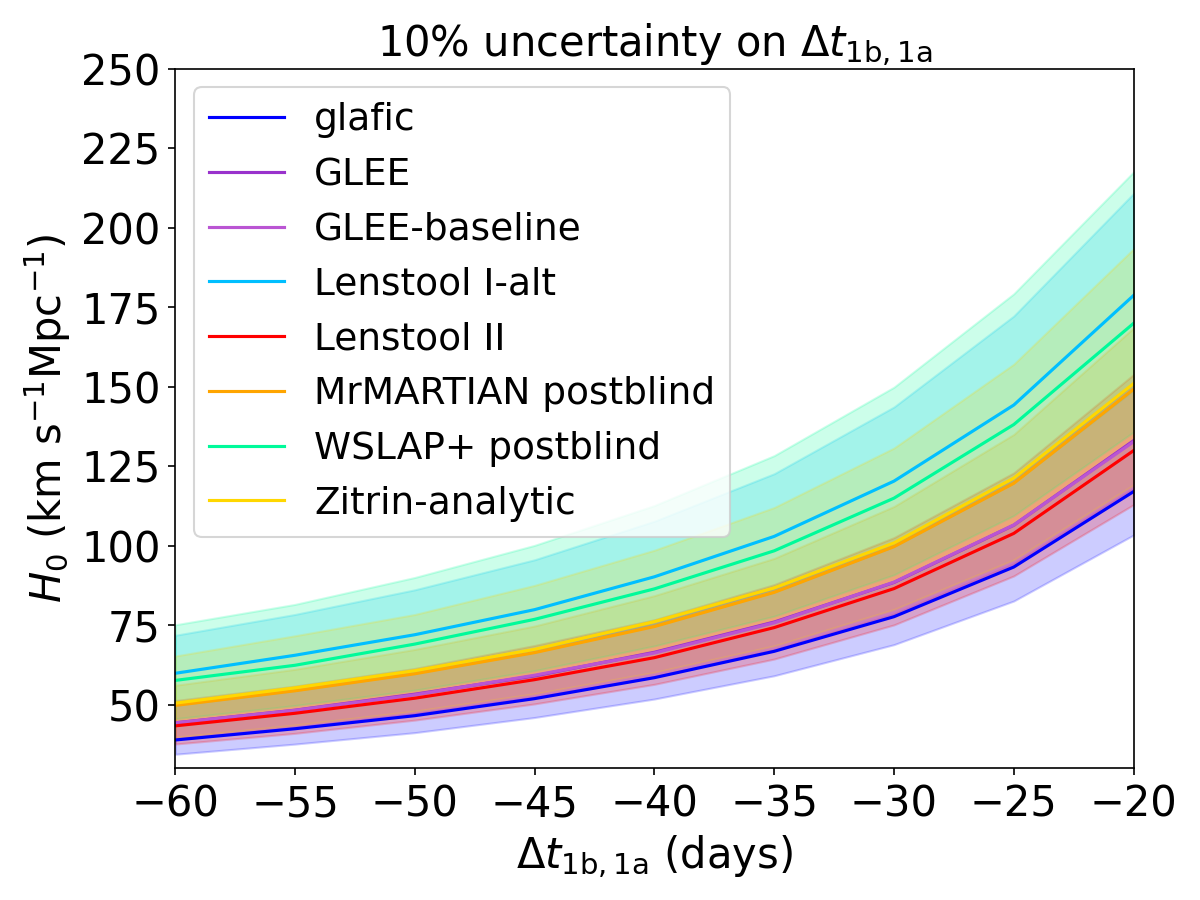}
\includegraphics[scale=0.3]{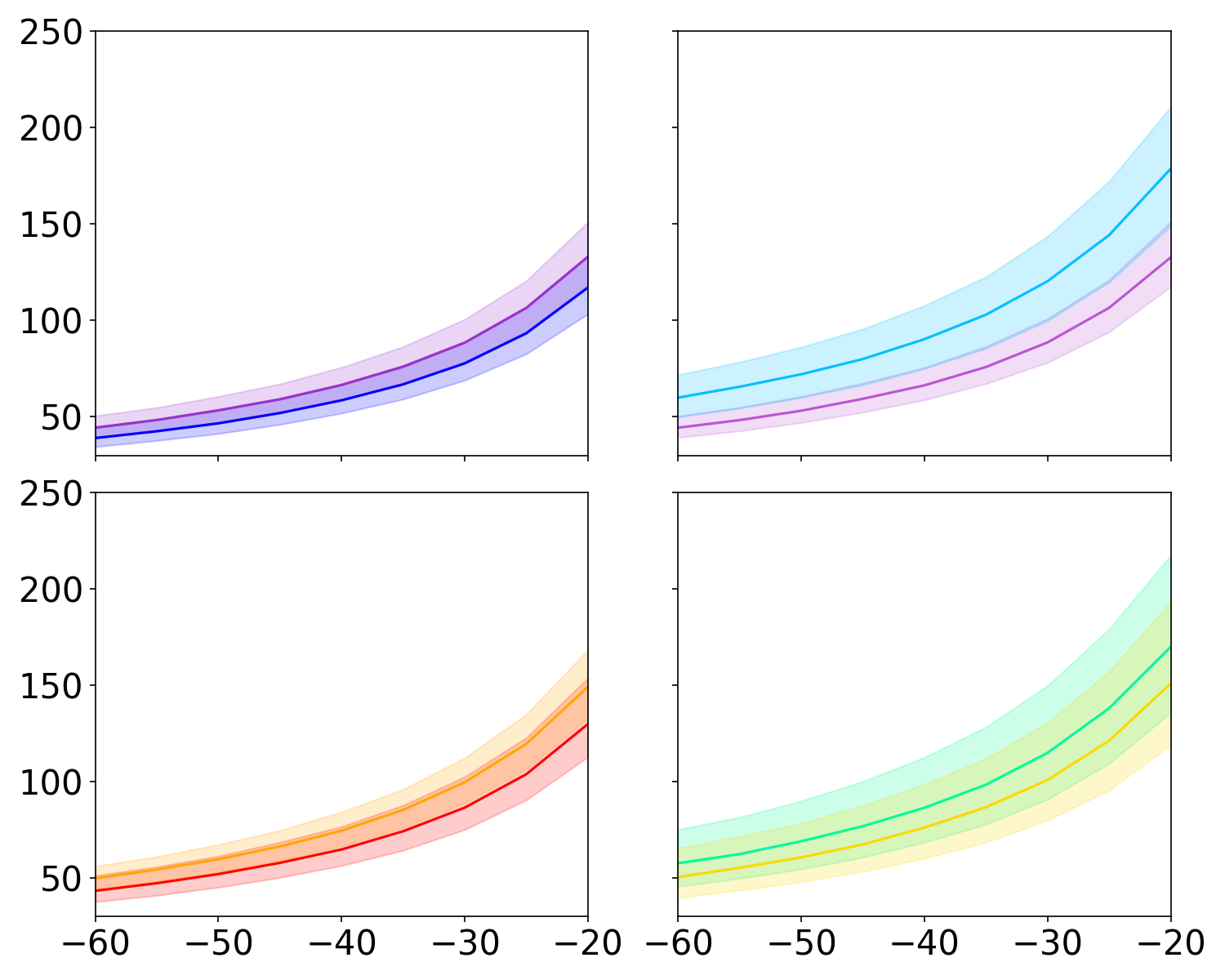}
\caption{Same format as Fig.~\ref{fig:H0_td_1b1a}, but including the post-blind alternative \lenstoolone\ model results, which illustrates improved agreement with the other models.}
\label{fig:H0_td_1b1a_10percent_altLenstoolI}
\end{figure*}

As discussed in Section \ref{sec:mass_model:kamieneski}, the original blinded \lenstoolone\ model has very large uncertainties in the model-predicted time delays (in particular, for the pair with the shortest delay, $\tbaEncore$). After the lens model unblinding, these large relative uncertainties were found to be the result of strong degeneracies between time delays and the ellipticity of the mass density profile describing JF-2\textemdash the jellyfish galaxy found to be the most massive in this model\textemdash with minimal covariance for other quantities such as the magnifications or image positions. An alternative postblind model (\lenstoolone-alt) was tested to examine the effect of parameterizing JF-2 as a circular SIS, thereby reducing the relative uncertainty on the model-predicted $\tbaEncore$ value (and similarly for $\tbaRequiem$). In contrast to the postblind results of \mrmartian\ and \wslap\ that were due to simple data post-processing issues, the alternate \lenstoolone\ model changed its model setup; therefore, it is considered as a new model constructed after unblinding, and is not independent for the comparison with the other models in Sects.~\ref{sec:comparison} and \ref{sec:H0-td}. The postblind \lenstoolone-alt model is shown in Fig. \ref{fig:H0_td_1b1a_10percent_altLenstoolI}, primarily to demonstrate the consistency between models and rule out errors in processing the \lenstoolone\ model results.
The alternative model yields only minimal change in the $\tdaEncore$ and $\tdaRequiem$ predictions (not shown); the latter remains an outlier as in Fig.~\ref{fig:H0_td_2d2a} primarily because only a small minority of posterior samples predict an image 2d for both the \lenstoolone\ and \lenstoolone-alt models.

\setcounter{figsave}{\value{figure}}

\begingroup
  \renewcommand{\thefigure}{F.\arabic{figure}}
  \renewcommand{\theHfigure}{F.\arabic{figure}}
  \setcounter{figure}{0}

\begin{figure*}
\centering
\includegraphics[scale=0.56]{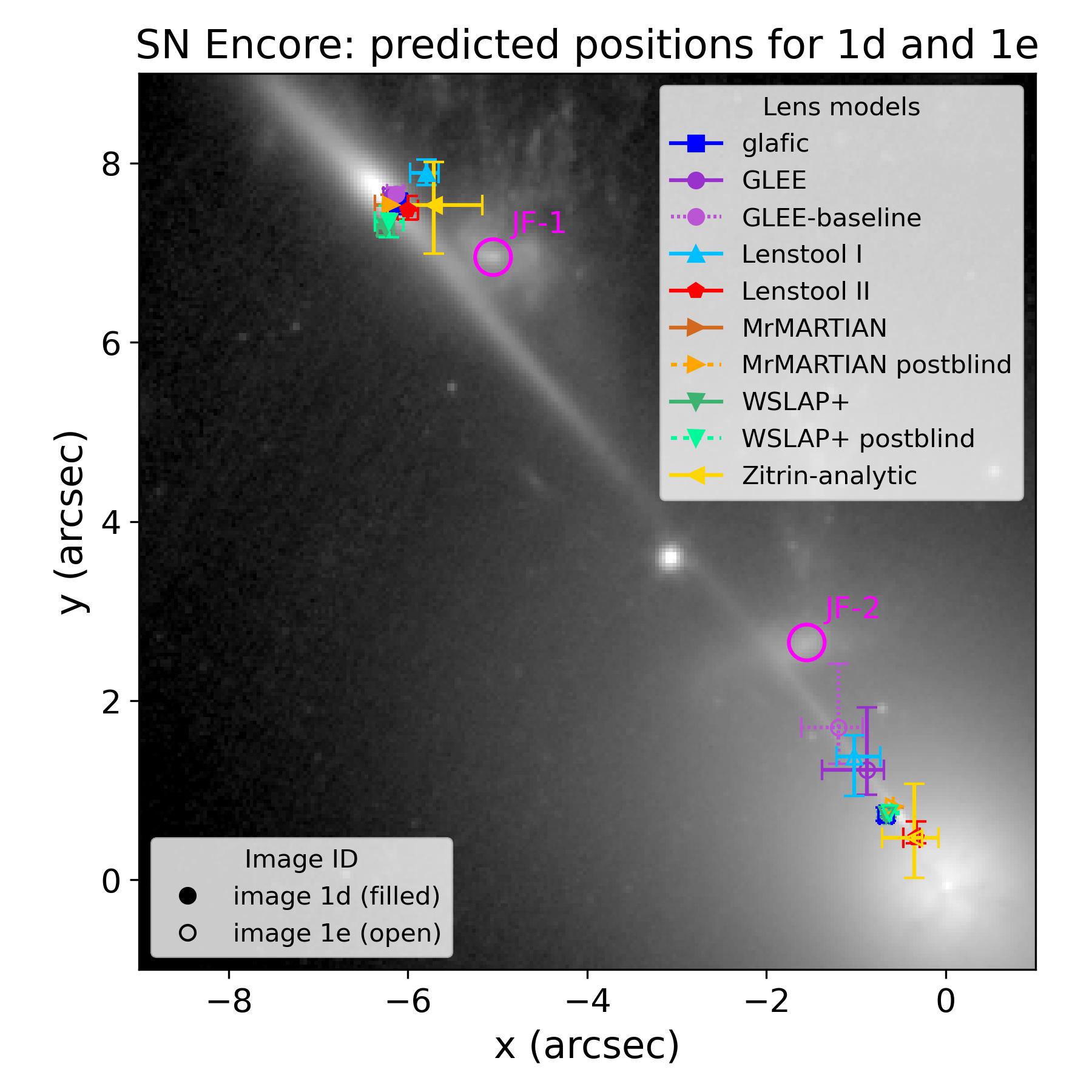}
\includegraphics[scale=0.56]{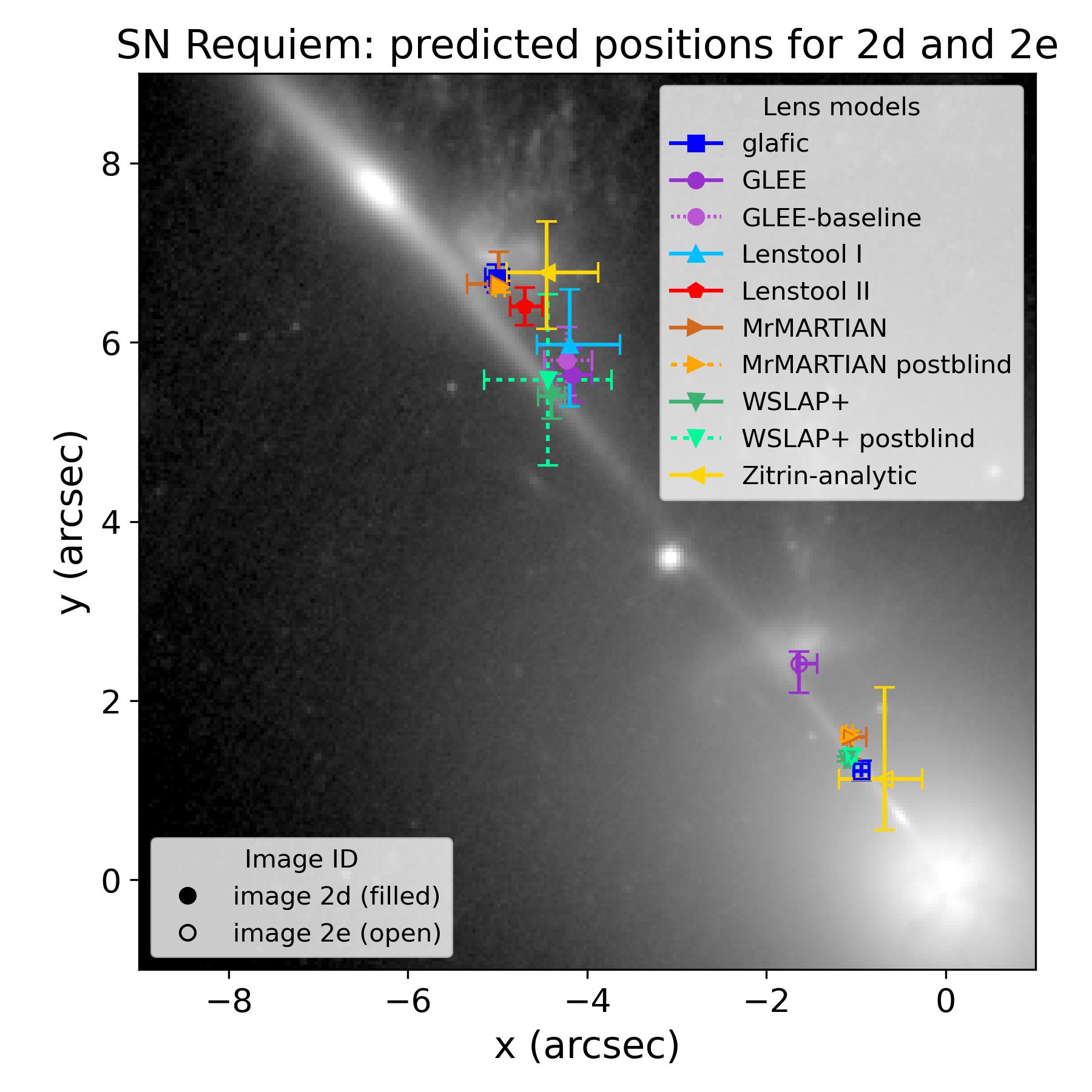}
\caption{Predicted positions of SN Encore images 1d and 1e (left panel) and SN Requiem images 2d and 2e (right panel), overlaid on the JWST F200W image. The 1d-1e and 2d-2e pairs of images are merging pairs along the radial arc. The jellyfish galaxies JF-1 and JF-2 (open magenta circles in the left panel) and the BCG alter the positions of the predicted images -- even changing image multiplicity, as not all models predict images 1e and 2e (see Tables \ref{tab:snencore} and \ref{tab:snrequiem}).}
\label{fig:Encore_Requiem_pos_radial_arc}
\end{figure*}

\endgroup

\setcounter{figure}{\value{figsave}}

\section{Model predictions of SN Encore and SN Requiem}
\label{app:predictions}

In Tables \ref{tab:snencore} and \ref{tab:snrequiem}, we list the model-predicted positions, magnifications and time delays of the multiple images of SN Encore and SN Requiem from the independent lens mass models.

\begin{sidewaystable*}
\centering
\caption{Model-predicted values of position, magnification, and time delay of the multiple images of SN Encore.}
\resizebox{\columnwidth}{!}{
\begin{tabular}{cccccccccccc}
\hline\hline \noalign{\smallskip}
 & & \glafic & \GLEE & \GLEE \texttt{-baseline} & \texttt{Lenstool} I & \texttt{Lenstool} II & \texttt{MrMARTIAN} & \texttt{MrMARTIAN} & \wslap & \wslap & \zitrin \\ 
  & &  &  &  &  &  &  & \texttt{postblind} &  & \texttt{postblind} & \\ 
\noalign{\smallskip} \hline \noalign{\smallskip}
$x_{\rm 1a}$$^{\mathrm{a}}$ & (arcsec) & $8.406^{+0.008}_{-0.008}$ & $8.40^{+0.04}_{-0.04}$ & $8.41^{+0.04}_{-0.04}$ & $8.40^{+0.00}_{-0.00}$ & $8.36^{+0.05}_{-0.06}$ & $8.367^{+0.006}_{-0.014}$ & $8.369^{+0.006}_{-0.005}$ & $8.07^{+0.03}_{-0.03}$ & $7.50^{+0.11}_{-0.11}$ & $8.39^{+0.03}_{-0.36}$ \\
$y_{\rm 1a}$$^{\mathrm{a}}$ & (arcsec) & $-16.10^{+0.04}_{-0.04}$ & $-16.07^{+0.01}_{-0.01}$ & $-16.07^{+0.01}_{-0.01}$ & $-16.07^{+0.00}_{-0.00}$ & $-16.07^{+0.02}_{-0.01}$ & $-16.084^{+0.003}_{-0.003}$ & $-16.083^{+0.003}_{-0.002}$ & $-16.26^{+0.03}_{-0.03}$ & $-16.50^{+0.04}_{-0.04}$ & $-16.13^{+0.03}_{-0.18}$ \\
$\mu_{\rm 1a}$ & & $-27.5^{+2.1}_{-2.3}$ & $-28.0^{+2.5}_{-3.7}$ & $-26.6^{+2.5}_{-2.7}$ & $-19.9^{+1.6}_{-1.7}$ & $-26.1^{+3.3}_{-4.4}$ & $-31.9^{+1.4}_{-7.0}$ & $-31.6^{+1.1}_{-1.3}$ & $-53^{+14}_{-14}$ & $-46.9^{+8.1}_{-8.1}$ & $-18.3^{+4.4}_{-4.4}$ \\
Occurrence$^{\mathrm{b}}$ & & 100 & 100 & 100 & 100 & 100 & 100 & 100 & 100 & 100 & 100\\
\noalign{\smallskip} \hline \noalign{\smallskip}
$x_{\rm 1b}$$^{\mathrm{a}}$ & (arcsec) & $0.230^{+0.007}_{-0.006}$ & $0.23^{+0.04}_{-0.04}$ & $0.22^{+0.04}_{-0.04}$ & $0.10^{+0.27}_{-1.69}$ & $0.24^{+0.05}_{-0.06}$ & $0.182^{+0.004}_{-0.011}$ & $0.183^{+0.004}_{-0.003}$ & $-0.09^{+0.03}_{-0.03}$ & $-0.72^{+0.02}_{-0.02}$ & $ 0.23^{+0.03}_{-0.21}$\\
$y_{\rm 1b}$$^{\mathrm{a}}$ & (arcsec) & $-17.85^{+0.04}_{-0.04}$ & $-17.862^{+0.007}_{-0.008}$ & $-17.861^{+0.008}_{-0.008}$ & $-17.80^{+0.10}_{-0.05}$ & $-17.87^{+0.01}_{-0.01}$ & $-17.859^{+0.001}_{-0.002}$ & $-17.859^{+0.001}_{-0.001}$ & $-17.86^{+0.03}_{-0.03}$ & $-17.79^{+0.03}_{-0.03}$ & $-17.89^{+0.03}_{-0.03}$ \\
$\mu_{\rm 1b}$ & & $48.4^{+7.6}_{-5.5}$ & $40.0^{+7.6}_{-6.3}$ & $39.6^{+5.9}_{-6.1}$ & $21.7^{+5.6}_{-3.6}$ & $31.8^{+6.1}_{-5.9}$ & $27.5^{+14.7}_{-2.3}$ & $26.7^{+2.2}_{-1.7}$ & $31.8^{+8.8}_{-8.8}$ & $21.3^{+3.8}_{-3.8}$ & $18.3^{+3.9}_{-4.1}$ \\
$\Delta t_{\rm 1b,1a}$ & (days) & $-32.4^{+2.2}_{-2.5}$ & $-37.1^{+2.7}_{-2.6}$ & $-36.9^{+2.6}_{-3.0}$ & $-75^{+55}_{-54}$ & $-35.6^{+3.1}_{-4.3}$ & $-40.6^{+9.8}_{-2.9}$ & $-41.8^{+2.7}_{-2.1}$ & $-112^{+32}_{-32}$ & $-47.2^{+9.6}_{-9.6}$ & $-40.2^{+7.5}_{-11.1}$ \\
Occurrence$^{\mathrm{b}}$ & & 100 & 100 & 100 & 100 & 100 & 100 & 100 & 100 & 100 & 100\\
\noalign{\smallskip} \hline \noalign{\smallskip}
$x_{\rm 1c}$$^{\mathrm{a}}$ & (arcsec) & $19.43^{+0.06}_{-0.06}$ & $19.43^{+0.03}_{-0.03}$ & $19.48^{+0.03}_{-0.04}$ & $19.54^{+0.09}_{-0.06}$ & $19.50^{+0.06}_{-0.04}$ & $19.406^{+0.006}_{-0.003}$ & $19.404^{+0.003}_{-0.003}$ & $19.57^{+0.03}_{-0.03}$ & $19.56^{+0.02}_{-0.02}$ & $19.56^{+0.15}_{-0.18}$ \\
$y_{\rm 1c}$$^{\mathrm{a}}$ & (arcsec) & $-2.54^{+0.19}_{-0.19}$ & $-2.62^{+0.13}_{-0.13}$ & $-2.64^{+0.10}_{-0.09}$ & $-2.63^{+0.31}_{-0.28}$ & $-2.80^{+0.17}_{-0.17}$ & $-2.59^{+0.02}_{-0.02}$ & $-2.59^{+0.02}_{-0.01}$ & $-2.32^{+0.03}_{-0.03}$ & $-2.31^{+0.10}_{-0.10}$ & $-2.14^{+0.36}_{-0.66}$ \\
$\mu_{\rm 1c}$ & & $11.4^{+0.6}_{-0.6}$ & $12.1^{+1.0}_{-0.9}$ & $11.0^{+0.8}_{-0.9}$ & $7.7^{+0.7}_{-0.5}$ & $9.5^{+1.0}_{-1.3}$ & $11.1^{+3.4}_{-0.4}$ & $11.0^{+0.3}_{-0.3}$ & $15.6^{+1.6}_{-1.6}$ & $15.3^{+1.0}_{-1.0}$ & $9.1^{+1.6}_{-1.5}$ \\
$\Delta t_{\rm 1c,1a}$ & (days) & $-325^{+21}_{-23}$ & $-326^{+21}_{-20}$ & $-340^{+16}_{-18}$ & $-381^{+81}_{-130}$ & $-352^{+20}_{-27}$ & $-307^{+75}_{-10}$ & $-310^{+10}_{-9}$ & $-410^{+32}_{-32}$ & $-268^{+19}_{-19}$ & $-368^{+69}_{-76}$ \\
Occurrence$^{\mathrm{b}}$ & & 100 & 100 & 100 & 100 & 99.8 & 100 & 100 & 100 & 100 & 100\\
\noalign{\smallskip} \hline \noalign{\smallskip}
$x_{\rm 1d}$$^{\mathrm{a}}$ & (arcsec) & $-6.12^{+0.11}_{-0.11}$ & $-6.18^{+0.10}_{-0.09}$ & $-6.14^{+0.09}_{-0.10}$ & $-5.79^{+0.13}_{-0.19}$ & $-6.00^{+0.11}_{-0.12}$ & $-6.195^{+0.009}_{-0.174}$ & $-6.190^{+0.007}_{-0.010}$ & $-6.25^{+0.09}_{-0.09}$ & $-6.21^{+0.16}_{-0.16}$ & $-5.71^{+0.54}_{-0.51}$ \\
$y_{\rm 1d}$$^{\mathrm{a}}$ & (arcsec) & $7.55^{+0.11}_{-0.12}$ & $7.63^{+0.09}_{-0.09}$ & $7.66^{+0.07}_{-0.09}$ & $7.89^{+0.15}_{-0.14}$ & $7.48^{+0.15}_{-0.11}$ & $7.53^{+0.12}_{-0.01}$ & $7.530^{+0.008}_{-0.008}$ & $7.28^{+0.09}_{-0.09}$ & $7.35^{+0.18}_{-0.18}$ & $7.53^{+0.48}_{-0.54}$\\
$\mu_{\rm 1d}$ & & $-4.7^{+0.3}_{-0.3}$ & $-2.3^{+0.3}_{-0.4}$ & $-2.6^{+0.5}_{-0.7}$ & $-1.7^{+0.5}_{-0.4}$ & $-3.3^{+0.6}_{-0.9}$ & $-4.8^{+0.2}_{-1.3}$ & $-4.7^{+0.1}_{-0.2}$ & $-613^{+1815}_{-613}$ & $-5.9^{+4.5}_{-4.5}$ & $-4.2^{+0.7}_{-0.7}$ \\
$\Delta t_{\rm 1d,1a}$ & (days) & $3035^{+74}_{-75}$ & $3175^{+57}_{-63}$ & $3143^{+73}_{-67}$ & $3465^{+202}_{-152}$ & $3193^{+97}_{-81}$ & $2994^{+33}_{-160}$ & $3009^{+24}_{-27}$ & $3544^{+242}_{-242}$ & $3160^{+106}_{-106}$ & $3125^{+294}_{-290}$ \\
Occurrence$^{\mathrm{b}}$ & & 100 & 100 & 100 & 100 & 100 & 100 & 100 & 100 & 100 & 100\\
\noalign{\smallskip} \hline \noalign{\smallskip}
$x_{\rm 1e}$$^{\mathrm{a}}$ & (arcsec) & $-0.67^{+0.07}_{-0.07}$ & $-0.88^{+0.19}_{-0.50}$ & $-1.20^{+0.27}_{-0.41}$ & $-1.02^{+0.29}_{-0.19}$ & $-0.33^{+0.04}_{-0.14}$ & $-0.582^{+0.004}_{-0.005}$ & $-0.584^{+0.003}_{-0.004}$ & $-0.66^{+0.03}_{-0.03}$ & $-0.63^{+0.03}_{-0.03}$ & $-0.35^{+0.27}_{-0.36}$ \\
$y_{\rm 1e}$$^{\mathrm{a}}$ & (arcsec) & $0.73^{+0.08}_{-0.07}$ & $1.23^{+0.70}_{-0.27}$ & $1.70^{+0.71}_{-0.41}$ & $1.38^{+0.24}_{-0.44}$ & $0.47^{+0.18}_{-0.06}$ & $0.815^{+0.009}_{-0.008}$ & $0.818^{+0.007}_{-0.005}$ & $0.73^{+0.03}_{-0.03}$ & $0.75^{+0.01}_{-0.01}$ & $0.47^{+0.60}_{-0.45}$\\
$\mu_{\rm 1e}$ & & $0.32^{+0.06}_{-0.05}$ & $2.8^{+4.3}_{-1.3}$ & $5.3^{+4.6}_{-2.4}$ & $-1.2^{+0.8}_{-0.3}$ & $-0.4^{+0.2}_{-0.1}$ & $0.65^{+0.05}_{-0.04}$ & $0.66^{+0.04}_{-0.03}$ & $0.39^{+0.03}_{-0.03}$ & $0.39^{+0.02}_{-0.02}$ & $0.5^{+0.4}_{-0.2}$ \\
$\Delta t_{\rm 1e,1a}$ & (days) & $3401^{+73}_{-72}$ & $3414^{+48}_{-52}$ & $3376^{+59}_{-67}$ & $3482^{+2800}_{-908}$ & $3497^{+74}_{-97}$ & $3325^{+30}_{-131}$ & $3336^{+19}_{-21}$ & $4036^{+93}_{-93}$ & $3436^{+47}_{-47}$ & $3628^{+250}_{-262}$ \\
Occurrence$^{\mathrm{b}}$ & & 100 & 35 & 13 & 4 & 14.8 & 100 & 100 & 100 & 100 & 89\\
\noalign{\smallskip} \hline
\end{tabular}
}
\tablefoot{
Median values and 68$\%$ confidence level intervals from the MCMC chains for a fixed cosmological model (flat $\Lambda$CDM with $\Hc = 70\,\kmsMpc$ and $\Om = 0.3 = 1-\OL$). 
\begin{list}{}{}
\item[$^{\mathrm{a}}$]With respect to the BCG luminosity center (RA$=24\fdg5157032$, Dec$=-21\fdg9254791$) and positive in the west and north directions.
\item[$^{\mathrm{b}}$]Percentage.
\end{list}
}
\label{tab:snencore}
\end{sidewaystable*}

\begin{sidewaystable*}
\centering
\caption{Model-predicted values of position, magnification, and time delay of the multiple images of SN Requiem.}
\resizebox{\columnwidth}{!}{
\begin{tabular}{cccccccccccc}
\hline\hline \noalign{\smallskip}
 & & \glafic & \GLEE & \GLEE \texttt{-baseline} & \texttt{Lenstool} I & \texttt{Lenstool} II & \texttt{MrMARTIAN} & \texttt{MrMARTIAN} & \wslap & \wslap & \zitrin \\ 
   & &  &  &  &  &  &  & \texttt{postblind} &  & \texttt{postblind} & \\ 
\noalign{\smallskip} \hline \noalign{\smallskip}
$x_{\rm 2a}$$^{\mathrm{a}}$ & (arcsec) & $11.28^{+0.05}_{-0.04}$ & $11.30^{+0.07}_{-0.07}$ & $11.29^{+0.07}_{-0.07}$ & $11.33^{+0.00}_{-0.00}$ & $11.39^{+0.08}_{-0.09}$ & $11.250^{+0.009}_{-0.007}$ & $11.250^{+0.007}_{-0.008}$ & $11.03^{+0.03}_{-0.03}$ & $12.75^{+0.42}_{-0.42}$ & $11.30^{+0.03}_{-0.21}$ \\
$y_{\rm 2a}$$^{\mathrm{a}}$ & (arcsec) & $-15.41^{+0.05}_{-0.05}$ & $-15.50^{+0.04}_{-0.04}$ & $-15.49^{+0.05}_{-0.05}$ & $-15.48^{+0.00}_{-0.00}$ & $-15.38^{+0.06}_{-0.06}$ & $-15.528^{+0.005}_{-0.006}$ & $-15.529^{+0.005}_{-0.005}$ & $-15.64^{+0.03}_{-0.03}$ & $-14.40^{+0.32}_{-0.32}$ & $-15.51^{+0.03}_{-0.12}$ \\
$\mu_{\rm 2a}$ & & $-30.1^{+2.7}_{-3.2}$ & $-39.8^{+5.4}_{-6.5}$ & $-33.8^{+3.5}_{-4.2}$ & $-25.4^{+2.7}_{-2.9}$ & $-30.9^{+3.5}_{-4.7}$ & $-47.5^{+6.3}_{-25.5}$ & $-45.6^{+6.0}_{-5.5}$ & $-215^{+101}_{-101}$ & $-139^{+75}_{-75}$ & $-26.2^{+7.0}_{-9.2}$ \\
Occurrence$^{\mathrm{b}}$ & & 100 & 100 & 100 & 100 & 100 & 100 & 100 & 100 & 100 & 100\\
\noalign{\smallskip} \hline \noalign{\smallskip}
$x_{\rm 2b}$$^{\mathrm{a}}$ & (arcsec) & $1.97^{+0.04}_{-0.04}$ & $1.99^{+0.08}_{-0.07}$ & $2.01^{+0.08}_{-0.07}$ & $2.46^{+0.32}_{-0.40}$ & $1.99^{+0.10}_{-0.12}$ & $2.055^{+0.007}_{-0.007}$ & $2.056^{+0.006}_{-0.006}$ & $2.25^{+0.03}_{-0.03}$ & $1.89^{+0.03}_{-0.03}$ & $2.46^{+0.90}_{-0.51}$\\
$y_{\rm 2b}$$^{\mathrm{a}}$ & (arcsec) & $-18.57^{+0.05}_{-0.06}$ & $-18.61^{+0.02}_{-0.02}$ & $-18.62^{+0.02}_{-0.02}$ & $-18.67^{+0.05}_{-0.05}$ & $-18.63^{+0.02}_{-0.02}$ & $-18.600^{+0.006}_{-0.002}$ & $-18.601^{+0.001}_{-0.001}$ & $-18.60^{+0.03}_{-0.03}$ & $-18.63^{+0.03}_{-0.03}$ & $-18.69^{+0.12}_{-0.06}$ \\
$\mu_{\rm 2b}$ & & $25.7^{+1.9}_{-1.8}$ & $24.3^{+2.9}_{-2.2}$ & $23.8^{+2.3}_{-2.3}$ & $19.2^{+1.7}_{-1.7}$ & $22.3^{+2.7}_{-3.4}$ & $19.5^{+7.0}_{-1.3}$ & $19.0^{+1.2}_{-0.8}$ & $48^{+18}_{-18}$ & $33.0^{+7.0}_{-7.0}$ & $18.5^{+4.1}_{-3.8}$ \\
$\Delta t_{\rm 2b,2a}$ & (days) & $-49.8^{+4.1}_{-4.4}$ & $-47.9^{+3.5}_{-3.6}$ & $-49.7^{+3.3}_{-3.4}$ & $-74^{+52}_{-57}$ & $-52.2^{+3.8}_{-5.5}$ & $-29.8^{+2.4}_{-2.2}$ & $-30.0^{+1.8}_{-2.1}$ & $-61^{+14}_{-14}$ & $-31^{+11}_{-11}$ & $-35.0^{+11.2}_{-14.2}$ \\
Occurrence$^{\mathrm{b}}$ & & 100 & 100 & 100 & 100 & 100 & 100 & 100 & 100 & 100 & 100\\
\noalign{\smallskip} \hline \noalign{\smallskip}
$x_{\rm 2c}$$^{\mathrm{a}}$ & (arcsec) & $18.74^{+0.06}_{-0.06}$ & $18.84^{+0.03}_{-0.03}$ & $18.86^{+0.03}_{-0.03}$ & $18.81^{+0.12}_{-0.09}$ & $18.88^{+0.03}_{-0.04}$ & $18.815^{+0.003}_{-0.019}$ & $18.816^{+0.002}_{-0.002}$ & $18.85^{+0.03}_{-0.03}$ & $18.63^{+0.03}_{-0.03}$ & $18.54^{+0.24}_{-0.36}$ \\
$y_{\rm 2c}$$^{\mathrm{a}}$ & (arcsec) & $-6.66^{+0.06}_{-0.05}$ & $-6.65^{+0.07}_{-0.07}$ & $-6.64^{+0.08}_{-0.08}$ & $-6.86^{+0.25}_{-0.28}$ & $-6.67^{+0.11}_{-0.12}$ & $-6.715^{+0.005}_{-0.059}$ & $-6.713^{+0.006}_{-0.005}$ & $-6.58^{+0.03}_{-0.03}$ & $-7.02^{+0.07}_{-0.07}$ & $-7.22^{+0.51}_{-0.75}$ \\
$\mu_{\rm 2c}$ & & $15.8^{+1.5}_{-1.2}$ & $15.7^{+1.9}_{-1.5}$ & $14.9^{+1.7}_{-1.6}$ & $10.6^{+1.4}_{-1.0}$ & $14.1^{+1.6}_{-2.3}$ & $15.7^{+13.0}_{-0.9}$ & $15.2^{+0.9}_{-0.6}$ & $30.5^{+8.1}_{-8.1}$ & $30.3^{+6.7}_{-6.7}$ & $13.4^{+4.2}_{-3.4}$ \\
$\Delta t_{\rm 2c,2a}$ & (days) & $-80.4^{+5.4}_{-6.0}$ & $-78.4^{+7.2}_{-8.7}$ & $-86.3^{+7.4}_{-7.1}$ & $-83^{+51}_{-84}$ & $-87.7^{+7.9}_{-8.2}$ & $-70.6^{+36.4}_{-6.2}$ & $-72.6^{+5.5}_{-6.2}$ & $-95^{+43}_{-43}$ & $-59^{+12}_{-12}$ & $-58.3^{+25.2}_{-33.9}$ \\
Occurrence$^{\mathrm{b}}$ & & 100 & 100 & 100 & 100 & 100 & 100 & 100 & 100 & 100 & 100\\
\noalign{\smallskip} \hline \noalign{\smallskip}
$x_{\rm 2d}$$^{\mathrm{a}}$ & (arcsec) & $-5.01^{+0.13}_{-0.13}$ & $-4.17^{+0.22}_{-0.21}$ & $-4.23^{+0.28}_{-0.25}$ & $-4.20^{+0.57}_{-0.36}$ & $-4.70^{+0.20}_{-0.16}$ & $-4.99^{+0.06}_{-0.35}$ & $-4.97^{+0.05}_{-0.05}$ & $-4.40^{+0.15}_{-0.15}$ & $-4.44^{+0.71}_{-0.71}$ & $-4.45^{+0.57}_{-0.45}$ \\
$y_{\rm 2d}$$^{\mathrm{a}}$ & (arcsec) & $6.72^{+0.15}_{-0.16}$ & $5.64^{+0.29}_{-0.30}$ & $5.80^{+0.37}_{-0.39}$ & $5.98^{+0.62}_{-0.69}$ & $6.40^{+0.21}_{-0.21}$ & $6.65^{+0.36}_{-0.09}$ & $6.62^{+0.07}_{-0.07}$ & $5.40^{+0.25}_{-0.25}$ & $5.58^{+0.96}_{-0.96}$ & $6.78^{+0.57}_{-0.63}$\\
$\mu_{\rm 2d}$ & & $-4.6^{+0.4}_{-0.4}$ & $-2.2^{+0.4}_{-0.5}$ & $-2.3^{+0.4}_{-0.7}$ & $-1.3^{+0.3}_{-0.3}$ & $-2.2^{+0.6}_{-0.6}$ & $-5.1^{+0.2}_{-0.3}$ & $-5.1^{+0.3}_{-0.2}$ & $-18^{+15}_{-15}$ & $-7.7^{+2.8}_{-2.8}$ & $-4.0^{+0.7}_{-1.1}$ \\
$\Delta t_{\rm 2d,2a}$ & (days) & $3711^{+80}_{-76}$ & $3867^{+64}_{-63}$ & $3863^{+64}_{-64}$ & $4516^{+1439}_{-1662}$ & $3957^{+135}_{-108}$ & $3788^{+49}_{-318}$ & $3806^{+36}_{-40}$ & $4177^{+300}_{-300}$ & $3826^{+60}_{-60}$ & $4111^{+268}_{-281}$ \\
Occurrence$^{\mathrm{b}}$ & & 100 & 100 & 100 & 16 & 99.8 & 100 & 100 & 100 & 100 & 98\\
\noalign{\smallskip} \hline \noalign{\smallskip}
$x_{\rm 2e}$$^{\mathrm{a}}$ & (arcsec) & $-0.94^{+0.08}_{-0.09}$ & $-1.63^{+0.20}_{-0.03}$ & & & & $-1.06^{+0.17}_{-0.05}$ & $-1.07^{+0.03}_{-0.04}$ & $-1.10^{+0.03}_{-0.03}$ & $-1.05^{+0.05}_{-0.05}$ & $-0.68^{+0.42}_{-0.51}$ \\
$y_{\rm 2e}$$^{\mathrm{a}}$ & (arcsec) & $1.21^{+0.12}_{-0.10}$ & $2.41^{+0.14}_{-0.32}$ & & & & $1.59^{+0.06}_{-0.24}$ & $1.62^{+0.04}_{-0.04}$ & $1.35^{+0.03}_{-0.03}$ & $1.38^{+0.07}_{-0.07}$ & $1.13^{+1.02}_{-0.57}$\\
$\mu_{\rm 2e}$ & & $0.57^{+0.12}_{-0.10}$ & $1.55^{+0.77}_{-1.15}$ & & & & $1.34^{+0.18}_{-0.20}$ & $1.41^{+0.12}_{-0.10}$ & $1.72^{+0.39}_{-0.39}$ & $1.86^{+0.55}_{-0.55}$ & $0.8^{+0.9}_{-0.4}$ \\
$\Delta t_{\rm 2e,2a}$ & (days) & $3891^{+80}_{-78}$ & $3876^{+55}_{-50}$ & & & & $3910^{+35}_{-260}$ & $3921^{+30}_{-31}$ & $4491^{+78}_{-78}$ & $3909^{+34}_{-34}$ & $4346^{+270}_{-277}$ \\
Occurrence$^{\mathrm{b}}$ & & 100 & 3 & 0 & 0  & 0 & 100 & 100 & 100 & 100 & 94\\
\noalign{\smallskip} \hline
\end{tabular}
}
\tablefoot{
Median values and 68$\%$ confidence-level intervals from the MCMC chains for a fixed cosmological model (flat $\Lambda$CDM with $\Hc = 70\,\kmsMpc$ and $\Om = 0.3 = 1-\OL$). 
\begin{list}{}{}
\item[$^{\mathrm{a}}$]With respect to the BCG luminosity center (RA$=24\fdg5157032$, Dec$=-21\fdg9254791$) and positive in the west and north directions.
\item[$^{\mathrm{b}}$]Percentage.
\end{list}
}
\label{tab:snrequiem}
\end{sidewaystable*}

\section{Predicted positions of future images of SN Encore and SN Requiem}
\label{app:SN_Encore_Requiem_de_pos}

To understand the origin of the scatter in the predicted positions of images 1d, 1e, 2d and 2e in Figs.~\ref{fig:SN_Encore_pos_mag_td} and \ref{fig:SN_Requiem_pos_mag_td}, we show in Fig.~\ref{fig:Encore_Requiem_pos_radial_arc} the same predicted positions of images 1d and 1e of SN Encore overlaid on the JWST F200W image in the left-hand panel and similarly for images 2d and 2e of SN Requiem in the right-hand panel.  The image pair 1d and 1e and the pair 2d and 2e are along the radial arc, where there are nearby foreground cluster galaxies, notably the jellyfish galaxies JF-1 and JF-2.  These two jellyfish galaxies and the BCG, depending on their mass distribution, can alter the critical curves and change the predicted image configuration (as noted in Sect.~\ref{sec:mass_model:lenstoolII}).  Given the limited number of image position constraints (only system 3) near these jellyfish galaxies, the jellyfish galaxies' mass distributions are currently not well constrained, leading to the different number of image multiplicity and the scatter in the predicted images near the radial arc of SNe Encore and Requiem from the different modeling teams.

\end{appendix}

\end{document}